\documentclass[12pt]{article}
\usepackage{amsmath,longtable,amssymb}
\usepackage{rotating}
\usepackage{graphicx}
\usepackage[authoryear, round, sort]{natbib}

\newcommand\T{\rule{0pt}{9 ex}}
\newcommand\B{\rule[-1ex]{0pt}{0pt}}
\newcommand{\mbold}[1]{\mbox{\boldmath $#1$}}

\begin{document}
\title{Optimal signal processing in small stochastic biochemical networks}
\author{
Etay Ziv$^{1,2}$
Ilya Nemenman,$^{5}$
Chris Wiggins,$^{3,4}$\\
\normalsize{
$^1$College of Physicians \& Surgeons,}\\
\normalsize{
$^2$Department of Biomedical Engineering,}\\
\normalsize{
$^3$Department of Applied Physics and Applied Mathematics,}\\
\normalsize{
$^4$Center for Computational Biology and Bioinformatics;}\\
\normalsize{
Columbia University, New York, NY 10027, USA}\\
\normalsize{
$^5$Computer, Computational and Statistical Sciences Division;}\\
\normalsize{Los Alamos National Laboratory, Los Alamos, NM 87545, USA}
}

\maketitle

\date{}

\begin{abstract}
  We quantify the influence of the topology of a transcriptional regulatory network
  on its ability to process environmental signals. By posing the
  problem in terms of information theory, we may do this without specifying the function performed by the
  network.  Specifically, we study the maximum mutual information
  between the input (chemical) signal and the output (genetic)
  response attainable by the network in the context of an analytic
  model of particle number fluctuations.  We perform this analysis for
  all biochemical circuits, including various feedback loops, that can
  be built out of 3 chemical species, each under the control of one
  regulator. We find that a generic network, constrained to low
  molecule numbers and reasonable response times, can transduce more
  information than a simple binary switch and, in fact, manages to
  achieve close to the optimal information transmission fidelity.
  These high-information solutions are robust to tenfold changes in
  most of the networks' biochemical parameters; moreover they are
  easier to achieve in networks containing cycles with an odd number
  of negative regulators (overall negative feedback) due to their
  decreased molecular noise (a result which we derive analytically).
  Finally, we demonstrate that a single circuit can support multiple
  high-information solutions. These findings suggest a potential
  resolution of the ``cross-talk'' dilemma as well as the previously
  unexplained observation that transcription factors which undergo
  proteolysis are more likely to be auto-repressive. 
  \end{abstract}
  
\section{Introduction}
\label{sec:intro}

Genetic regulatory networks act as biochemical computing machines in
cells, measuring, processing, and integrating inputs from the cellular
and extracellular environment and producing appropriate outputs in the
form of gene expression.  The behavior of these networks is not
deterministic; many of the molecules involved in genetic regulation
(e.g., DNA, mRNA, transcription factors) are found in low copy
numbers, and are thus subject to severe copy number fluctuations. In
living cells, the origins and consequences of stochasticity are
well-studied \citep{walters95,hume94,hume00,levin03,hasty00,arkin97};
one can analyze propagation of noise through cellular networks
\citep{pedraza} and disambiguate noise from different sources (e.g.,
{\em intrinsic} vs.\ {\em extrinsic}
\citep{elowitz02a,elowitz02b,oshea04}).  Surprisingly, cells function
in the presence of noise remarkably well, often performing close to
the physical limits imposed by the discreteness of the signals and the
signal processing machinery \citep{berg77,bialek05}. 

From an information-theoretic perspective \citep{shannon}, noise
intrinsic to the gene network presents an obstacle for signal
transduction and biochemical computation: with too much noise, the
information about the state of the environment (the {\em signal}) may
be lost. While it may be possible to build stable biochemical switches
even in the presence of just tens of copies of a transcription factor
\citep{bialek00}, networks often need to exhibit a more detailed
response.  As an example, the well-studied {\em p53} module responds
to ionizing radiation in a ``digital'' manner
\citep{lahav04,gustavo05}, initiating a number of disparate cellular
responses, including cell cycle arrest, apoptosis, and induction of
cellular differentiation, among others \citep{vogelstein00}. The
\emph{p53} module must not only transduce a simple binary answer (was
there DNA damage or not?), but also more specific information (What
was the damage? How severe? What should be done about it?) It is not
evident that a few tens of molecules, whose abundance is subject to
intrinsic copy number fluctuations, can successfully perform this
task.  Of note, a series of recent papers studying the effect of
single allele loss in various tumor supressor genes, including {\em
  p53}, challenge the classic two-hit model of tumorigenesis
\citep{knudson} by demonstrating dosage-dependent modulation of
phenotype (see \citep{fodd_dose, hohenstein_dose, ghosh_dose} and
references therein).

The above example is just one of many instances of ``cross-talk''--a
perplexing dilemma observed across many cellular signaling systems in
which a single noisy biomolecular species, presumably existing in just
two states (active/inactive), is able to transmit complicated
information. Perhaps the most well-studied example of cross-talk
occurs in the protein signaling mitogen-activated protein kinase
(MAPK) pathways.  MAPK cascades transduce multiple stimuli from the
environment into distinct genetic programs. Many of these signals are
transmitted by common components \citep{fink98}. Specificity can be
established by sequestration including cell type, subcellular localization, temporal, or with scaffold proteins \citep{schwartz,bardwell01,fink97,fink98,brunner94}. In some cases sequestration mechanisms are not available and specificity is achieved via signaling kinetics. For example, in mammal pheochromoctyoma cells, ligands triggering distinct programs (proliferate or differentiate) activate the same receptor tyrosine kinase pathway but with different amplitudes \citep{marshall95}. In fact, by increasing or decreasing receptor expression, the wrong program may be also be initiated \citep{schlessinger} implying that tight control of kinetics may be required and that uncontrolled noise may be disastrous. Cross-talk can be exploited by
cancer cells to initiate uncontrolled cell growth \citep{garcia_cancer}
even in the presence of chemotherapeutic agents targeting individual
signaling pathways.

In this Chapter we present an information-theoretic measure of circuit
quality which is independent of an assumed network's function and
demonstrate that generic small networks under biological constraints
can transduce more information than a simple binary switch, often
coming close to the \emph{optimal} transmission fidelity, which we
calculate numerically and analytically from physical constraints. In
choosing a general information-theoretic quality measure, we obviate
the problem of requiring prior knowledge of the function of the
network
\citep{leibler_osc,leibler_rob,cluzel,serrano,elowitz_rep,kollmann05,wagner_rob,alon02a},
which is obviously network-specific and often unknown, and the related
problem that a given network may perform multiple functions
\citep{arkin05,wall_ffl,ingram06}.  We also demonstrate that the
presence of an odd number of negative regulators in a feedback loop
confers an advantage to the circuit in terms of noise regulation and
thus information transmission. Finally, we show that the ability to
transduce information reliably is insensitive to most large (tenfold)
deviations of a network's kinetic parameters.

\subsection{Measure of quality of a biochemical computation}
\label{sec:guet}
To motivate our approach, consider the experimental set-up of Guet et
al.\ \citep{calin02}.  Probing experimentally the relationship between
structure and function in transcriptional networks, Guet and coworkers
built a combinatorial library of $3$-gene circuits and looked at the
steady-state expression $G$ of a reporter gene (GFP), coupled to one
of the genes in the circuit, in response to four different chemical
inputs $C$, namely two binary states of two different chemicals. The
circuits acted as transducers, converting chemical signals into
genetic response.  They found that some networks could perform
different behaviors (that is, behave as different logic gates), while
others could achieve only one particular function. Of note, while some
circuits responded differently to different inputs, for other
circuits, the reporter expression did not depend on the chemical input
state. The latter are clearly ``broken circuits,'' transducing no
information about the inputs.

The actual number of GFP reporters in each cell
clearly is not repeatable due to the stochastic nature of the involved
cellular machinery. For this reason, the input-output relation for a
circuit should be described in terms of the probability distribution
$P(c,g)\equiv P(C=c,G=g)$, where $c$ stands for particular chemical
states, and $g$ measures the number of reporter molecules. Then a
natural measure of a circuit's quality is the \emph{mutual
  information} between its inputs and outputs \citep{shannon}
\begin{equation}
I(C,G)=\int dc\,dg\, P(c,g) \log \frac{P(c,g)}{P(c)P(g)},
\label{mutual}
\end{equation}
where $\log$ is taken with base 2. This dimensionless, nonmetric
quantity measures in bits the extent to which $C$ and $G$ are
dependent (complete independence implies $p(C,G)=p(C)p(G)$, and thus
$I(C,G)=0$). The mutual information is bounded, $0\leq I(C,G)\leq
\min[H(C),H(G)]$, where $H(X)$ is the entropy, $H(X)=-\sum_x p(x)\log
p(x)$. In \citep{calin02}, there were $||C||=4$ possible input states
$c\in\{1,2,3,4\}=\{c_i\}$ and two possible output states, GFP {\em on}
or {\em off}. For a circuit with a constant $g$, $H(G)=0$, and then
$I(G,C)=0$.  At the other extreme, if the reporter gene is {\em on}
for exactly two of the four equiprobable chemical inputs, then each
reporter state has $p=1/2$, and $I(C,G)=1$ bit.  Similarly, for
multinomial distributions of $g$, the mutual information seamlessly
takes into account all possible relations between $g$ and $c$.

A crucial advantage in adopting mutual information as a quality
measure is that it can be evaluated independently of the function of
the circuit. For steady state responses considered here, the only
reasonable way to define a qualitative {\em function} of the circuit,
or to characterize the computation performed by it, is to consider how
$\langle g(c_i)\rangle$ are ordered. As long as all $||C||$ responses
are sufficiently resolved, the mutual information will be $\sim
\log||C||$, irrespective of the ordering. Thus the mutual
information-based circuit quality measure is insensitive to the type
of computation performed by the circuit, but only to whether the
computation assigns a different output to each input. Furthermore, due
to the {\em data processing inequality} \citep{shannon}, high $I(C,G)$
is a sufficient condition for a high-quality realization of {\em any}
computational function that depends (stochastically or
deterministically) on $P(c,g)$. High $I(C,G)$ is especially important
for sensory stages in biochemical signaling, where the same
biomolecular species may control responses of many different
biochemical modules, requiring high quality information about many
different properties of the signal at the same time.

\subsection{Proposal}
\label{sec:proposal}
We propose to investigate how the topology of a regulatory circuit
affects its computational and information transmission properties, as
measured by the steady state signal-response mutual information,
Eq.~(\ref{mutual}). While the results of \citep{calin02} may be
interpreted as revealing that some circuits may perform better than
others, this effect can be caused in part by operating at suboptimal
kinetic parameters, some of which are biologically easy to adjust to
improve the information transmission fidelity.  In fact in \citep{calin02}, several
networks identical in topology but different in their kinetic
properties, performed markedly different functions. To avoid the
problem, we study instead the maximum mutual information attainable by
the circuit under realistic conditions. Specifically, for a regulatory
circuit $t$, with a set of kinetic parameters
$\mbold{\vartheta}=\{\vartheta_1,\vartheta_2,\dots\}$, which responds
to inducer (input) concentrations $C=\{c_i\}=\{c_1,c_2,\dots\}$ with
different genetic (output) expression levels
$P(g|c,\mbold{\vartheta})$, we propose to investigate numerically
$\mbold{\vartheta_{*}}^{t}=\arg\max_{\mbold{\vartheta}} I^t(C,G)$.

As in \citep{calin02}, we limit ourselves to $3$-gene topologies where
each gene is regulated by exactly one transcription factor (see Tables
\ref{table2} and \ref{table1} for the list of these topologies, and
Sec.~\ref{sec:mm_tops} for their more detailed description). We
measure the output of the circuit in terms of the steady-state
expression of the reporter gene, which is always downregulated by
another gene denoted $Z$ (see Tables \ref{table2} and \ref{table1}). This
limits us to $24$ possible circuits, cf.~Sec.~\ref{sec:mm_tops}. The
kinetics associated with these topologies are described in
Sec.~\ref{sec:mm_model}. Note in particular that even though we use
the {\em genetic regulation} terminology throughout the paper, the
kinetic model is general enough to account for {\em protein signaling}
and other regulatory mechanisms as well.

For each of the chosen topologies, we need to find stable fixed points
of the dynamical systems that describe the circuit,
cf.~Sec.~\ref{sec:mm_stablept}, evaluate the fluctuations around them,
$P(g|c)$, estimate the corresponding mutual information, and then
optimize it with respect to the kinetic parameters. Note that all of
the parameters of the system that we treat as variable can, in fact,
be easily adjusted by the cell over its lifetime by means of many
biological mechanisms, cf.~Sec.~\ref{sec:mm_model}.

Rather than discretizing the reporter output, as in \citep{calin02}, we
take into account the actual numbers of the reporter
molecules. Assuming mesoscopic copy numbers, we use the linear noise
approximation, cf.~Sec.~\ref{sec:mm_lna}, to derive the reporter gene
distribution as a sum of Gaussians with means at the stable fixed
points. Under this assumption, the mutual information between the two
random variables, $C$ representing the \emph{discrete} chemical
(input) states, and $G$ measuring the \emph{continuous} reporter
expression (output), is
\begin{equation}
  I(C,G) = \frac{1}{M} \sum_{c=1}^{||C||} \sum_{i=1}^{M_c} \int dg\,
  {\cal{N}}(g_i^c,\sigma_i^c)\; \log \frac{\frac{1}{M}\sum_{j=1}^{M_c}
  {\cal{N}}(g_j^c,\sigma_j^c)} {\frac{M_c}{M}\frac{1}{M}\sum_{d=1}^{||C||} \sum_{k=1}^{M_d}
{\cal{N}}(g_k^d,\sigma_k^d)}.
\label{eqn:mi}
\end{equation}
Here $M$ is the total number of fixed point calculations performed for
the circuit, and $M_c$ is the number of those done with $C=c$;
${\cal{N}}(g_i^c,\sigma_i^c)$ denotes the output response for the
$i$'th calculation with $C=c$, which is a Gaussian distribution with
mean $g_{i}^{c}$ and variance $(\sigma_{i}^{c})^2$. Many calculations
at each point in the parameter space, $\mbold{\vartheta}$, are needed
to explore multiple stable fixed points of the dynamical system (see
Sec.~\ref{sec:mm_stablept}).  Finally, we choose each chemical state
with equal probability, $P(c)={\rm const}$.

When optimizing Eq.~(\ref{eqn:mi}) with respect to $\mbold{\vartheta}$
(see Sec.~\ref{sec:mm_opt}), we need to consider two computationally
trivial (and biologically unrealistic) ways of achieving high
$I(C,G)$. First, given discrete $c$ and an infinite range of $g$,
achieving the upper bound $I(C,G)=H(C)$ is easy: as the number of
molecules of the reporter $ g^c$ increases, the magnitude of its
fluctuations, as measured by its standard deviation $\sigma^c$, grows
slower as $\sigma^c \sim \sqrt{ g^c}$, so the responses to all $c$'s
can be well-separated if we allow for an infinite number of
molecules. This, however, is not biologically relevant, since
producing many copies of a molecule takes time and energy, both of
which are limited. In fact, here we are interested only in solutions
that involve low copy numbers as this is precisely the regime in which
gene regulatory networks function. We note also that many apparently
deterministic, high copy number systems may actually fall into this
regime if the threshold of the system can be overcome with only a few
molecules \citep{schneidman,cluzel_ultra,huang_ultra,kholodenko}. 

Second, and perhaps less obvious, a trivial solution can also be
obtained if we allow for multi-scale ({\em stiff}) systems. For
example, if the response time of the reporter $\tau_G$ is very large
relative to the upstream regulators $\tau_{Z}$, then all of the noisy
upstream fluctuations will be filtered \citep{berg77,bialek05}:
effectively, the reporter takes $N_Z\tau_G/\tau_{Z}\gg1$ samples of
$Z$ per its response time (here $N_Z$ is the mean number of $Z$
molecules), and fluctuations are small. However, living cells must
respond in a timely manner to changes in the environment, so infinite
response times are also not biologically relevant. 

These observations suggest that our objective function to be maximized
requires some biologically reasonable constraints. For this reason, we
have investigated many different realizations of the constraints, and,
instead of maximizing the mutual information, we chose to maximize the
following {\em constrained mutual information}
\begin{equation}
L=I(G,C)-\lambda \langle N \rangle - \gamma \langle q \rangle,
\label{eqn:L}
\end{equation}
where $\lambda$ and $\gamma$ are chosen such that the average number
of molecules of all of the components in the system $\langle N
\rangle=\frac{1}{||C||M_cN_s}\sum_{c=1}^{||C||}
\sum_{i=1}^{N_s}\sum_{j=1}^{M_c} N^{c}_{ij}$ (where $N^{c}_{ij}$ is
the average number of molecules of species $i$ for fixed point $j$
given $C=c$ and $N_s=4$ is the total number of species in the system)
is on the order of $100$, and the average stiffness of the system
$\langle q \rangle=\sum_i r_i/r_G$ (where $r_i$ are the decay rates of the
transcription factors, and $r_G$ is the decay rate of the reporter) is
approximately $3$ orders of magnitude. 

\section{Methods}
\subsection{Topologies}
\label{sec:mm_tops}
As in the experimental set-up in \citep{calin02}, we consider $3$-node
circuits, in which genes are regulated by exactly one gene (including
the possibility of auto-regulation). This also reduces the assumptions
we would otherwise need to make about the dynamics associated with
combinatorial regulation.  The $3$ genes ($X$, $Y$, and $Z$) in each
circuit are interconnected by exactly $3$ edges. There are only $3$
such non-redundant topological structures, which, when we include the
possibility of either excitatory or inhibitory interactions, results
in $2^3=8$ possible configurations per structure, for a total of $24$
topologies (see Tables \ref{table2} and \ref{table1}). The fourth
(reporter) gene $G$ is always down-regulated by $Z$, as in
\citep{calin02}. Extensions to other topology classes are easily
implemented.

\subsection{Model and parameters}
\label{sec:mm_model}
The dynamics of transcription and translation have been modeled with
remarkable success for small circuits by avoiding the translation step
completely and coupling the genes to each other directly by means of
simple rational functions $\alpha_j$
\citep{collins_toggle,hasty,elowitz_rep}. In general, each of the
species ${\bf X}=\{X,Y,Z,G\}$ in the circuit is subject to a degradation and a
creation process
\begin{eqnarray}
X & \overset{r_x}{\rightarrow} & \emptyset, \\
\emptyset &\overset{\alpha_x}{\rightarrow} & X.
\end{eqnarray}
The macroscopic, deterministic description of the system is governed
by the system of ordinary differential equations
\begin{equation}
d\phi_j / dt = -r_j \phi_j + \alpha_j (\phi_{\pi_j}),
\label{dynamics}
\end{equation}
where $\{ \phi_1, \dots, \phi_N \}$ is the concentration vector of the
$N$ chemical components, $r_j$ is the degradation rate of $\phi_j$,
and $\alpha_j$ is a production term that depends on the concentration of a
regulator (parent) molecule of $j$, namely $\pi_j$. The production is
modeled as a constitutive expression (the {\em leak}) plus a Hill
activation or inhibition,
\begin{equation}
\alpha(\phi)=a_0 + a \frac{(\phi/s_i)^n}{K^n + (\phi/s_i)^n},
\end{equation}
or 
\begin{equation}
\alpha(\phi)=a_0 + a \frac{K^n}{K^n + (\phi/s_i)^n}, 
\end{equation}
where $a_0$ describes the leakiness of the promoter, $a$ specifies its
dynamic range, $K$ is the concentration of the regulator at
half-saturation (the {\em Michaelis} constant), $n$ is the Hill
coefficient, and $s_i$ is the modulating effect of the $i$'th input
molecule on the regulator protein. $s_i$ can be modeled equivalently
by rescaling $K$. One can think of this as the chemical signal binding
to the protein, changing its conformation, and influencing various
affinities. The two equations correspond to an excitatory and
inhibitory regulation, respectively. For this dynamics, there is no
distinction between the protein and the mRNA of a gene species, and we
use the terms interchangeably. As in \citep{calin02}, we allow each input to take two binary states (either the input molecule is present or not). We have a total of $3$ inputs and $2^3$ input states, and each input modulates the expression of one of the three transcription factors. For a chemical state $c$ where an input molecule $i$ is not present, we set $s_i=1$. We set the units of measurements such
that volume $\Omega$ of a cell is $1$, so that concentration of $1$ is
equivalent to $1$ molecule per cell.

In all we have
\begin{enumerate}
\item $3$ decay rates $r_X,r_Y,r_Z$ corresponding to decay rates for
  the $3$ transcription factors. We set $r_G$ corresponding to a
  response time of approximately a half hour.
\item $4$ {\em Michaelis} constants $K_X,K_Y,K_Z,K_G$ and $4$ range
  parameters $a_X, a_Y, a_Z,a_G$ describing the regulation function
  for each component of the circuit.
\item $3$ input parameters $s_X,s_Y,s_Z$ modulating the effect of each
  input on the $3$ transcription factors
\item $1$ leak parameter $a_0$.
\end{enumerate}
For simplicity we assume $n=2$. In total we have $15$ parameters.

Notice that all of these parameters can be easily adjusted by the cell
by means of a variety of biological mechanisms, thus validating our
proposal to study the dependence of the signal transduction, optimized
with respect to the parameter values. Below is a non-exhaustive list
of such regulatory mechanisms.
\begin{enumerate}
\item All protein/mRNA decay rates can be adjusted independently of
  each other by microRNA expressions or by regulated proteolysis, such
  as using ubiquitin tagging.
\item Michaelis constants depend on structural properties of proteins
  and the DNA, as well as on the abundance of the proteins near a DNA
  binding site compared to the overall protein concentration. Thus
  they can be adjusted by chromatin rearrangement, or by controlling
  the nuclear pore transport.
\item Effects of chemical inputs on transcription factors depend on
  the chemical-protein affinity and on the abundance of the chemicals
  near the relevant proteins. The former can be changed by modulating
  chemical-protein binding reaction by means of expression of various
  enzymes, while the latter can be achieved by controlling transport
  processes.
\item The leak depends on the concentration of the RNA polymerase,
  ribosome, as well as the DNA accessibility. All are easy to adjust
  in a living cell.
\end{enumerate}

\subsection{Determining Stable Fixed Points}
\label{sec:mm_stablept}
All of our circuits incorporate some feedback mechanism (e.g., the
``feedback dyad'' \citep{mishra02}) and, therefore, may have multiple
stable steady state solutions. We find these by numerically solving
the macroscopic chemical kinetics system (\ref{dynamics}) describing
the circuit using MATLAB's {\tt ode15s} with the parameters as
described in section \ref{sec:mm_model}.  We randomly sample different
initial conditions for the time-evolution to obtain a set of (almost
all) fixed points for each chemical state and each
topology. Additionally, since {\em in vivo} the system will be
flipping between different input states, the steady-state solution of
one input state is the potential initial condition for the
time-evolution of the other inputs. To include these potential initial
conditions, we first randomly choose $10$ initial conditions for each
$c_i$, and then we take the resulting stable solutions and use them as
the initial conditions for each $c_{j\neq i}$.

When a time-evolution of the system results in
oscillations or chaotic behaviors, we neglect these solutions since, under our
assumptions, they will result in multiple genetic outputs
corresponding to the same chemical input and hence in a low mutual
information. 

\subsection{Linear Noise Approximation (LNA)}
\label{sec:mm_lna}
For excellent reviews and discussions of the linear noise technique,
we refer the reader to
\citep{elf1,elf2,paulsson,kampen}. Here we briefly review one
particular formulation that simplifies the analysis.

Given a system with volume $\Omega$ and $N$ different particles, we
denote the particle concentrations as $\mbold{\phi} = \{\phi_1, \dots,
\phi_N\}$, and the copy numbers as ${\mbold{n}}=\Omega
\mbold{\phi}$. The state of the system is defined by ${\mbold{n}}$, and it
changes when an elementary reaction $j$, $j=1,\dots,R$ takes
place. When reaction $j$ occurs, the copy number $n_i$ changes by $S_{ij}$,
which is the $N\times R$ stochiometric matrix. Then the evolution of
the joint probability distribution $P({\mbold{n}},t)$ is given by the
following master equation
\begin{equation}
  \frac{dP({\mbold{n}},t)}{dt} = \Omega \sum_{j=1}^R
  ( \Pi_{i=1}^N E^{-S_{ij}}-1) f_j({\bf \phi},\Omega)P({\mbold{n}},t),
\end{equation}
where $E^{-S_{ij}}$ is the {\em step operator}, which acts by removing
$S_{ij}$ molecules from $n_i$, and $f_j$ is a  rate for $j$.

While this equation is usually mathematically intractable, a Monte
Carlo algorithm exists to solve it numerically (the {\em Gillespie
  algorithm}) \citep{gillespie,gibson}. To generate a particular
stochastic trajectory, this method draws random pairs $(\tau, e)$ from
the joint probability density function $P(\tau,e|{\mbold{n}})$, where
$\tau$ is the time to the next elementary reaction, and $e$ is its
index.  Multiple trajectories allow to estimate the necessary moments
of $P({\mbold{n}},t)$.  However, this approach is computationally
intensive, and quickly becomes infeasible if one wants to explore
multiple system parameterizations, or if $f_j$ span multiple scales.

Alternatively, one can expand the master equation in orders of
$\Omega^{-1/2}$. Introducing $\xi$, such that $n_i=\Omega \phi_i +
\Omega^{1/2} \xi_i$ and treating $\xi$ as continuous, the first two
terms in the expansion yield the macroscopic rate and the linear
Fokker-Plank equations, respectively:
\begin{eqnarray}
\Omega^{1/2}&:& \sum_{i=1}^N \frac{\partial \phi_i}{\partial t}
 \frac{\partial \Pi (\xi,t)}{\partial \xi_i}  = 
\sum_{i=1}^N\sum_{j=1}^R S_{ij}f_j(\phi)\frac{\partial
 \Pi(\xi,t)}{\partial \xi_j}, \label{classical}\\
\Omega^{0}&:&  \frac{\partial \Pi(\xi,t)} {\partial t} =
 -\sum_{i,k} A_{i,k} \frac{\partial (\xi_k \Pi)}{\partial \xi_i} 
+ \frac{1}{2} \sum_{i,k} \lbrack B \rbrack_{ik} 
\frac{\partial^2 \Pi}{\partial \xi_i \partial \xi_k},
\label{fokker}
\end{eqnarray}
where $A_{jk}= \sum_{j=1}^R S_{ij}\frac{\partial f_j}{\partial
  \phi_k}$ and $\lbrack B \rbrack_{ik}=\sum_{j=1}^R S_{ij} S_{kj}
f_j (\mbold{\phi})$. The steady-state solution of Eq.~(\ref{fokker})
is a multivariate Gaussian
\begin{equation}
P(\xi)=\left[(2\pi)^{N} \det \Xi\right]^{-1/2} \exp \left(-\frac{\xi^T\, \Xi\, \xi}{2}\right),
\end{equation}
where the covariance matrix $\Xi$ is given by the matrix Lyapunov equation 
\begin{equation}
A \Xi + \Xi A^T + B=0. 
\label{Lyapunov}
\end{equation}
This system is solved using the standard matrix Lyapunov equation
solvers (MATLAB's {\tt lyap}). In order to assess the validity of
LNA for our system we compared the steady state
solutions to multiple Gillespie runs. We found that, even at very low
copy numbers ($\sim10$), LNA performed well as measured by the
Jensen-Shannon divergence (see Sec.~\ref{sec:lna_v_gill} for details).  Based on these results, we approximate the steady-state
distribution as a sum of multivariate Gaussians with means at the
stable fixed points of Eq.~(\ref{classical}), and with covariances as
in Eq.~(\ref{Lyapunov}). 

We note that both the LNA and the Gillespie algorithm are derived
assuming that the reactions $j$ are truly elementary.  
In our system, a single particle
creation, $\alpha$, encapsulates all processes, starting from the
protein-DNA binding and ending with the translation. Justification for using
``elementary complex'' reactions is provided in
\citep{rao_stoch,elf1, bundschuh,min,weinan, ball}). However, the complex nature of the reactions has acomparatively small influence on the low frequency components of thestochastic response \citep{misha06}, which is our focus here. For this reason we believe that approximating terms in Eq.~(\ref{dynamics}) as elementary and using LNA is a less important approximation than merging transcription and translation into a single step. 
Generalizations to LNA with elementary reactions is
straight-forward, provided the reaction system is known (which is more
complicated).

\subsection{LNA Performance}
\label{sec:lna_v_gill}
Here we quantify LNA accuracy as a function of the mean molecule numbers in the system. We approximate the steady-state probability distribution $p_l$ using LNA and compare this to the steady-state distribution $p_g$ obtained with multiple trajectories of the Gillespie algorithm. As a measure of the similarity between the two probability distributions, we use the symmetric \emph{Jensen-Shannon divergence} $JS_{\Pi}$ 
\begin{equation}
JS_{\Pi}[p_l,p_g]=\pi_1D_{KL}[p_l||(\pi_1p_l+\pi_2p_g)] + \pi_2D_{KL}[p_g||(\pi_1p_l+\pi_2p_g)]
\end{equation}
where $D_{KL}[pl|q]=\sum_i p_i\log_2 (p_i/q_i)$ is the Kullback-Leibler divergence and we set $\pi_1=\pi_2=1/2$. We note that $D_{KL}\geq0$ and in the case that the two distributions are equal $p_l=p_g$, then $D_{KL}=JS_{\Pi}=0$.   

\begin{figure}[htbp]
\begin{center}
  {\includegraphics[width=5in] {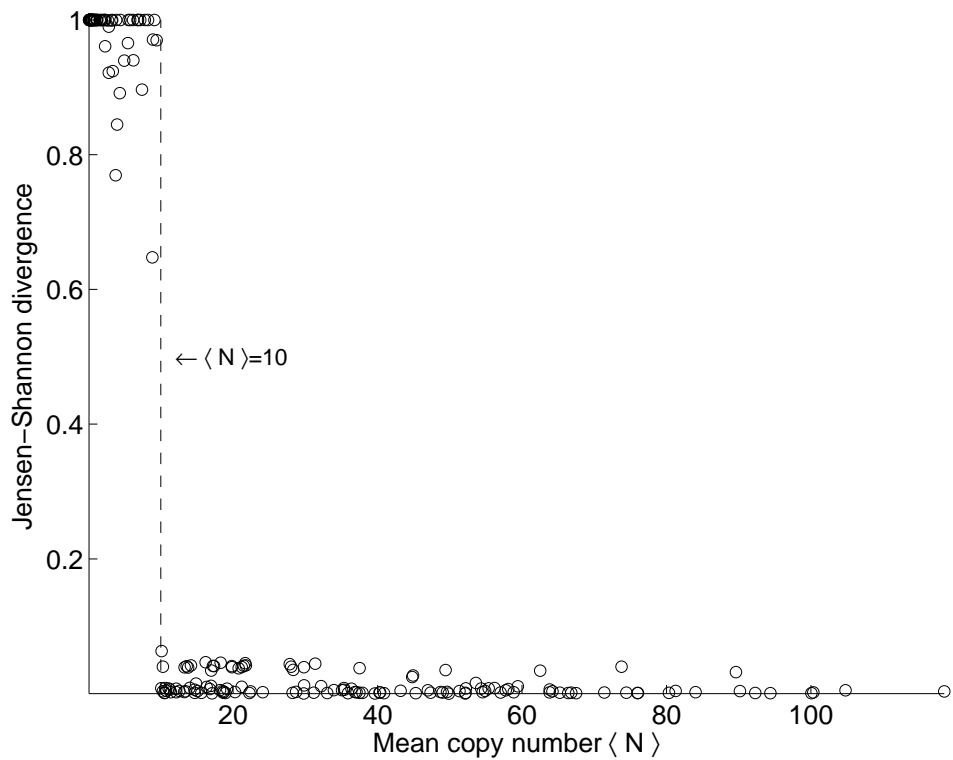} }
\caption[Accuracy of Linear Noise Approximation]{Jensen-Shannon divergence $JS_{\Pi}$ between distributions obtained by LNA and the Gillespie algorithm for multiple circuits and multiple parameterizations plotted as a function of mean copy number. At $JS_{\Pi}=0$, the distributions are identical. There appears to be a sharp threshold at $10$ molecules, below which the LNA does poorly, but above which, LNA does well. }
\label{lna_v_gill_js}
\end{center}
\end{figure}

We calculated $JS_{\Pi}$ for multiple circuits and multiple parameterizations and found excellent consistency between LNA and Gillespie (see Fig.~\ref{lna_v_gill_js}). Below $10$ molecules, the two distributions were easily distinguishable, but above $10$ molecules we consistently found large overlap. 

\subsection{Optimization}
\label{sec:mm_opt}
  We employ a simplex optimization (using MATLAB's {\tt
  fminsearch}) to maximize $L=I(G,C)-\lambda \langle N \rangle -
\gamma \langle q \rangle$ over the $\log_{10}$ of the $15$
parameters where $\lambda$ and $\gamma$ are chosen to accommodate
biologically relevant molecule numbers and stiffness. For example, for
an average of approximately $100$ molecules for each transcription
factor and a stiffness of order $3$ we choose $\lambda=0.01$ and
$\gamma=0.001$.  To explore the parameter space for each topology, we
uniformly randomly select biologically relevant starting points
(protein half-life near $10$ minutes, promoter leakiness near $0.01$
proteins/sec, promoter range near $10$ proteins/sec, regulator at
half-saturation near $100$ proteins/sec, and input molecule modulation
of regulator near $2$). To make the search for maxima more efficient
we only maximize random points that start already above a certain
threshold ($L\ge0$).

\subsection{Maximum mutual information for a fixed copy number}
\label{sec:mm_maxmi}
Suppose a molecular species $G$ with concentration $g$, $\int dg P(g)
g = N_G$ is used as a reporter species for a cascade of biochemical
computations, so that the species is not allowed to participate in any
feedback loops. Then its stochasticity is limited from below by a
Poisson noise. That is, if $\bar{g}$ is the deterministic value of $g$
produced by some biochemical reaction kinetics, and $\bar{g}\gg1$,
then
\begin{eqnarray}
&&g = \bar{g} +\nu,\\
&&\langle \nu \rangle=0, \,\, \langle \nu\nu\rangle=\bar{g}.
\end{eqnarray}
Furthermore, $\bar{g}$ itself is distributed probabilistically
according to $P(\bar{g})$, $\int d\bar{g} P(\bar{g}) \bar{g}=N_{\bar
  G}$, due to stochasticity of inputs to and of the internal dynamics
of the biochemical system.  We are interested in 
 the maximum number of bits that can be transmitted reliably by this
reporter species (that is, is its channel capacity) at fixed
$N_G$.

Intuitively, the noise in this system is $\sim \sqrt{N_{\bar
    G}}\approx\sqrt{N_G}$, so the number of distinguishable states
of the reporter is also $\sim \sqrt{N_G}$, and one should be able to
transmit about $1/2\log_2 N_G$ bits reliably. This argument has been
used extensively (e.g., \citep{detwiler}). However, it fails
(a) to establish the correct constant of proportionality in front of
the number of distinguishable states and (b) to take into the account
the $\bar{g}$ dependence of the noise variance (which leads to a
higher resolution at smaller $\bar{g}$). Both of these effects are
likely to contribute only $O(1)$ bits to the channel capacity, but,
for $N_G\lesssim100$ considered in this work, this might be an
important correction. We are unaware of a prior derivation of the
channel capacity for this system up to $o(1)$, and we present it here.

We write:
\begin{eqnarray}
  I(G,\bar{G}) &=& H(G)-H(G|\bar{G}) = H(\bar{G}) - H(G|\bar{G}) +
  O\left(\frac{1}{{N_{\bar G}}}\right)\label{eq:def}\\
  &=& -\int d\bar{g}\, P(\bar{g}) \log_2 P(\bar{g}) -\frac{1}{2} \int
  d\bar{g} P(\bar{g}) \log_2 2\pi e\bar{g} + O\left(\frac{1}{N_{\bar{G}}}\right)\label{eq:lna}.
\end{eqnarray}
Eq.~\ref{eq:def} is valid if ${\rm var}\, (G|\bar{G})\sim N_{\bar
  G}\ll {\rm var}\, (\bar{G})\sim N_{\bar G}^2$, and
Eq.~\ref{eq:lna} holds for a Poisson noise in the reporter.

To find the channel capacity of the reporter species, we maximize
$I(G,\bar{G})$ with respect to $P(\bar{g})$ subject to
\begin{equation}
\int d\bar{g}\, P(\bar{g})=1,\quad\quad
\int d\bar{g}\, P(\bar{g}) \bar{g} = N_{\bar G}.
\end{equation}
This results in 
\begin{equation}
P(\bar{g})\approx \frac{1}{(2\pi\bar{g}N_{\bar G})^{1/2}}e^{-\bar{g}/2N_{\bar
    G}}\label{eq:input},
\end{equation}
where $\approx$ is due to the  approximation involved in replacing
$H(G)$ by $H(\bar{G})$. Plugging $P(\bar{G})$ into the equation for
$I$, we get the channel capacity
\begin{eqnarray}
I_0(G,\bar{G}) &=& \int d\bar{g}\, P(\bar{g}) \left[\frac{1}{2}\log_2 2\pi
N_{\bar G} \bar{g} +\log_2e \frac{\bar{g}}{2N_{\bar{G}}} -
\frac{1}{2}\log_22\pi e\bar{g}\right] +O(N_{\bar{G}}^{-1})\\
&=& \frac{1}{2}\log_2N_{\bar{G}}+O(N_{\bar{G}}^{-1}).\label{eq:capacity}
\end{eqnarray} 

Thus, for the optimal distribution of inputs, as in
Eq.~\ref{eq:input}, the naive estimate of $I_0=1/2\log_2N_{\bar{G}}$
for a biochemical reporter is correct up to terms non-vanishing with
$N_{\bar{G}}^{-1}$. For the distribution of inputs analyzed in this
work (up to 8 discrete input states), the maximum possible
$I(G,\bar{G})$ is clearly less than this channel capacity. One can
obtain the maximum information for such input distributions by
numerical optimization of $I$ with respect to the values of the
$\bar{g}$ input states, assuming a Poisson distribution of $g$ around
$\bar{g}$.

\subsection{Network noise analysis using LNA}
\label{sec:mm_noise}
Consider a regulatory network of $N$ transcription factors indexed by
$i\in\{1,2,\dots,N\}$. The macroscopic behavior of the system is
determined by
\begin{eqnarray*}
\dot{\phi_1} & = & f_1(\phi_1,\dots, \phi_N) \\
\dot{\phi_2} & = & f_2(\phi_1,\dots, \phi_N) \\
&\dots &  		\\
\dot{\phi_N} & = & f_N(\phi_1,\dots, \phi_N) 
\end{eqnarray*}
where $\phi_i$ is the concentration of the $i$'th transcription
factor. Let $\mbold{n}=\Omega \phi$ be the vector of molecule copy
numbers with volume $\Omega$. Using LNA
\citep{elf2}, we can calculate the covariance matrix $C=\langle
(\mbold{n} -\langle \mbold{n} \rangle)(\mbold{n}-\langle \mbold{n}
\rangle)^T\rangle=\Xi\Omega$ by solving Eq.~\ref{Lyapunov}: 
\begin{eqnarray}
  B_{ii} &=&- 2\sum_{j\in\pi_i} \frac{\partial f_i}{\partial \phi_j} C_{ij} \\
  &=& -2\left(\frac{\partial f_i}{\partial \phi_i} C_{ii} + \sum_{j \neq i,j\in\pi_i}
    \frac{\partial f_i}{\partial \phi_j} C_{ij}\right),\\
  C_{ii}&=&\frac{-1}{2\frac{\partial f_i} 
    {\partial \phi_i}} \left ( B_{ii} + 2 \sum_{j \neq i, j \in \pi_i} 
    \frac{\partial f_i}{\partial \phi_j} C_{ij} \right ).
\label{eqn:var}
\end{eqnarray}
This suggests that the topology or structure of the network can also
play a role in controlling noise. Specifically, the variance of the
$i$'th transcription factor $C_{ii}$ can be reduced by decreasing the
product $\frac{\partial f_i}{\partial \phi_j} C_{ij}<0$, where
$j\in\pi_i$.  

The covariance $C_{ij}$ is a more complicated function of the other
covariances:
\begin{eqnarray}
B_{i,j} & = & - \left( \sum_{k\in\pi_i} \frac{\partial f_i}{\partial \phi_k} C_{jk}
 + \sum_{k\in\pi_j} \frac{\partial f_j}{\partial \phi_{k}} C_{ik} \right ) \\
& = & - \left( \frac{\partial f_i}{\partial \phi_i} C_{ij} + 
\frac{\partial f_j}{\partial \phi_j} C_{ij} +  \sum_{k \neq i, k \in \pi_i} 
\frac{\partial f_i}{\partial \phi_k} C_{jk} + \sum_{k\neq j, k \in \pi_j} 
\frac{\partial f_j}{\partial \phi_{k}} C_{ik} \right), \\
C_{ij} & = & \frac{-1}{\frac{\partial f_i}{\partial \phi_i} + 
\frac{\partial f_j}{\partial \phi_j}} \left ( B_{ij} + 
\sum_{k \neq i, k \in \pi_i} \frac{\partial f_i}{\partial \phi_k} C_{kj} +
 \sum_{k\neq j, k \in \pi_j} \frac{\partial f_j}{\partial \phi_{k}} C_{ik} \right ).
\label{eqn:covar}
\end{eqnarray}
If $j\in\pi_i$, then from Eq.~\ref{eqn:covar} we see that $C_{ij}$ is
a function of the covariances between $i$ and the regulators of its
regulators ($C_{ik}$, where $k\in\pi_j$, and $j \in \pi_i$). We
can write these covariances in turn as functions of covariances
between $i$ and the regulators of regulators of regulators of $i$, and
so on. This implies a recursion which will end when we either reach a
regulator which has no other regulators or, in the case of a cycle, we
reach $i$ again.

In the latter case, the recursion will end back with $C_{ii}$, and the
last term in Eq.~\ref{eqn:covar} will have the form
\begin{equation}
C_{ii}\prod_{j\in {\rm cycle}} \frac{\partial f_j} {\partial \phi_{\pi_j}}.
\label{product}
\end{equation}
Since $C_{ii}\geq0$, this implies that one way to reduce $C_{ij}$ (and
hence $C_{ii}$ itself) is to have the product in Eq.~\ref{product}
be {\em negative}.  Crucially, the only way to achieve this is
if the cycle contains an odd number of negative regulators.

\subsection{Some simple examples of sub-Poisson noise}
The transcription factors in the network may participate in various
feedback loops. In some cases, this allows the usual Poisson noise
lower bound to be overcome, resulting in a sub-Poisson noise
($C_{ii}<\bar{\phi_i}$). Below we give some simple examples for
$1$-,$2$-,and $3$-cycles.

The set-up of \citep{calin02}, which we use in this work, simplifies
the analysis since we only consider one promoter transcription
factors, so that $f_i = -r_i \phi_i + \alpha_i(\phi_{\pi_i})$, where
$\pi_i$ includes just one gene.  In steady-state, $\bar{\phi_i} =
\alpha_i/r_i$. Finally, all of our reactions are enzymatic, so the
diffusion matrix $B$ will only have diagonal nonzero elements. Then,
since $B_{ii} = r_i \bar{\phi}_i + \alpha_i$, we use the expression
for $\bar{\phi_i}$ to find $B_{ii}=2r_i\bar{\phi_i}$.

\subsubsection{{Auto-repression}}
For the auto-repressive case there are no covariance terms and
$\frac{\partial f_i}{\partial \phi_i} = -r_i + \alpha'_i$, so we can
rewrite
\begin{equation}
C_{ii} = \frac{\bar{\phi_i}}{1 - \frac{\alpha'_i}{r_i}}.
\end{equation}
Auto-repression implies $\alpha'_i <0$. Thus $C_{ii} < \bar{\phi_i}$,
resulting in a sub-Poisson noise. A similar derivation using LNA is given for
regulated degradation in \citep{el_samad} and regulated synthesis in \citep{thattai}.

\subsubsection{{A $2$-cycle}}
In this case, $\pi_i=i-1$, and $\pi_{i-1}=i$. Assuming no
auto-regulation, let $\frac{\partial f_i} {\partial \phi_{i-1}} =
\alpha'_i$ and $\frac{\partial f_{i-1}} {\partial \phi_{i}} =
\alpha'_{i-1}$. Now we write,
\begin{eqnarray}
C_{ii} &=& \bar{\phi_i} + \frac{1}{r_i}\alpha'_i C_{i,i-1},\\
C_{i,i-1} &=& \frac{1}{r_i + r_{i-1}} \left ( \alpha'_i C_{i-1,i-1} + \alpha'_{i-1} C_{ii} \right).
\end{eqnarray}

To reduce $C_{ii}$ we can reduce the magnitude of $C_{i,i-1}$. One way
to achieve this is to have opposite signs for $\alpha'_i$ and
$\alpha'_{i-1}$. Moreover, the sub-Poisson noise is possible if
$\alpha'_i C_{i,i-1}<0$, which is possible only if $\alpha'_i\alpha'_{i-1}<0$. Thus the presence of a negative and
positive regulator in a $2$-cycle is a necessary, but not sufficient
condition for achieving sub-Poisson noise. For sufficiency, we also
need $|\alpha'_i C_{i-1,i-1}| < | \alpha'_{i-1} C_{ii}|$.

\subsubsection{{A $3$-cycle}}
In this case, $\pi_i=i-1$, $\pi_{i-1}=i-2$, and $\pi_{i-2}=i$. The
variance equation stays the same
\begin{equation}
C_{ii} = \bar{\phi}_i + \frac{1}{r_i} \alpha^{'}_i C_{i,i-1}. 
\end{equation}
However, now we have
\begin{equation}
C_{i,i-1} = \frac{1}{r_i + r_{i-1}} \left ( \alpha^{'}_i C_{i-1,i-1} + \alpha^{'}_{i-1} C_{i,i-2} \right).
\end{equation}
and
\begin{equation}
C_{i,i-2} = \frac{1}{r_i + r_{i-2}} \left ( \alpha^{'}_i C_{i-1,i-2} + \alpha^{'}_{i-2} C_{i,i} \right).
\end{equation}

Combining the above into a single expression for $C_{ii}$, we have
\begin{equation}
C_{ii}=\bar{\phi}_i + \frac{1}{r_i(r_i+ r_{i-1})}\left\{ \alpha^{'}_i 
\left[\alpha^{'}_i C_{i-1,i-1} + \alpha^{'}_{i-1}\frac{1}{r_i+r_{i-2}} \left(
 \alpha^{'}_i C_{i-1,i-2} + \alpha^{'}_{i-2}C_{i,i} \right) \right] \right\}. 
\end{equation}
The last term gives us a product of the derivatives, $\alpha^{'}_i
\alpha^{'}_{i-1} \alpha^{'}_{i-2}$. If this product is negative (that
is, if we have an odd number of repressors in the cycle) then we can
reduce the overall magnitude of the variance $C_{ii}$. Note here that
we have two extra terms in the variance. One,
$(\alpha^{'}_i)^2C_{i-1,i-1}$, is always positive, while the other can
be of either sign. Thus the overall negative regulation is not a
guarantee of a sub-Poisson noise in this case.

Ultimately, noise regulation can be improved with cycles with odd
number of negative regulators. However, as cycles get larger and the
network becomes more complex, the achievability of sub-Poisson noise
becomes more limited. This may be related to the observation that,
whereas small cycles are over-represented in a metabolic network,
large cycles occur less frequently than one would expect given several
different possible null models \citep{gleiss}.

\section{Results}
\subsection{Transmitting more than $1$ bit at low copy number}
\label{sec:results_1bit}
\begin{figure}
  \begin{center}
    \begin{tabular}{ll}
      {\includegraphics[width=1.8in] {top19} } & {\includegraphics[width=1.8in] {top11} }  \\
      {\includegraphics[width=3in] {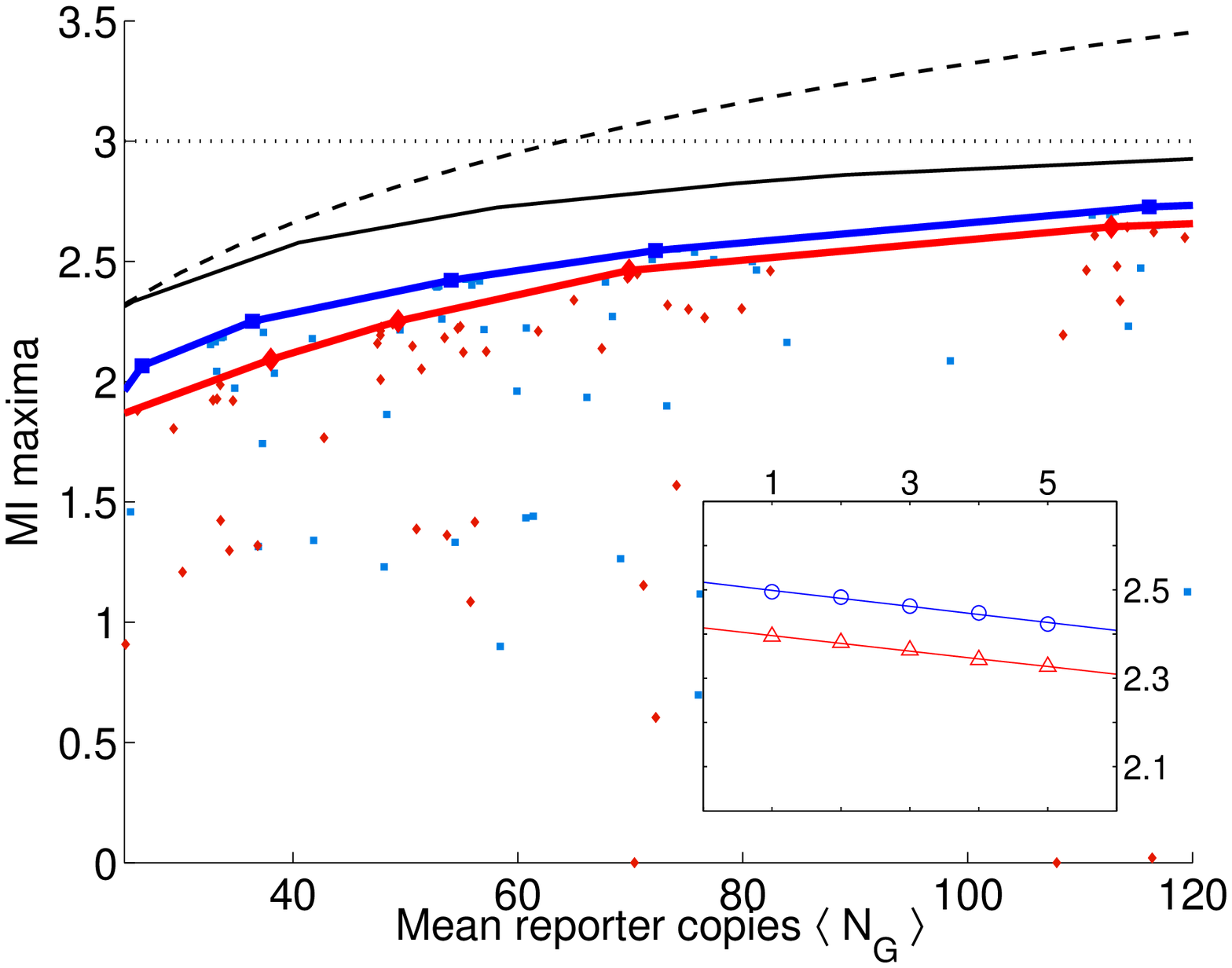} } & {\includegraphics[width=3in] {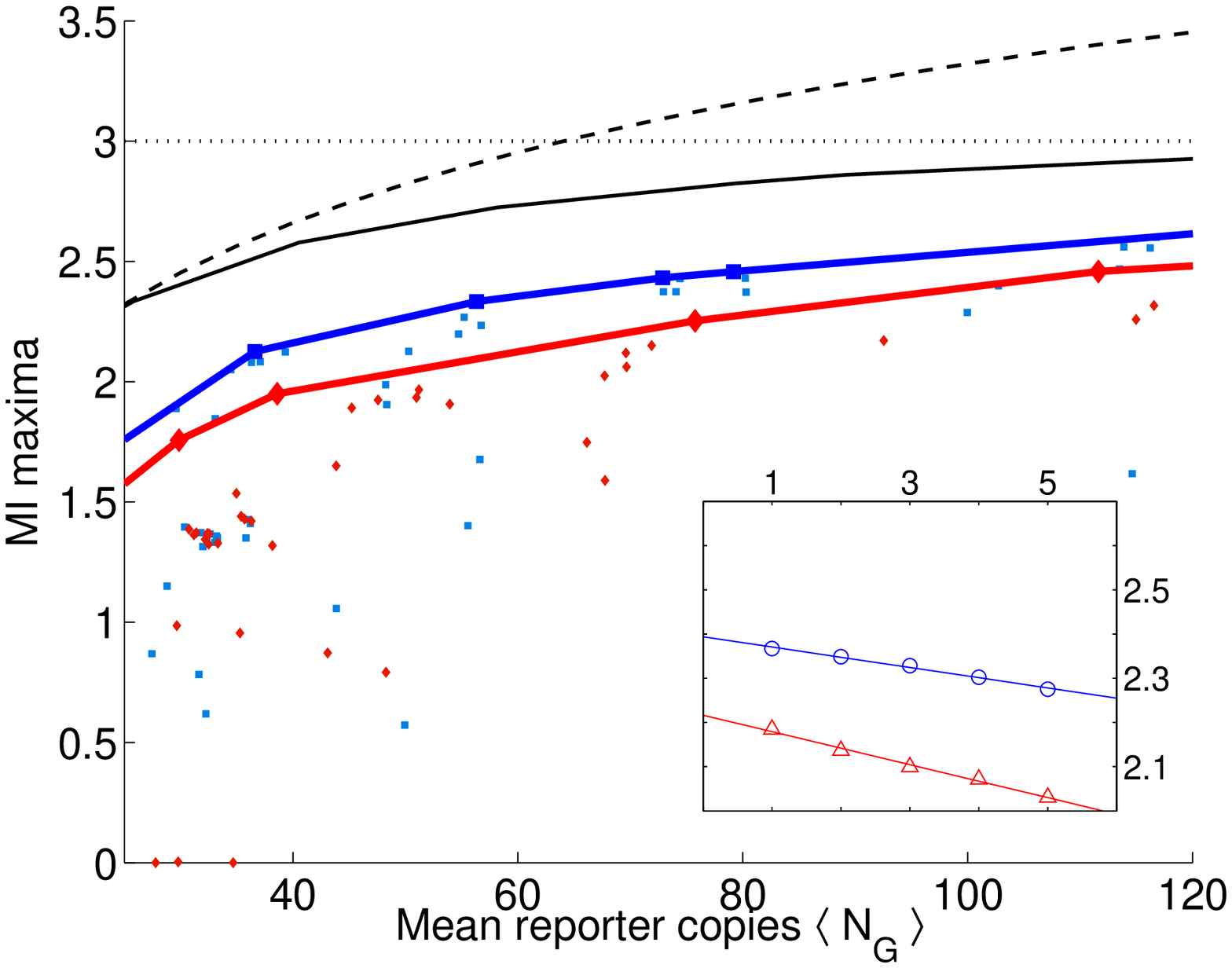} }
    \end{tabular}
  \end{center}
  \caption[Optimality in $1$-cycles]{We ran multiple optimizations $\mbold{\vartheta_*} =
    \mbox{argmax}_{\mbold{\vartheta}} \mbox{ }L$. For each
    optimization run, we plot the mutual information
    $I_*=I(C,G|\mbold{\vartheta_*})$ vs. the mean number of molecules
    of the reporter protein $\langle N_{G}\rangle$. Input distribution
    $p(c)=1/||C||$ and $||C||=8$ so that $I(C,G)\leq H(C)=3 \mbox{
      bits}$. Blue squares and red triangles are for $\gamma=0.001$
    and $\gamma=0.01$, respectively. The blue and red linearly
    interpolated lines correspond to the convex hull for each
    respective $\gamma$ value. The black solid curve gives the numerically
    evaluated optimal bound (cf. Sec.~\ref{sec:results_opt}) and dashed curve gives analytic bound for any input distribution (cf. Sec.~\ref{sec:mm_maxmi}). 
    {\bf{Inset:}} $\langle I \rangle$ as a function of a fraction of data $m$
    included in the analysis (cf. Sec.~\ref{sec:results_opt}). Blue and red correspond to two different
    $\gamma$ values. Linear regression extrapolated to case of
    infinite data (y-intercept). Results for two typical circuits with
    $1$-cycles: {\bf{(a)}} Circuit 19 with odd number of negative
    regulators in cycle and {\bf{(b)}} Circuit 11 with even number of
    negative regulators in cycle. Note here as in Fig.~\ref{Ivnfigs2}
    and Fig.~\ref{Ivnfigs3} circuits on left have larger $\langle I
    \rangle $ values as well as smaller differences between the two $\gamma$
    values than circuits on right.}
\label{Ivnfigs1}
\end{figure}

We tested the ability of each of the $24$ different circuits to
reliably transduce input signals.  For each circuit, we numerically
optimized Eq.~\ref{eqn:L} at different $\lambda$ and $\gamma$. The results of a
single optimization thus give us a local maximizer
$\mbold{\vartheta_*} (\lambda,\gamma)$ of $L$. For each numerically
obtained $\mbold{\vartheta_*}$ we then plot the corresponding mutual
information $I_*$ (as calculated by Eq.~\ref{eqn:mi}) as a function of
the actually observed average number of reporter molecules $\langle
N_G \rangle=\frac{1}{||C||M_c}\sum_{c=1}^{||C||}\sum_{i=1}^{M_c}
g_i^c$. For example, in Fig.~\ref{Ivnfigs1} we show the results of
multiple maximizations for two typical circuits. Each point on the
plot corresponds to a $\mbold{\vartheta_*}({\lambda,\gamma})$. The
blue squares and the red diamonds correspond to the two different
$\gamma$ values and the solid lines correspond to the ``best''
solutions which we determine by finding the convex hull of the set of
all maxima. 

\begin{figure}
\begin{tabular}{ll}
{\includegraphics[width=1.4in] {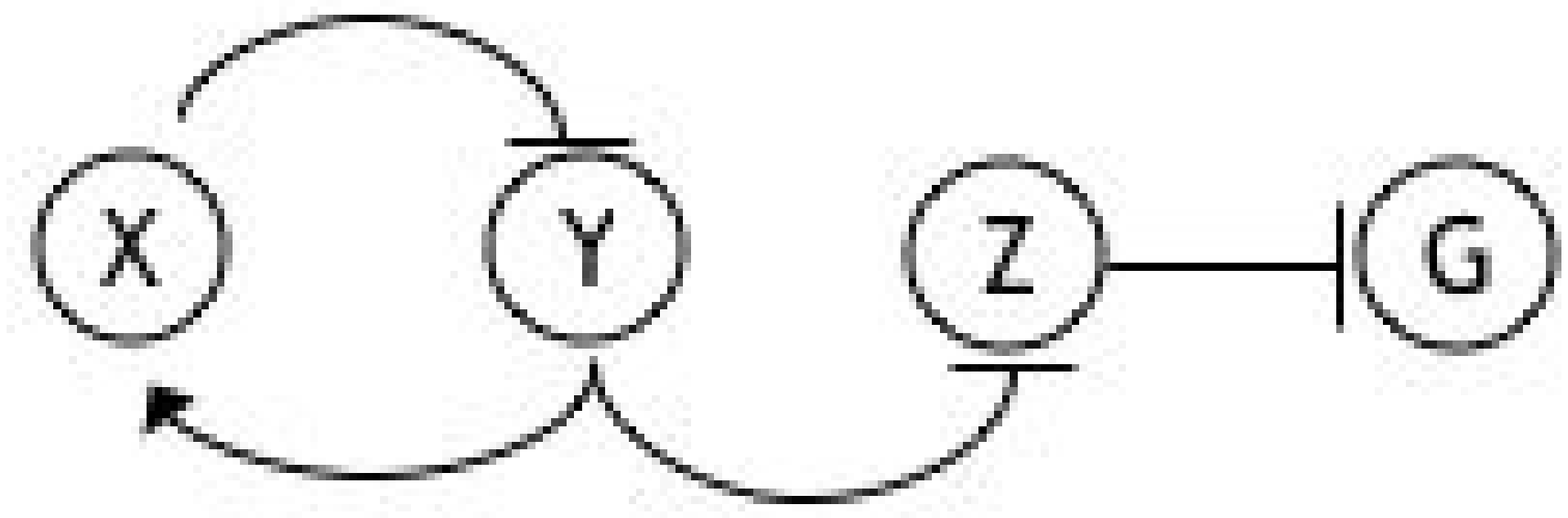} } & {\includegraphics[width=1.4in] {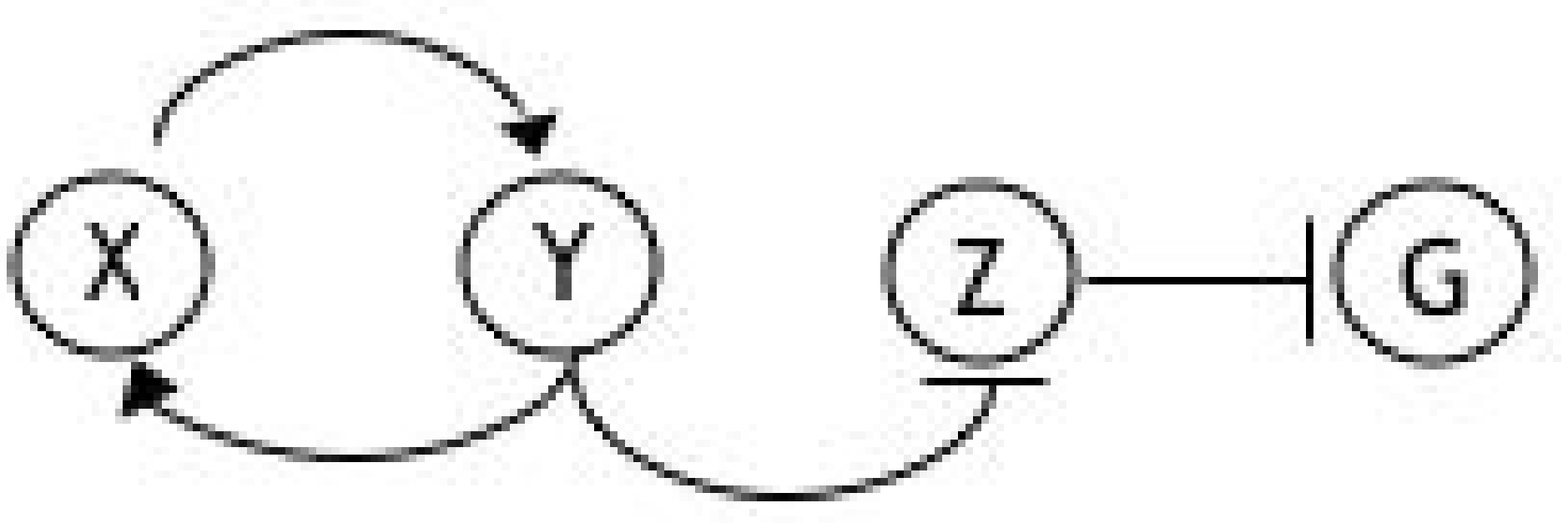} }  \\
{\includegraphics[width=3in] {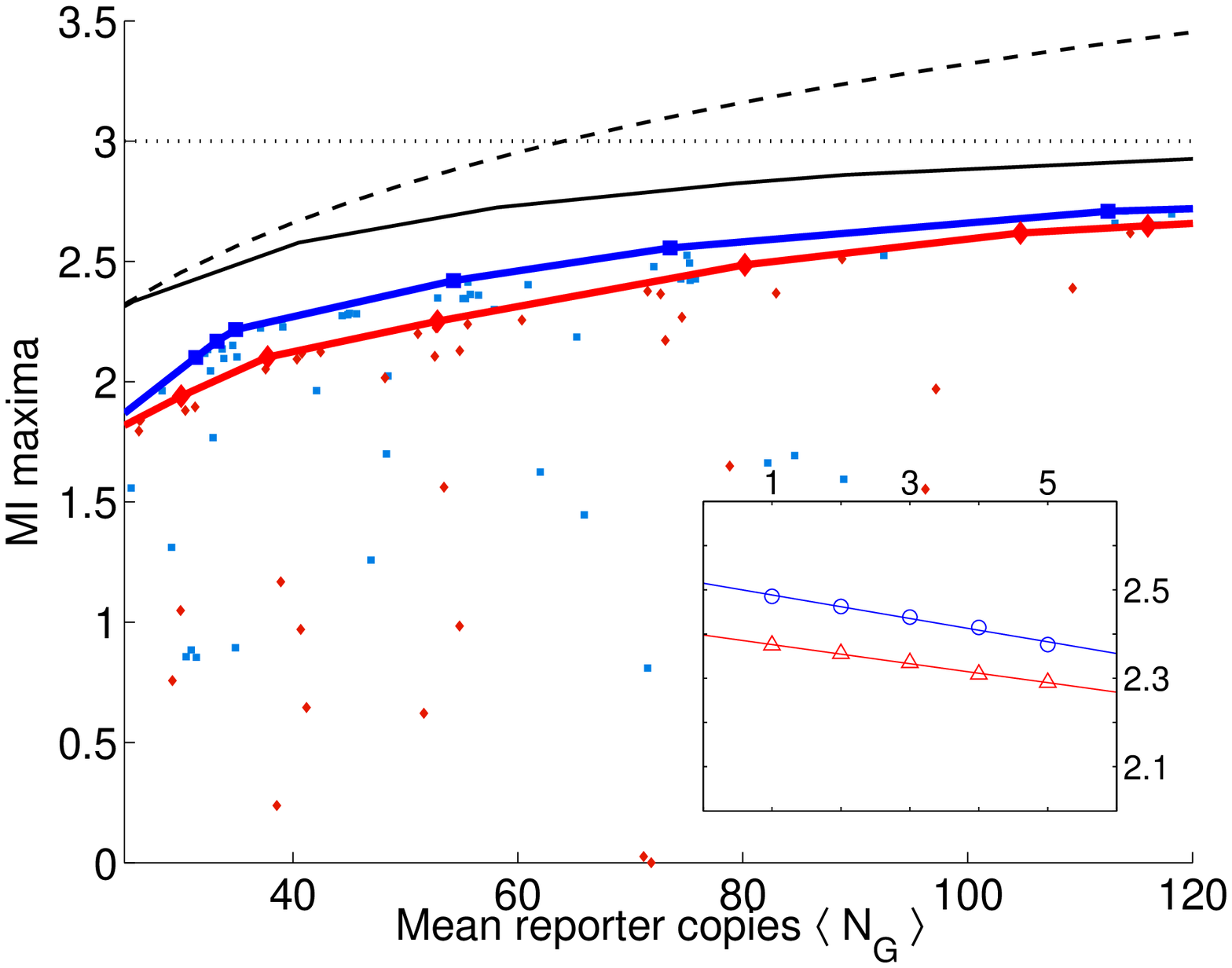} } & {\includegraphics[width=3in] {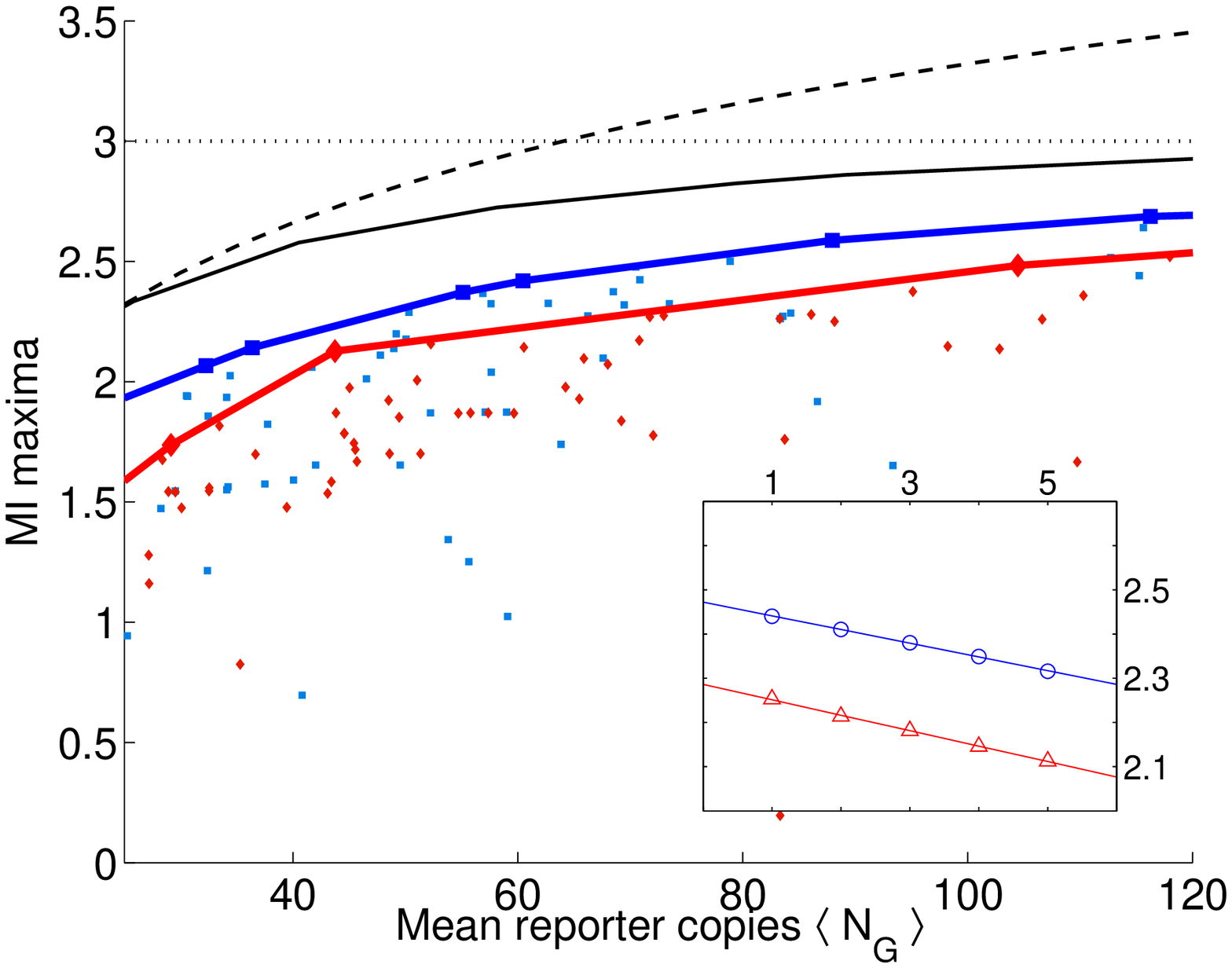} }  
\end{tabular}
 \caption[Optimality in $2$-cycles]{Same as in Fig.~\ref{Ivnfigs1} for two circuits with $2$ cycles: {\bf{(a)}} Circuit 23 with odd number of negative regulators in cycle and {\bf{(b)}} Circuit 5 with even number of negative regulators in cycle.}
\label{Ivnfigs2}
\end{figure}

\begin{figure}
\begin{tabular}{ll}
{\includegraphics[width=1.1in] {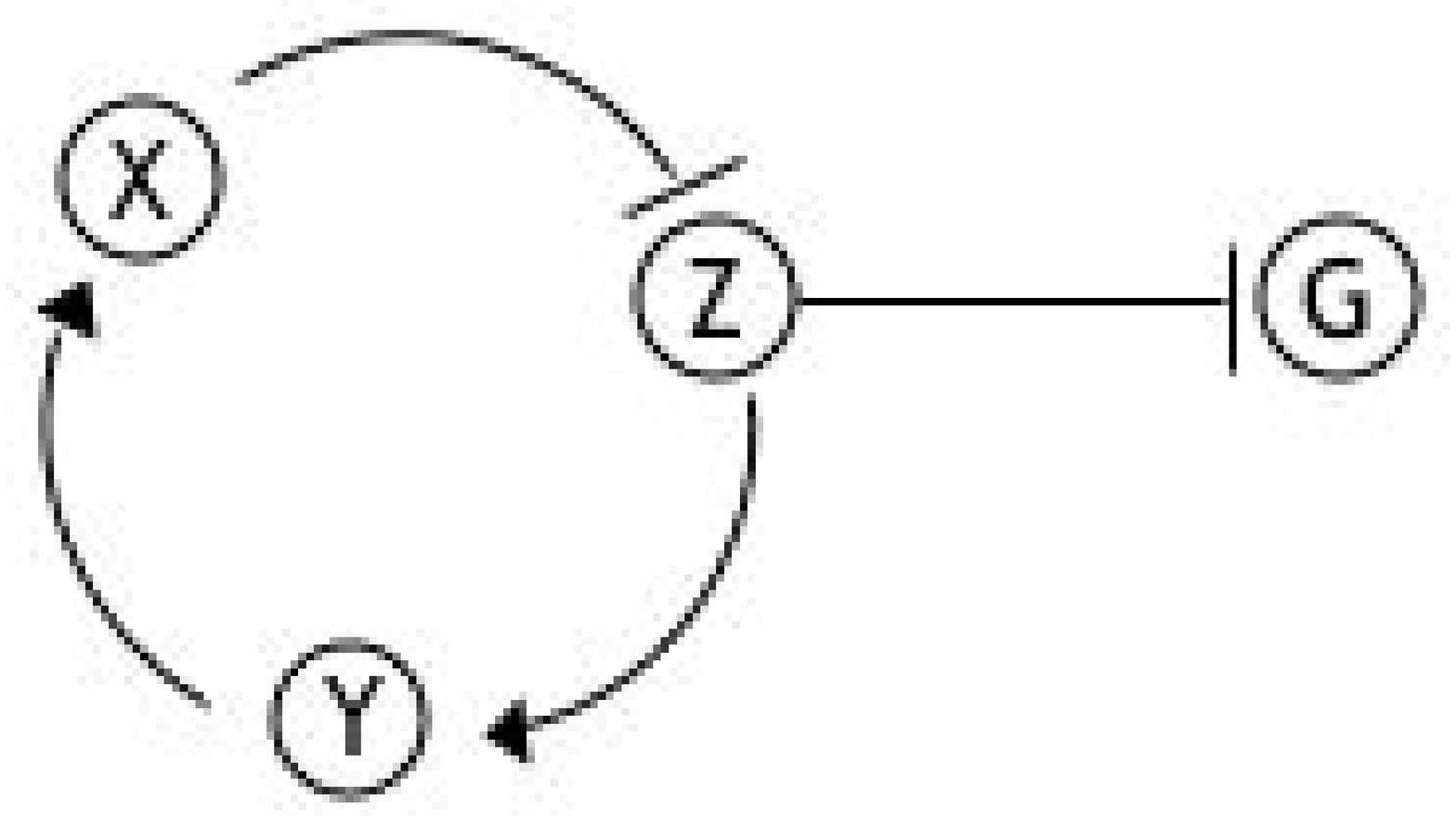} } & {\includegraphics[width=1.1in] {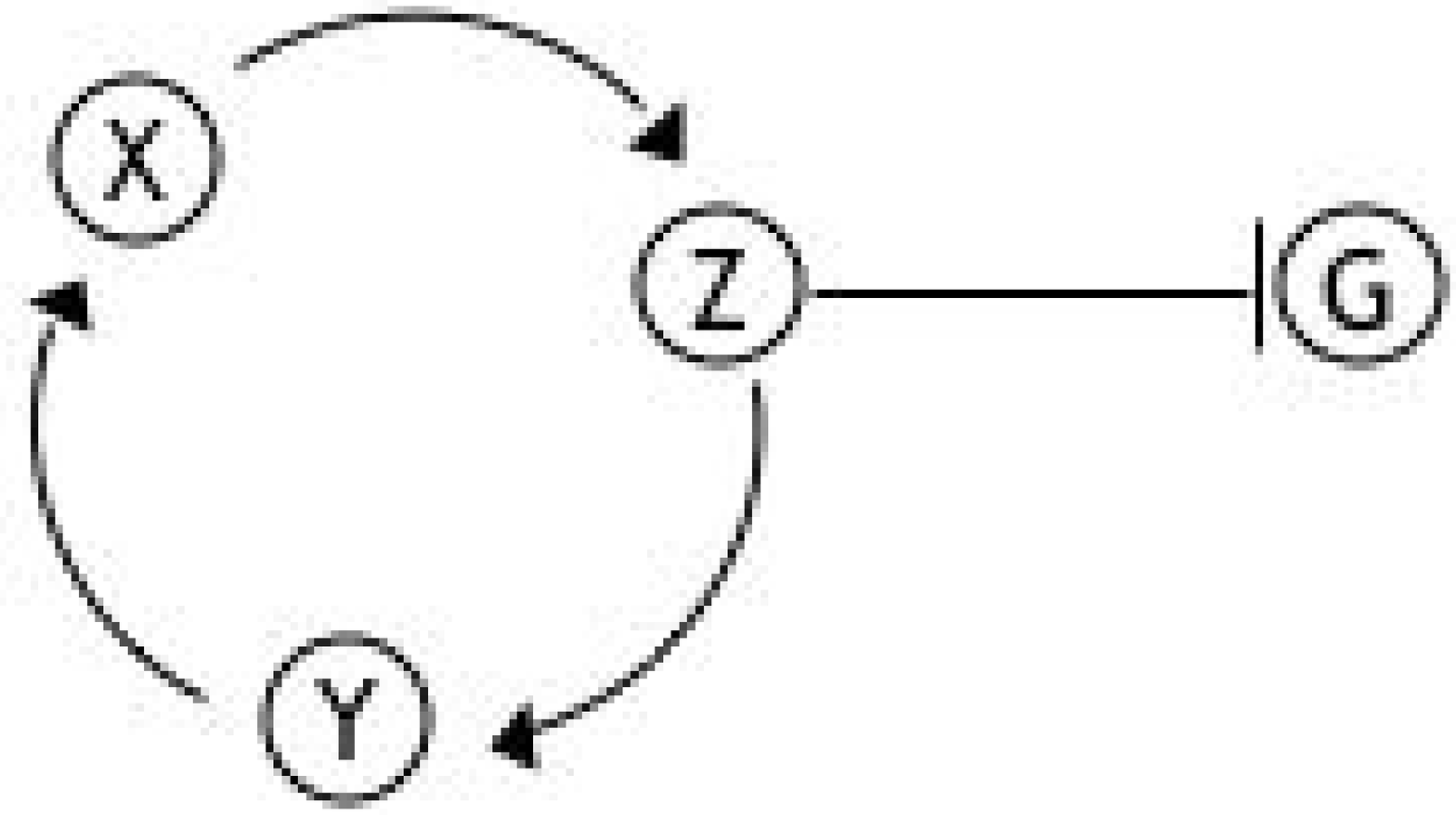} }  \\
{\includegraphics[width=3in] {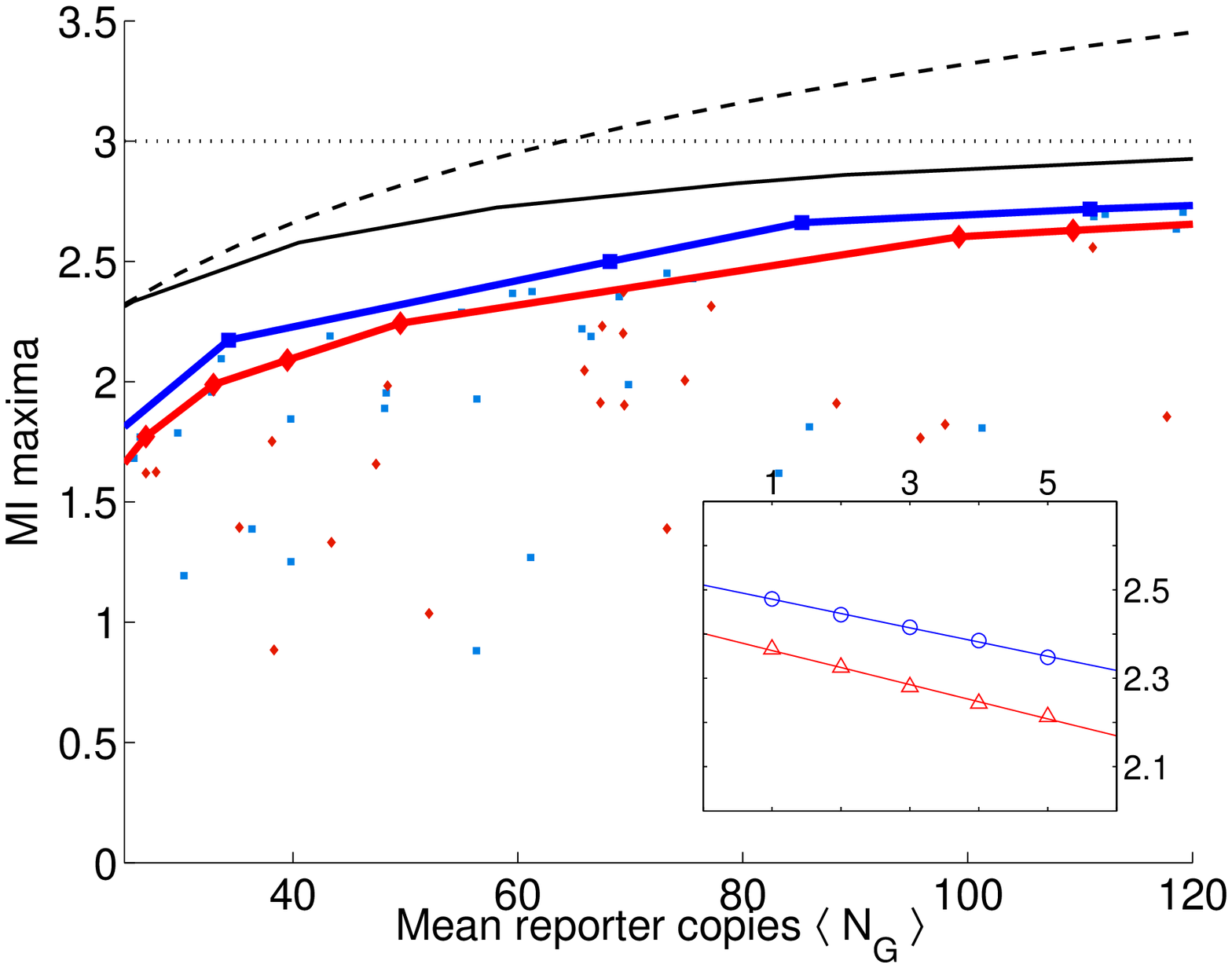} } & {\includegraphics[width=3in] {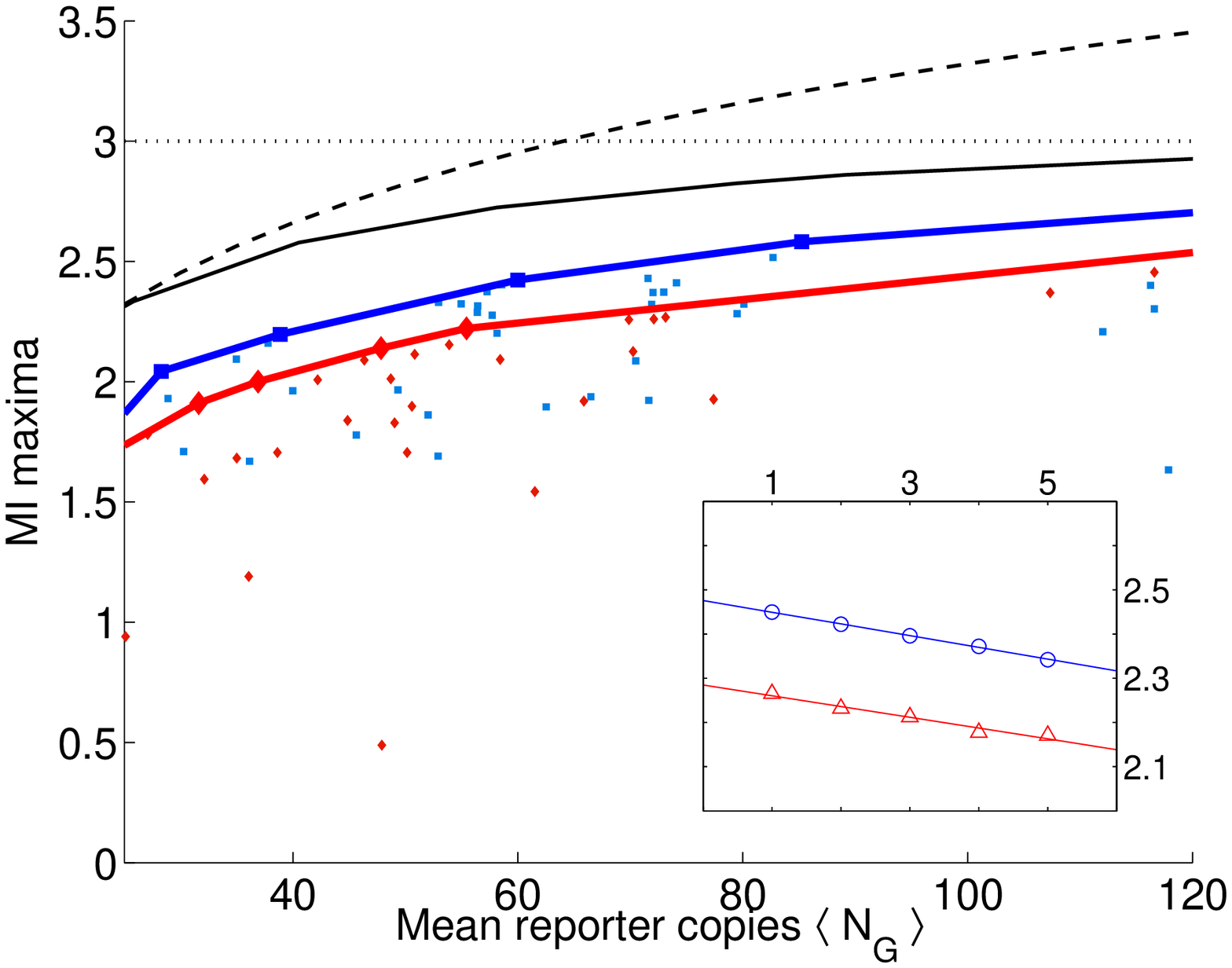} }  
\end{tabular}
 \caption[Optimality in $3$-cycles]{Same as in Fig.~\ref{Ivnfigs1} for two circuits with $3$ cycles: {\bf{(a)}} Circuit 13 with odd number of negative regulators in cycle and {\bf{(b)}} Circuit 17 with even number of negative regulators in cycle.}
\label{Ivnfigs3}
\end{figure}

Not surprisingly, as the $\lambda$ constraint is weakened, and higher
molecule numbers are allowed, more information is transduced on
average (blue and red curves always increase monotonically), 
Similarly, as the $\gamma$ constraint is increased, and the stiff solutions are
constrained, less information is transduced (red curve is always less
than or equal to blue curve). We report that all $24$ topologies can
pass more than $1$ bit of information with molecule numbers far
smaller than $100$. In fact, at $25$ molecules, most circuits can pass
nearly $2$ bits of information. In short, generic topologies under
biological constraints of response time and molecule numbers can still
transduce more information than a simple binary switch.

\subsection{Determining optimal bounds}
\label{sec:results_bounds}
To determine how well the circuits performed compared to the optimal
solution, we first note that all solutions are upper bounded by the
entropy of the input distribution, which in our case is $H(C)=3$
bits. Next, recall that the reporter protein, $G$, must be at least
subject to its own intrinsic noise, and the variance of this noise
must be at least that of a Poisson distribution
($p(x)=\exp(-\lambda)\lambda^x/x!$) with mean $\lambda=g^c$ (since the
reporter does not have any feedback) \citep{paulsson}.  Given this
noise lower bound and a probability distribution over inputs $C$ (in
this case a uniform distribution of 8 delta functions), we can
numerically calculate an optimal transduction curve. That is, we
optimize
\begin{equation}
\tilde{L}=I(C,\tilde{G})-\lambda \langle N_{\tilde{G}}\rangle,
\end{equation}
where $\langle
N_{\tilde{G}}\rangle=\frac{1}{||C||}\sum_{c=1}^{||C||}\tilde{g}^c$ and
we denote $\tilde{g}^c$ to emphasize that the means $g^c$ are
optimally separated and are described by a Poisson distribution. For
different values of $\lambda$, we can define an optimal curve
$\tilde{I}$ vs.\ $ \langle N_{\tilde{G}}\rangle$, where $\tilde{I}$ is
the mutual information at the maximum of $\tilde{L}$. All $24$
topologies are upper bounded by the resulting curve, since the
reporter gene must be at least subject to its own $g^c$ noise plus any
noise translated from upstream factors.  Finally we note that
$\tilde{I}$ is always bounded by the channel capacity $I_0$ which is
defined to be the maximum over all input distributions and can be
approximated analytically as in Eq.~\ref{eq:capacity} (see section
\ref{sec:mm_maxmi}).  For $\langle N_G \rangle=25$ molecules,
$I_0=2.32$ bits and for $\langle N_G \rangle=100$ molecules $I_0=3.32$
bits.

\subsection{(Almost) optimal circuits}
\label{sec:results_opt}
We find that all $24$ circuits are able to achieve close to the
optimal transmission fidelity, implying that they are able to tune the
noise from the upstream factors to almost negligible values (see
Figs.~\ref{Ivnfigs1}-\ref{Ivnfigs3} and
Table~\ref{ivn_all}). To quantify how well the
circuits perform compared to the optimal bound and to each other, we
define the statistic
\begin{equation}
\langle I \rangle = \frac{1}{|b-a|}\int_a^b I(x)\,dx,
\end{equation} 
where $I$ is the linearly interpolated convex hull and $a$ and $b$ are
set to $25$ and $120$ molecules, respectively. Note that for our
discrete input distribution we can upper bound $\langle I \rangle
\leq 2.75$ bits, where we use the linearly interpolated curve derived
numerically $\tilde{I}$, as described in
Sec.~\ref{sec:results_bounds}; similarly, for \emph{any} input
distribution we can upper bound $\langle I \rangle \leq 3.03$ bits,
where we use the analytic approximation $I_0$ derived in section
\ref{sec:mm_maxmi}.

Since the convex hull area can only grow with the number of
optimizations we run, there is a bias in our calculated statistic
$\langle I \rangle$. That is, with $k$ optimization runs, $\langle I
\rangle_k \geq \langle I \rangle_{k-1}$. We are interested in $\langle
I\rangle=\langle I \rangle_\infty$, but this is clearly
unattainable. Moreover, for different topologies, $\langle I\rangle$
may be approached with different speeds as a function of $k$, making
comparisons between topologies suspect.  We use jackknifing to
estimate the bias. That is, in the spirit of \citep{strong,treves}, instead of
the total number of optimization runs $N_{\rm opt}$, we use only
$N_{\rm opt}/m$ of them to calculate $\langle I\rangle$. Then one can
estimate $\langle I\rangle_\infty$ by fitting
\begin{equation}
  \langle I(m) \rangle= \langle I \rangle + \frac{A_1}{m} + \frac{A_2}{m^2} + \dots.
\end{equation}
where $A_i$ are some constants. In the insets of Figs.~\ref{Ivnfigs1}-\ref{Ivnfigs3} we show the dependence of $\langle I\rangle$ on $m$,
the inverse fraction of data included. We see that for the most part
$\langle I(m) \rangle$ is well fit by a straight line, and
contributions from the higher order corrections are insignificant. The
results of extrapolating $\langle I \rangle$ to $m=0$ for each circuit
are reported in Tables \ref{table2} and \ref{table1}. The average
$\langle I \rangle$ over all circuits is $2.48\pm0.05$ and
$2.32\pm0.09$ bits for $\gamma=0.001$ and $\gamma=0.01$
respectively. We find the circuits are within $10\%$ of the optimal
transduction capacity.

\subsection{Ranking circuits}
The optimality measure $\langle I \rangle$ provides a ranking of the
topologies (see Tables \ref{table2} and \ref{table1}). While,
strikingly, all of the circuits perform close to the optimal bound,
systematic differences are revealed.  Consider for example the
linear chains with autoregulation (circuits $1$, $19$, $14$, and $4$
with negative feedback and circuits $21$, $15$, $16$, and $11$ with
positive feedback). We note that the negative feedback circuits all
have higher $\langle I \rangle$ values than their positive feedback
counterparts. Moreover, the difference (\emph{gap}) between the $\gamma=0.001$ and
$\gamma=0.01$ curves is smaller for the negative feedback circuits
than for the positive feedback circuits. That is, even when the
stiffness is constrained, these circuits still do well, whereas the
other circuits are more reliant on stiff dynamics.  These results are
consistent with findings in \citep{serrano} that autorepressive
circuits can help regulate noise. Interestingly, this trend can be
generalized to the circuits with longer cycles as well. For example,
we also find that for circuits with $2$-cycles, those that
perform best are those that have opposite regulations (one repressive,
one activating) rather than two activating or two repressing
regulators. For the case of $3$-cycles, those circuits with $1$ or $3$
negative regulators have on average higher values of $\langle I
\rangle$. In Figs.~\ref{Ivnfigs1}-\ref{Ivnfigs3}
we display curves for typical $1$-, $2$-, and $3$- cycles,
respectively with both odd (left column) and even (right column)
number of negative regulators.

\begin{figure}
\begin{center}
{\includegraphics[width=6in] {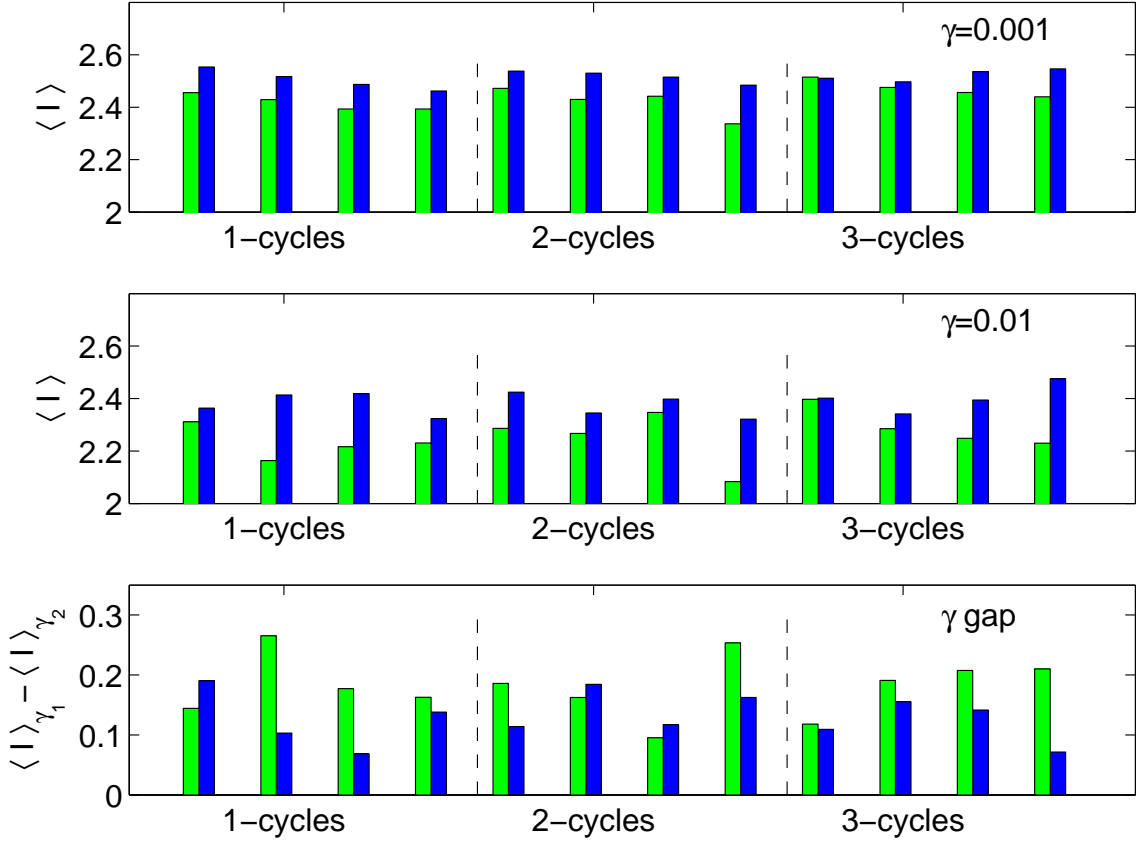} } 
 \caption[Optimality scores for odd and even cycles]{Bar graphs for $\langle I \rangle$ values for the two classes of circuits: odd (blue) includes circuits with cycles containing an odd number of regulators and even (green) includes circuits with cycles containing an even number of regulators. Top $\gamma_1=0.001$, middle $\gamma_2=0.01$, and bottom $\gamma_1-\gamma_2$. For all $3$ measures, there is a statistically significant difference between the two classes of circuits as calculated by the U Test (top $p=0.0002$, middle $p=0.0003$ and bottom $p=0.01$).}
\label{bar_visual}
\end{center}
\end{figure}

These findings imply that there are some structural constraints that
impart small, but measurable limitations to the circuit's transduction
capacity. In particular, those circuits with an odd number of
regulators (an overall negative feedback) in their cycles are
generally ranked higher than those circuits with an even number of
regulators (an overall positive feedback), see Tables \ref{table2} and
\ref{table1}. In Fig.~\ref{bar_visual} we show a bar graph of the
values of $\langle I \rangle$ for the two classes of circuits (odd and
even number of regulators in the cycle) for different $\gamma$ values
and for different length cycles. The average mutual information for
the circuits with an odd number of negative regulators is $2.51\pm
0.03$ and $2.39\pm0.05$ for the two $\gamma$ values, whereas for the
circuits with an even number of negative regulators it is $2.44\pm.03$
and $2.26\pm.05$ for the two $\gamma$ values. Between the two classes,
these values are more than one standard deviation apart. To test the
significance of this observation, we perform the non-parametric
Mann-Whitney U (Wilcoxon) Test \citep{mann_whitney,wilcoxon}, which measures the
difference in medians between two samples. We find for $\gamma=0.001$,
$U=8$ and $p=0.0002$, and for $\gamma=0.01$, $U=10$ and
$p=0.0003$. That is, the null hypothesis that the optimality measures
for the two classes of circuits (odd and even number of regulators,
or, alternatively, overall negative and positive feedback) are drawn
from the same distribution and, therefore, have the same medians, is
highly unlikely.

\subsubsection{Noise analysis}
Since circuits containing cycles with an odd number of negative
regulators are better signal transducers, we might expect that they
are able to control the noise variance better. In fact, using
LNA (cf. Sec.~\ref{sec:mm_lna}), we prove this
assertion for a generic transcriptional network in
Sec.~\ref{sec:mm_noise}.  Furthermore, for simple networks we
demonstrate that the overall negative feedback is a necessary and, in
one case, even a sufficient condition to achieve sub-Poisson noise
(variance less than the mean).

For example, let $d\phi_i/dt=f_i\equiv -r_i\phi_i +
\alpha_i(\phi_{\pi_i})$ describe the deterministic dynamics of gene
$i$ (see Sec.~\ref{sec:mm_model} for explanation of the notation). In
a steady-state $\bar{\phi}_i=\alpha_i(\phi_{\pi_i})/r_i$. Then, for a
$1$-cycle where $\pi_i\equiv i$, Eq.~\ref{eqn:var} for a species
variance reduces to
\begin{equation}
  C_{ii} = \frac{\phi}{1-\frac{\alpha'}{r}}
\end{equation}
where $\alpha'$ is the derivative of the gene expression function, and
the above is evaluated at the deterministic steady state. In the case
of an auto-repression, $\alpha'<0$, and the variance $C_{ii}$ is
less than the mean \citep{nir,serrano}.

Similarly, Eq.~\ref{eqn:var} can be reduced for a $2$-cycle,
$i,j=\{1,2\}$
\begin{eqnarray}
C_{ii}&=&\phi_i + \frac{1}{r_i}\alpha'_iC_{ij} \\
C_{ij}&=&\frac{1}{r_i+r_j} (\alpha'_iC_{jj} + \alpha'_{j}C_{ii}).
\end{eqnarray}
Since $C_{ii}>0$, here a necessary (but not sufficient) condition for
sub-Poisson noise is $\alpha'_1\alpha'_2<0$.

This analysis (as well as the derivation in Sec.~\ref{sec:mm_noise})
also illustrates that it is easier to obtain smaller variance (and
hence larger mutual information), for cycles of shorter length. This is
in agreement with \citep{gleiss} where it was found that short cycles
are over-represented in a metabolic network, but large cycles occurred
less frequently than one would expect given several different possible
null models.

\subsection{Reliance on large $\langle q \rangle$}
The difference (gap) between the two $\gamma$ curves also suggests a statistic
to compare the circuits. Presumably, a large difference implies that the
circuit relies on large stiffness $\langle q \rangle$ to regulate
noise. Indeed, for large $\langle q \rangle$, the objective
function $L$ decreases, though this decrease is moderated by the value
of $\gamma$ such that smaller $\gamma$ values allow larger $\langle q
\rangle$ values. Stiff solutions have the advantage of allowing the
reporter protein to effectively act as a low-pass filter, slowly
sampling and responding to fluctuations in the circuit components. A
reliance on small values of $\gamma$ implies that the circuit has more
difficulty regulating noise. We therefore expect the circuits with
an odd number of negative regulators to have smaller gaps. Consistent
with this prediction, while the average gap over all circuits was
$0.16\pm0.05$, the average gap for the negative feedback circuits was
$0.13 \pm 0.04$, and for the rest it was $0.18 \pm0.05$ (see
Fig.~\ref{bar_visual}). The U Test using the gap measure gives $U=28$
and $p=0.01$, indicating a statistically significant difference.

Evidence from a database of transcription factors in prokaryotes
supports the finding that circuits with negative feedback can suppress
noise \citep{regulondb}.  In \emph{E. coli}, many
transcription factors do not undergo active degradation via
proteolysis, but are instead only passively degraded via dilution. The
half-lives of such proteins are on the order of the lifetime of the
cell, allowing them to respond only slowly to fluctuations in the mRNA
concentrations. As in the case of the stiff solutions with large
$\langle q \rangle$ in our circuits, these slowly responding
transcription factors have an advantage in noise control
\citep{peskin_rna}. Therefore we might expect that transcription
factors that do not undergo proteolysis will have no auto-repression,
or even positive auto-regulation. On the other hand, transcription
factors that do undergo proteolysis and cannot, therefore, filter mRNA
fluctuations as well would be more likely to require negative
auto-regulation. To test this hypothesis, we analyzed $145$
transcription factors of the \emph{E. coli} regulatory network (see supplemental material in \citep{bsp_supp}). For
each transcription factor we correlated whether the factor is
auto-repressive \citep{regulondb} with whether it potentially undergoes
proteolysis (by noting if the peptide sequence had any known cleavage
sites) \citep{MEROPS}. We found that of the $13$ transcription factors
which are likely to undergo proteolysis, $9$ are negative
auto-repressors, and out of the $132$ transcription factors which do
not, $88$ are not auto-repressors (see Table~\ref{fishertable}).  A Fisher exact probability test
revealed a statistically significant positive association between proteolysis and negative
feedback ($p=0.01$).

\begin{table}[htdp]

\begin{center}
\begin{tabular}{|c|c|c|}
 \hline
 	& & \\
 	& {\bf{Negative Feedback}} & {\bf{No Negative Feedback}}  \\  \hline
	& & \\
{\bf{Proteolysis}}& 9	& 4 \\  \hline
& & \\
{\bf{No Proteolysis}}  & 44 & 88 \\ 
& & \\
\hline
\end{tabular}
\caption[Table of proteolysis and auto-repression correlation]{Table comparing presence or absence of proteolysis to presence or absence of negative auto-regulation. Fisher exact probability test reveals signficant ($p=0.01$) positive association. This supports our prediction that transcription factors which undergo proteolysis, and therefore have faster response times, are less able to regulate noise using the slow response, filtering solution and require the presence of negative auto-regulation to help control their noise.}
\label{fishertable}
\end{center}
\end{table} 

\subsection{Robust, adaptive maxima}
\label{sec:results_robust}
An important consideration in further assessing the quality of our
circuits is the extent to which these high information maxima are
\emph{robust} to perturbations in the system. Qualitatively, we define
a maximum as robust if, in its vicinity, the cost function $L$ does
not change significantly in response to perturbation of the parameters
$R,K,a,a_0$, and $s$ (see Sec.~\ref{sec:mm_model} for parameter
definitions).  Relatedly, we would also like to consider the ability
of our circuits to \emph{adapt} their behavior to such
perturbations. An adaptive maximum does not significantly decrease its
function value $L$ in response to parameter perturbation, but does
alter its \emph{behavior}. Here we characterize the behavior of the
circuit at each constrained mutual information maximum by ordering
$g^c$.
That is, two maxima have different behaviors if the permutation
yielding the sorted sequence of $g^c$ is different.

\begin{figure}[tbp]
\begin{center}
\begin{tabular}{ll}
{\includegraphics[width=2.8in] {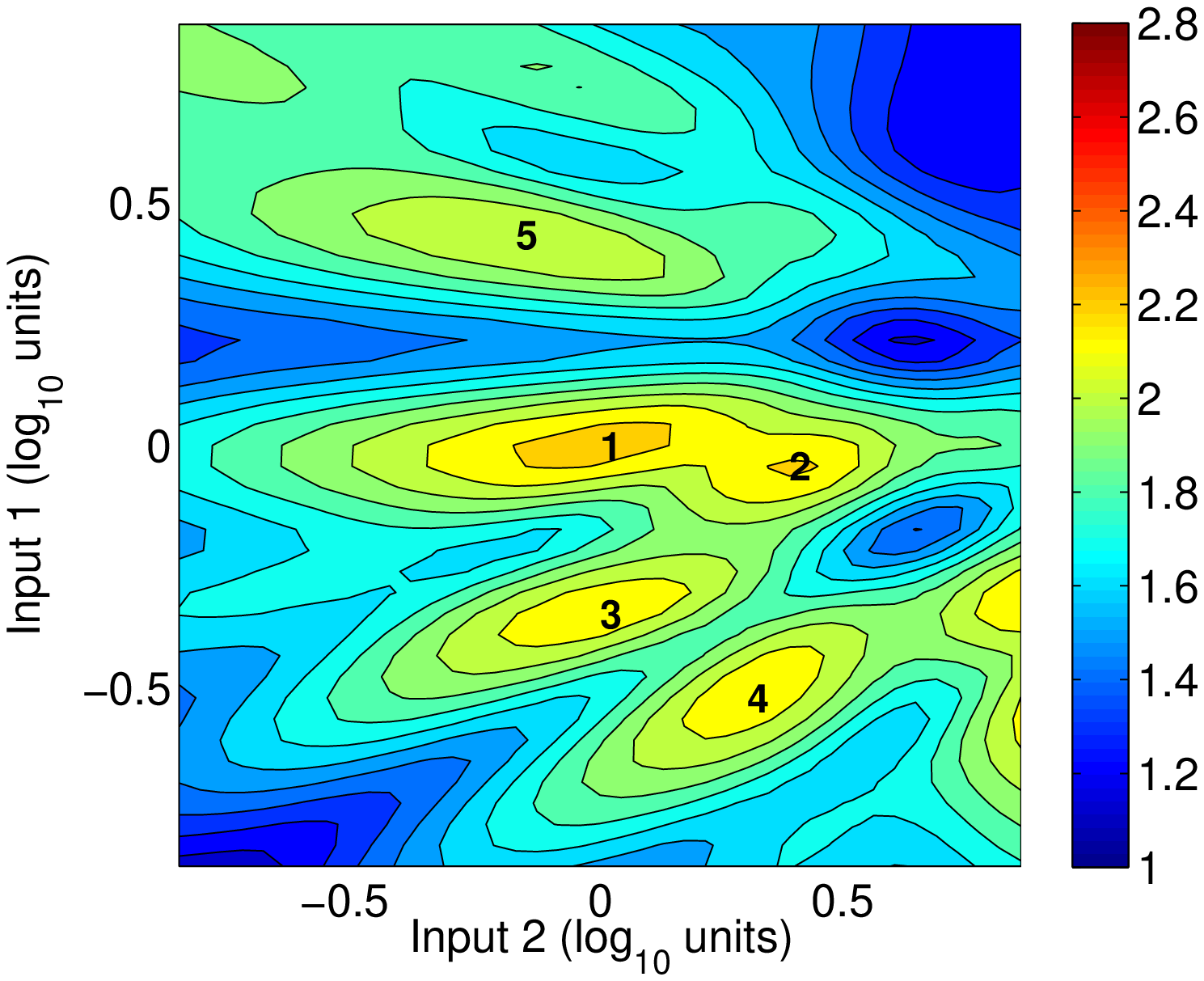} } &
{\includegraphics[width=2.8in] {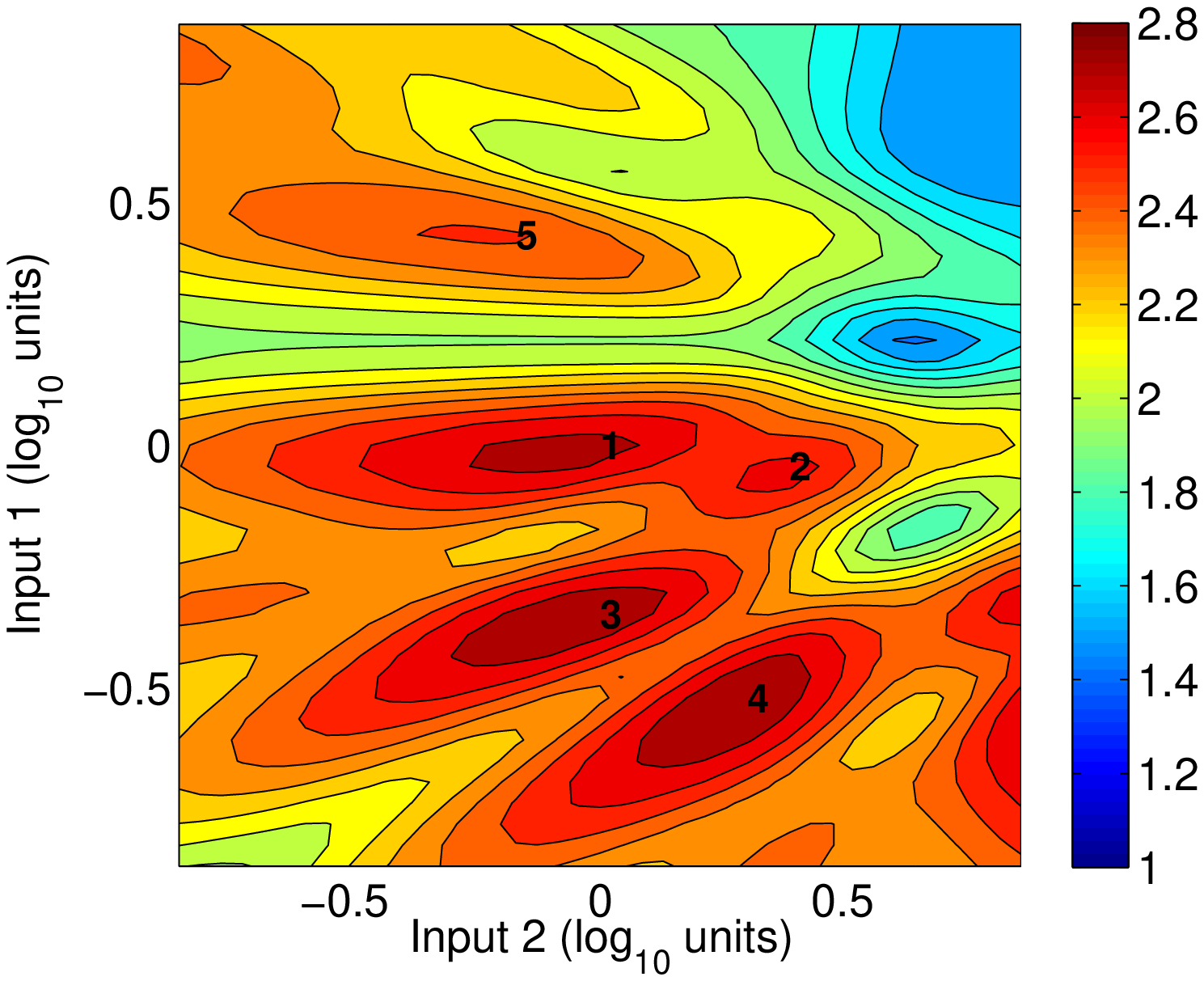} }\\ 
{\bf{(a)}} & {\bf{(b)}} 
\end{tabular}
 \caption[Cluster of mutual information maxima]{({\bf{a}}) The objective function $L$ and ({\bf{b}}) the mutual information $I$  as a function of the first two input parameters (input 1 is $s_X$ and input 2 is $s_Y$) for circuit $2$. The rest of the parameters are held constant for this figure. The five labeled peaks correspond to $5$ distinct behaviors or unique signal encodings (cf.~Fig.~\ref{spectra} and Table~\ref{fn_table}).}
\label{contours}
\end{center}
\end{figure}

As a preliminary investigation, we analyzed the functional $L$ of
circuit $2$ near one of its randomly selected maximum. In addition to
the original maximum, we found four other distinct nearby peaks as
displayed in Fig.~\ref{contours}. The circuit alters its behavior as a
result of changes along the $2$ displayed dimensions (cf.~Eq.~\ref{dynamics}), input $1$
($s_X$) and input $2$ ($s_Y$), so that at each
maximum the ordering of responses is distinct, and the signal is
encoded in a different way (see Table~\ref{fn_table}). Note that $4$ of the maxima are separated
by valleys no deeper than $2.3$ bits. In other words, by a change in
$s_X$ and $s_Y$ only, the circuit can alter its behavior, while
maintaining a high transmission fidelity. In this sense, we consider
these maxima to be adaptive.

\begin{table}[htdp]
\begin{center}
\begin{tabular}{||c||c|c|c|c|c|c|c|c||}
\hline
{\bf{Chemical State}} &{\bf{ 000}} & {\bf{001}} & {\bf{010}}& {\bf{011}}& {\bf{100}}& {\bf{101}} &{\bf{110}}&{\bf{111}} \\
\hline \hline
{\bf{Peak 1}} &     2 &    6  &   1  &   5  &   4  &   8 &    3  &   7 \\
\hline
{\bf{Peak 2}} &    2  &   6   &  4   &  1  &   5   &  8  &   3  &   7 \\
\hline
{\bf{Peak 3}} &    2  &   1   &  4  &   6  &   3  &   5  &   8  &   7 \\
\hline
{\bf{Peak 4}} &    2  &   1   &  6 &    4  &   5  &   3  &   8  &   7 \\
\hline
{\bf{Peak 5}} &    6  &   2   &  5 &    1   &  8  &   4  &   7  &   3 \\
\hline \hline
\end{tabular}
\end{center}
\caption[Table of behaviors]{Table of behaviors corresponding to the five peaks shown in Fig.~\ref{contours}. Behavior is defined as the ordering of $g^c$, where $g^c$ is the deterministic steady-state solution for given chemical input $c$ and $c\in\{000,001,010,011,100,101,110,111\}$. Each row describes the behavior of the circuit at one of the five maxima.}
\label{fn_table}
\end{table}

\begin{figure}[tbpl]
\begin{center}
{\includegraphics[width=6in] {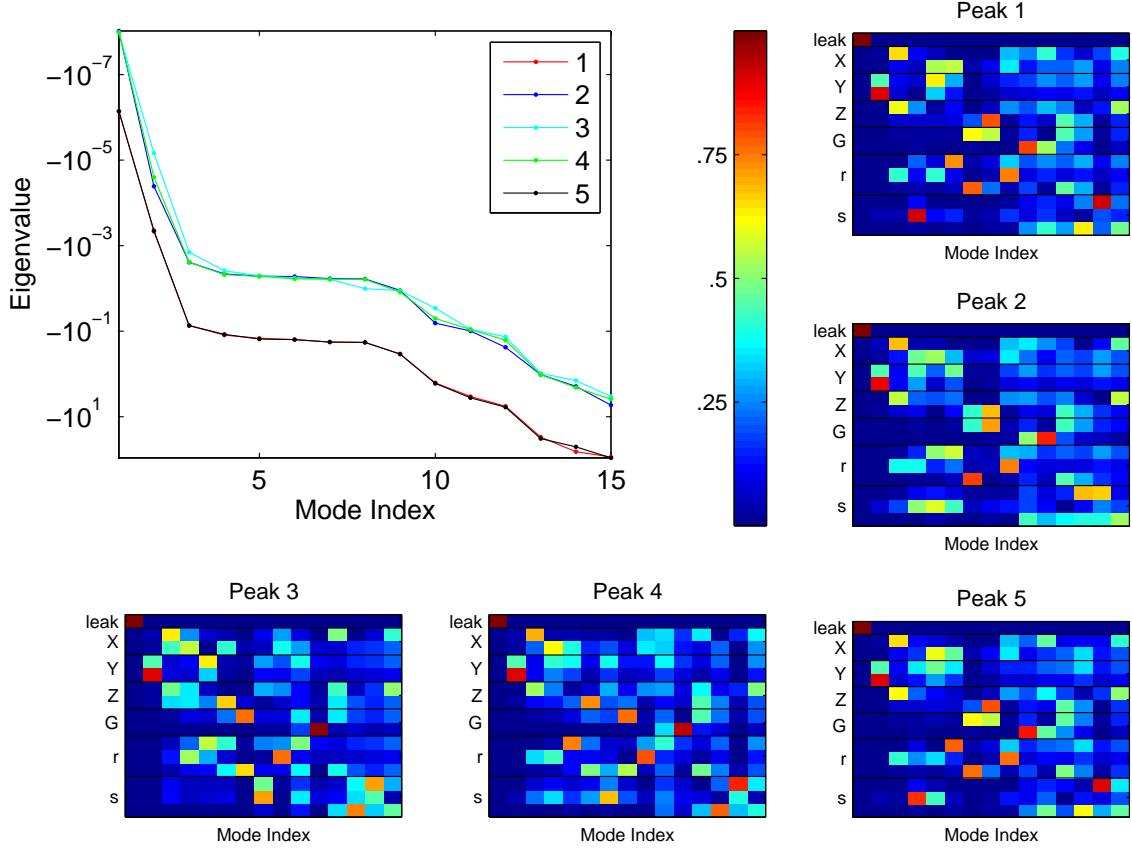} } 
\caption[Eigendecomposition of Hessian at maxima of cost function $L$]{{\bf{Top-left:}} Spectra for the numerically calculated Hessian at each of the corresponding
  $5$ peaks labeled in Fig.~\ref{contours}. Soft modes ($\to 0$) are directions in which $L$ has small curvature, hard modes ($\to -\infty $) are directions in which $L$ has large curvature. Many eigendirections exhibit small curvature (greater than $-10^{-2}$ for peaks 2-4 and $-10^{-1}$ for peaks 1 and 5), demonstrating that the maxima are robust to large deviations in parameter space. {\bf{Colored panels:}} Magnitude of contribution from each parameter to each eigenvector for each of the $5$ Hessians. Mode index is sorted as in top-left figure (from least curvature to greatest curvature). Row labeled leak corresponds to parameter $a_0$. Paired rows labeled $X$, $Y$, $Z$, and $G$ correspond to the two parameters, $K$ and $a$, describing the gene regulation function for each transcription factor ($X$, $Y$, $Z$) and reporter protein ($G$). Rows labeled $r$ correspond to the decay rates of each of the $3$ transcription factors. Rows labeled $s$ correspond to the input parameters modulating the three transcription factors. For all $5$ peaks, the two most soft modes correspond to $a_0$ and a mixture of $K_Y$, $a_Y$, respectively. $s_X$ and $s_Z$ contribute mostly to the hard modes.}
\label{spectra}
\end{center}
\end{figure}

We next numerically calculated the Hessian at each of the $5$
peaks. In Fig.~\ref{spectra}, we plot the Hessian spectra along with
the corresponding eigenvectors. By treating $L$ as locally quadratic
near each maximum we use the Hessian (evaluated with respect to
$\log_{10}$ of the parameters) to analyze how sensitive the maximum is
to deviations in the parameters. For example, for an eigenvalue equal
to $-1$, moving $10$-fold in the corresponding eigendirection would
result in a loss of 0.5 for the objective function.  Alternatively, an
eigenvalue equal to $-0.1$ means that we can move $10$-fold in that
direction while decreasing the objective function only by 0.05. This
should be compared with the typical values near maxima of $L,I \sim 2$
bits. We find that for most directions for all $5$ peaks
eigenvalues are less than $-0.1$. In this sense, we consider these
maxima to be robust.  

We can identify three different regimes for the spectra: an extremely
``soft'' regime corresponding to the first two modes, a second
soft regime, where modes 3 to 9 are basically equivalent, and
a third regime (modes 10 through 15), where the eigenvalues
become more negative. We note that the spectra for peaks $1$ and $5$ overlap
almost entirely, as do the ones for $2,3$ and $4$, and that the latter
appear to be more robust (eigenvalues are closer to $0$). Interestingly, all five spectra in Fig.~\ref{spectra} are
similar, due to the fact that the ${\mbold{\vartheta_*}}$ are
themselves quite similar --- that is, the maxima are closely arranged
not just in the $2$-dimensions displayed in Fig.~\ref{contours}, but
over all $15$ dimensions. This underscores the circuit's adaptability.

In Fig.~\ref{spectra}, we have also displayed the contributions from
each parameter to each eigenvector for all $5$ peaks. It is clear that
the first mode corresponds entirely to the leak parameter, which for
all $5$ peaks is being driven to $0$ as the optimization proceeds. The
second mode is also consistent for all $5$ peaks, and it corresponds
to the parameters $a_Y$ and $K_Y$ (cf. Sec.~\ref{dynamics}), governing
creation for the transcription factor $Y$. Essentially the range of
the gene activation function, $a_Y$, is increased while the
Michaelis constant $K_Y$ is decreased, so that $Y$ is squeezed to low
copy numbers. This is a reasonable strategy since maximizing the
information in the output signal requires that most of the energy
spent on building molecules is expended on the reporter protein. 

Another consistent finding for all of the peaks is that inputs $1$
($s_X$) and $3$ ($s_Z$) are more critical (less robust) than input $2$
($s_Y$). That is, in general, $s_X$ and $s_Z$ contribute to the
highest $3$ modes, whereas $s_Y$ contributes mostly to modes in the
``soft'' regime. Qualitatively, this is consistent with
Fig.~\ref{contours}, where  most of the peaks are ellipses with
major axes roughly in line with $s_Y$. Note that this is most evident
for peaks $1$ and $5$ for which $s_X$ and $s_Y$ correspond almost
exclusively to modes $14$ and $4$, respectively.

\section{Discussion}
\label{sec:discussion}
We have presented an information-theoretic, function-independent measure of circuit quality. We have demonstrated that generic, small networks can transduce more information than a simple binary switch; moreover, such generic topologies can achieve close to \emph{optimal} transmission fidelity, even under low copy number and fast response time (non-stiff) constraints. High information solutions can be robust to $10$-fold deviations in parameters. 

That such simple, stochastic systems can act as high-fidelity signal transducers suggests a possible explanation for the cross-talk dilemma, in which multiple ligands trigger the same signaling pathway, and yet reliably produce distinct genetic outputs. We have demonstrated that multiple discrete input states can be transduced by the same signaling pathway and even the same molecule if the encoding is in molecule numbers. Moreover, when the trivial solutions to this problem are constrained (high copy number and slow response time) the input state can still be transduced reliably, implying that these simple circuits have enough flexibility to regulate noise. To our knowledge, this is the first demonstration that a simple, stochastic system can overcome cross-talk, without invoking the traditional sequestering argument \citep{schwartz} (where signal specificity is achieved by spatially or temporally sequestering pathways with shared components). 

It may be possible to correlate some of the solutions of the circuits with experimental data to investigate to what extent is transduction optimality an essential goal of these systems. For example, a common generic solution of our circuits was to decrease the decay rate and increase the average molecule number of the reporter protein or the proteins that appear near the end of the circuit. The slower decay rate meant the reporter protein could temporally average the fluctuations of the proteins at the beginning of the circuit. Similarly, since the total number of molecules is constrained, it is best to expend energy building reporter molecules which need to encode the entire input signal, rather than wasting molecules on proteins in the beginning of the circuit. One well-known example of this is in the transcription-translation cascade from DNA to protein. Typically, mRNA degradation rates are faster than protein degaradation rates, and mRNA molecule numbers are smaller than protein molecule numbers. 

Other more subtle correlations may also be identified. Proteins that do not undergo proteolysis undergo decay by dilution; the characteristic time scale for these proteins is on the order of an hour. This implies that one can have noise filtering due to mRNA-protein scale mismatch without requiring negative feedback. One would expect to find a positive association between proteolysis and negative feedback. Consistent with this prediction, a Fisher exact probability test revealed a significant positive association ($p=0.01$) based on transcription factors from the \emph{E. coli} gene regulatory network. Rosenfeld et al. \citep{uri02jmb} have argued that an auto-repressive circuit averts the need for proteolysis, since the negative auto-regulation shortens the rise time of the circuit. Closer analysis of transcription factor databases for other organisms may help distinguish between these two putative roles of auto-repression. Interestingly, in prokaryotes the percent of transcription factors undergoing proteolysis is less than in eukaryotes, and the percent of transcription factors which are auto-repressive is less than in eukaryotes.

In their analysis of the phototransduction cascade, Detwiler et al. \citep{detwiler} emphasize that the signal processing characteristics of the cascade can be tuned simply by altering the concentrations of proteins, rather than changing the genetic sequence. That is, the parameters of the system can be optimized on a time scale far shorter than evolution. So too, in our simple circuits, all of the kinetic constants can be regarded as functions of concentrations of proteins extrinsic to the circuit, meaning the parameters may also be tuned, even independently, on a time scale shorter than the response time of the system. We highlight that circuits supporting multiple, distinct maxima should be able to flip between behaviors in response to different stimuli, and that theoretically such adaptation can be as rapid as changes in protein concentration. Importantly, based on our findings, such adaptation can be smooth and still occur without significant loss in transduction capacity. 

The fact that the $5$ peaks we analyzed collapsed onto two categories of spectra underscores a somewhat paradoxical finding. Namely, the maxima are \emph{robust} in that they can withstand $10$-fold perturbations in most of their parameters without significant loss in transmission fidelity, and yet they are \emph{adaptive} in that the circuit can flip between the different maxima (and different behaviors), again without significant loss in transmission fidelity. Intuitively, one might expect a tradeoff between robustness and adaptability. Our findings suggest that the circuits can avert this tradeoff by clustering the maxima in a general region of high transmission fidelity. Certainly a closer and more quantitative analysis of this tradeoff is warranted. For example, it is now well-established that a
single circuit can support multiple functions \cite{wall_ffl}. In this
vein, one interesting research direction would be to enumerate the
functions that a particular circuit can achieve and quantify how
easily the circuit can flip between these functions. Whereas our
circuits can all be regarded as ``optimal'' in the sense that they can
tune their parameters to transduce the optimal amount of input
information, it is evident that subtle distinctions in information
processing exist among them. Our setup is well-suited to
systematically explore these distinctions (e.g., varying the input
distribution, quantifying the mutual information between time-varying input
and output signals, and quantifying other statistics of the mutual
information landscape rather than
optimality).

\newpage

\begin{center}
\begin{longtable}{cc}
\caption[Mutual Information vs. the mean reporter copy number] {Table of mutual information $I$ as a function of mean reporter copy number $\langle N_G\rangle$. {\bf{Insets:}} $\langle I \rangle$ as a function of inverse data $m$ (cf.~\ref{sec:results_opt}). Note that circuits $5,11,13,17,19,23$ are shown in Figs.~\ref{Ivnfigs1}-\ref{Ivnfigs3}.}\\ Circuit 1 & Circuit 2 \\
{\includegraphics[width=1.8in] {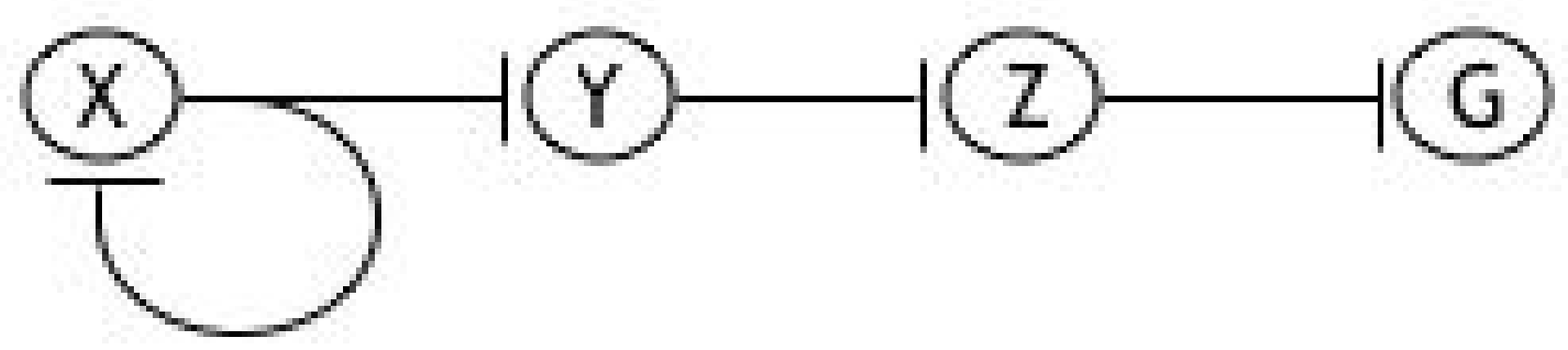} } & {\includegraphics[width=1.4in] {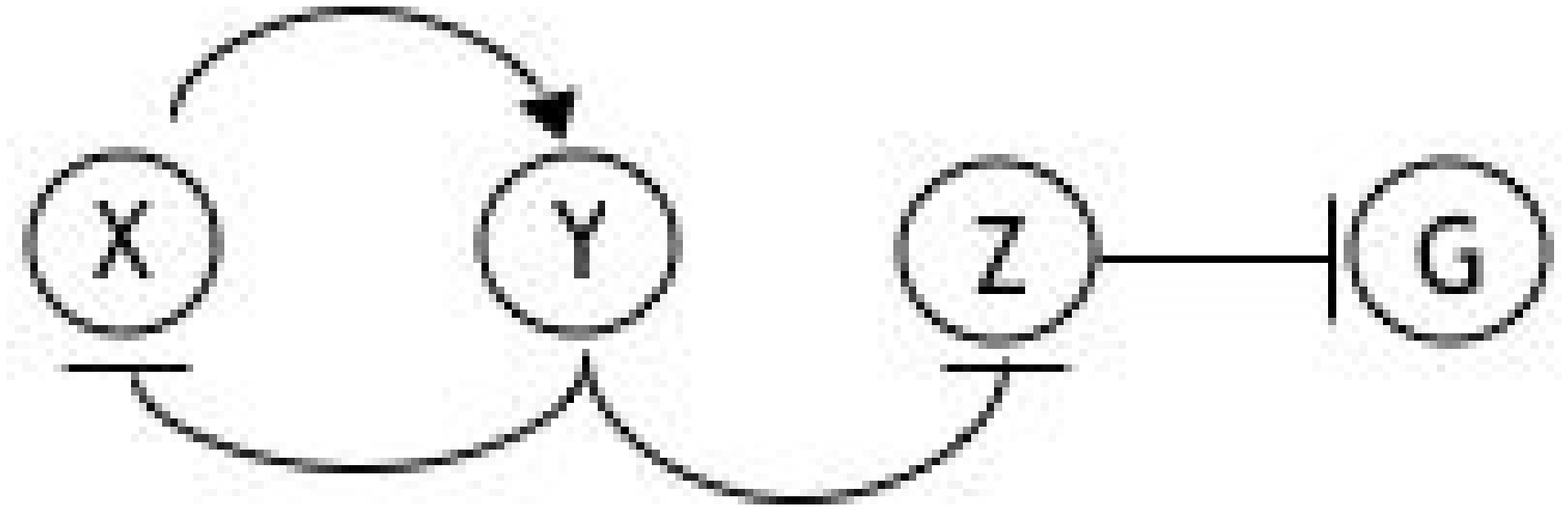} }\\
& \\
{\includegraphics[width=3in] {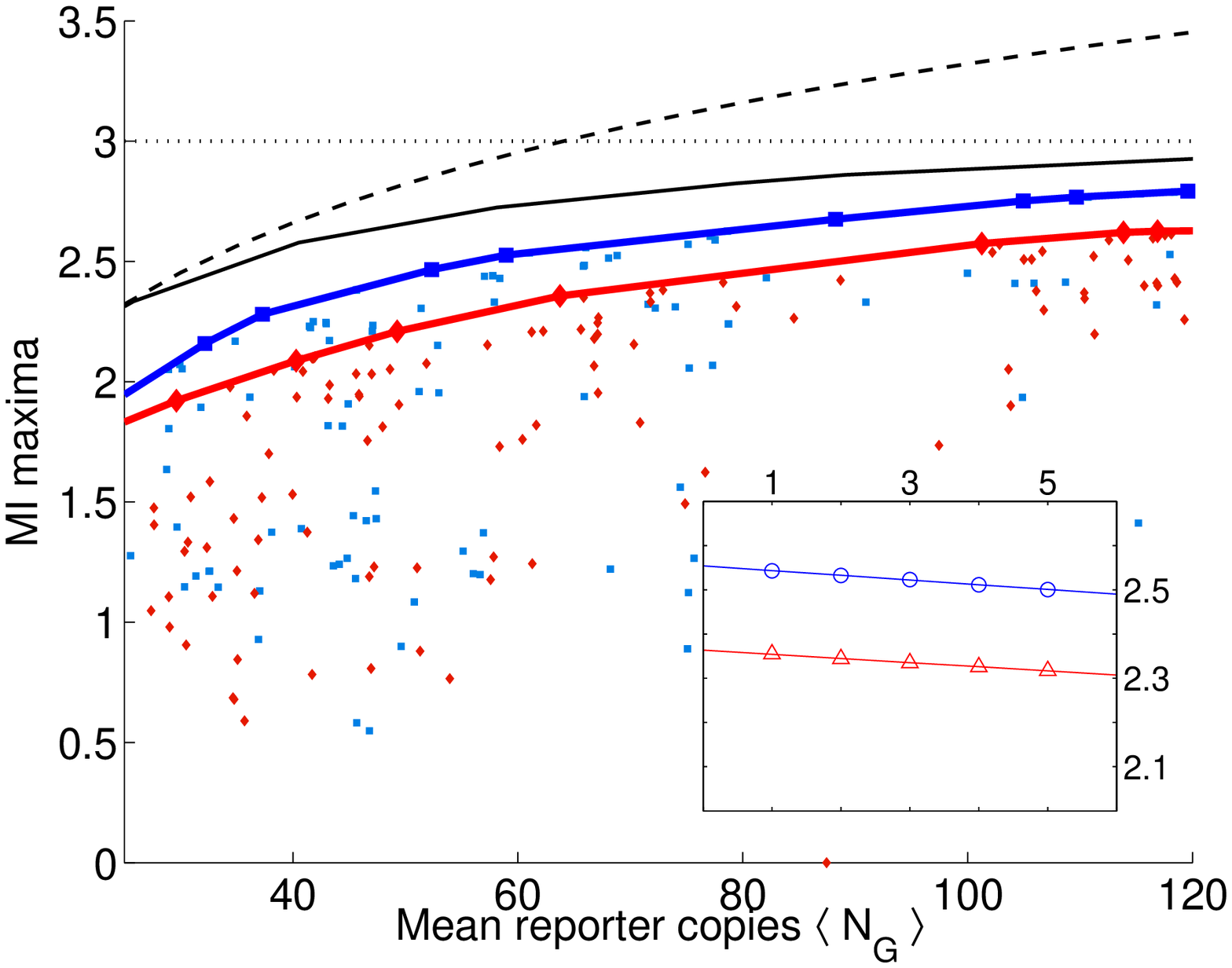} } & {\includegraphics[width=3in] {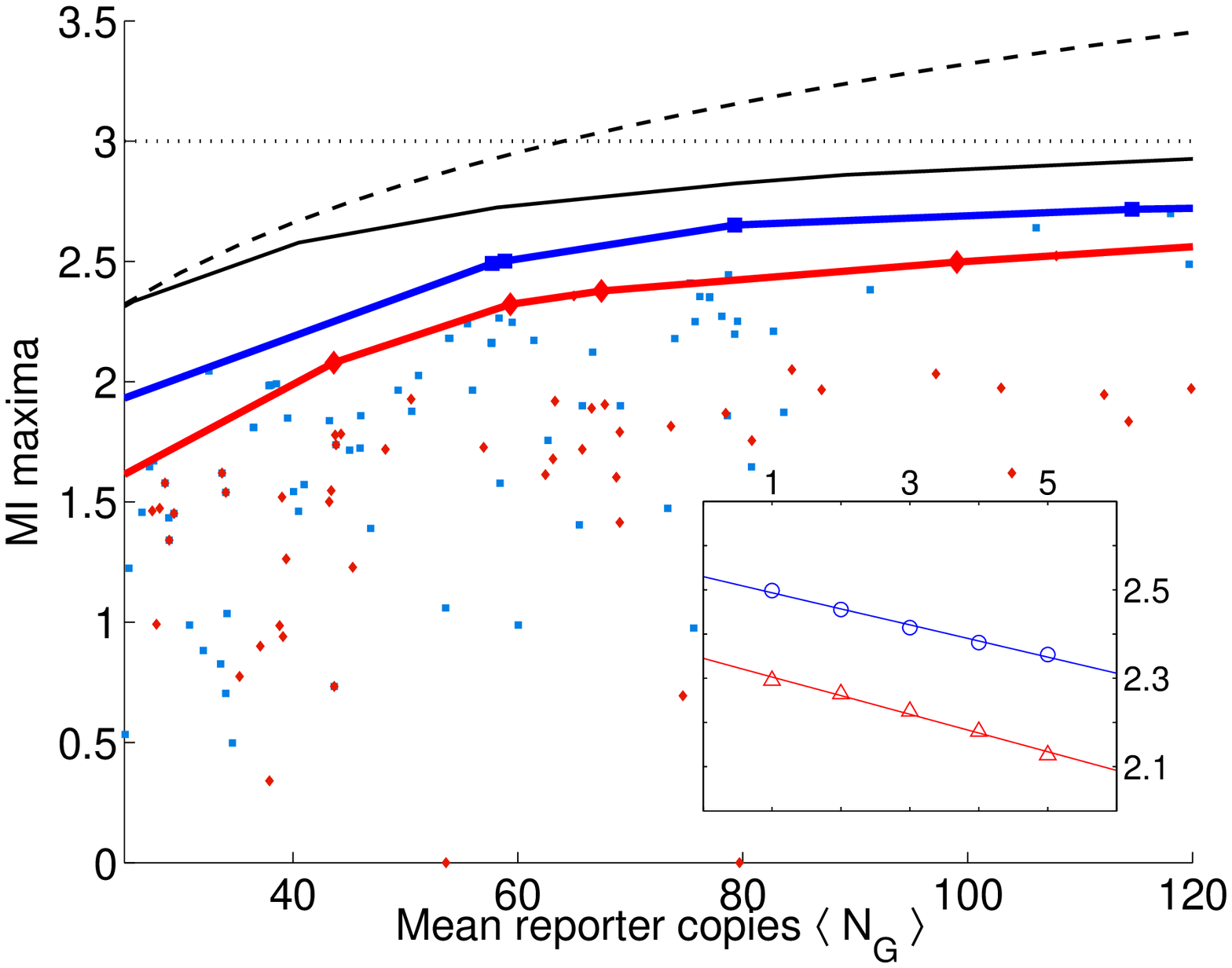} }\\
& \\

 Circuit 3 & Circuit 4\\
{\includegraphics[width=1.1in] {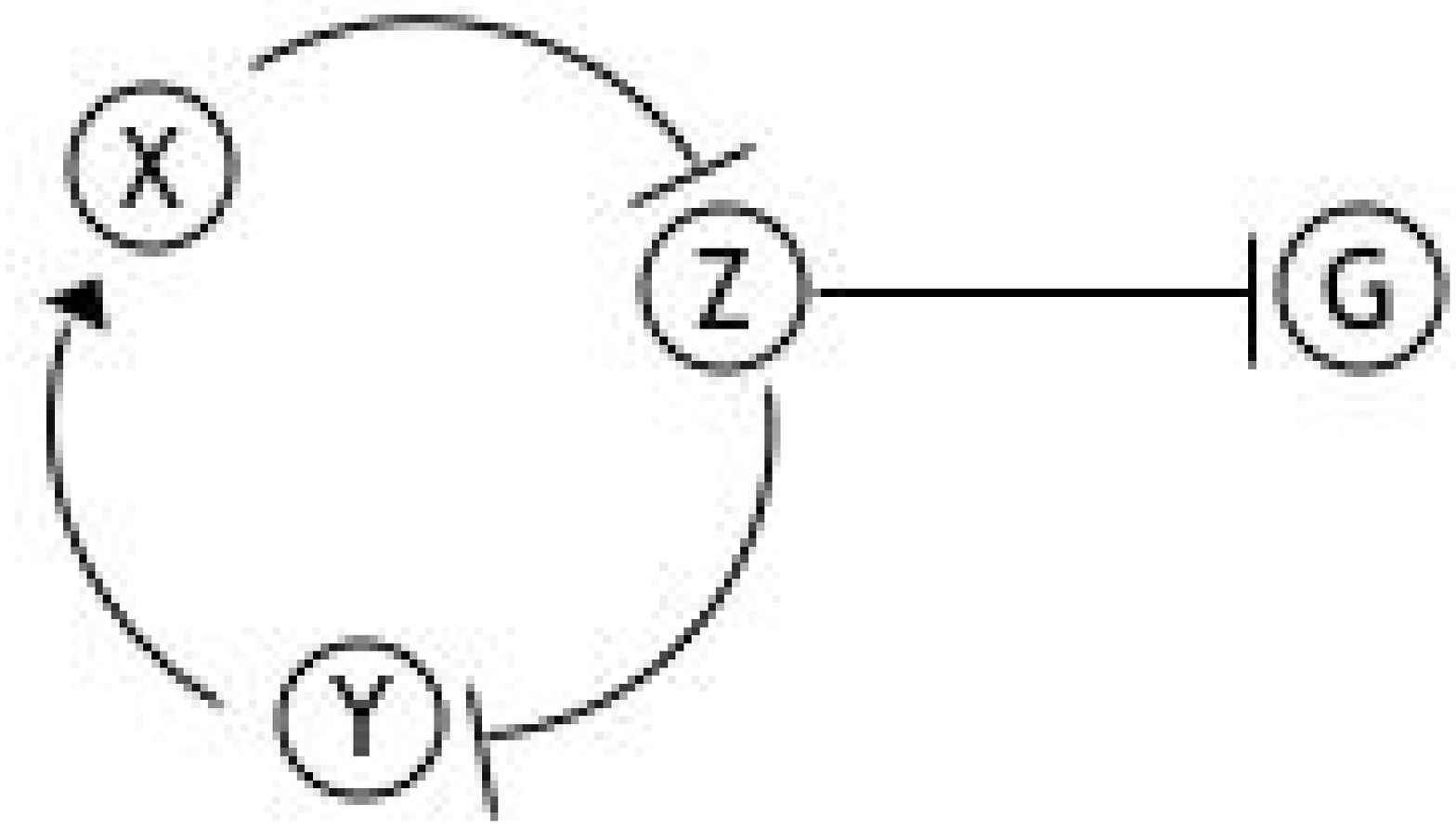} } & {\includegraphics[width=1.8in] {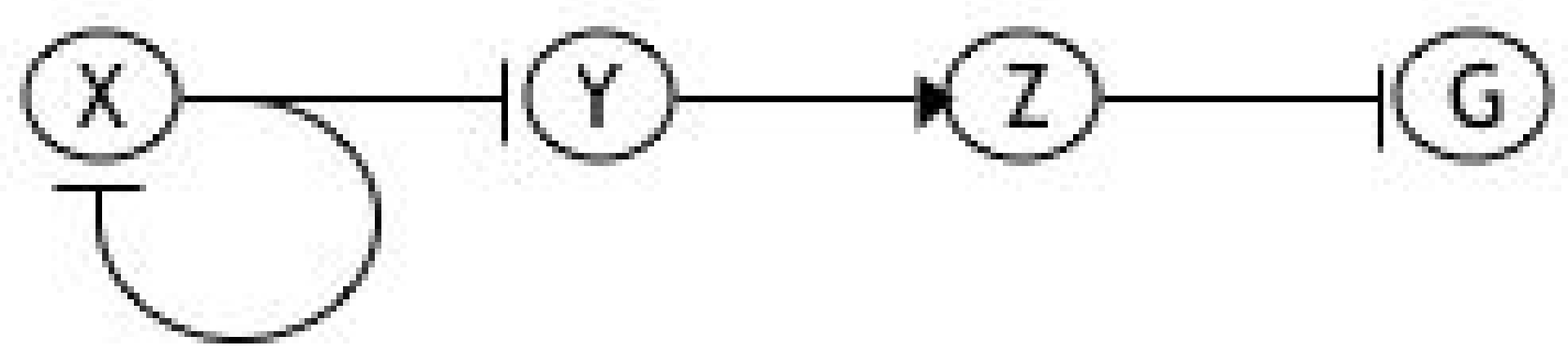} }\\
 & \\
 {\includegraphics[width=3in] {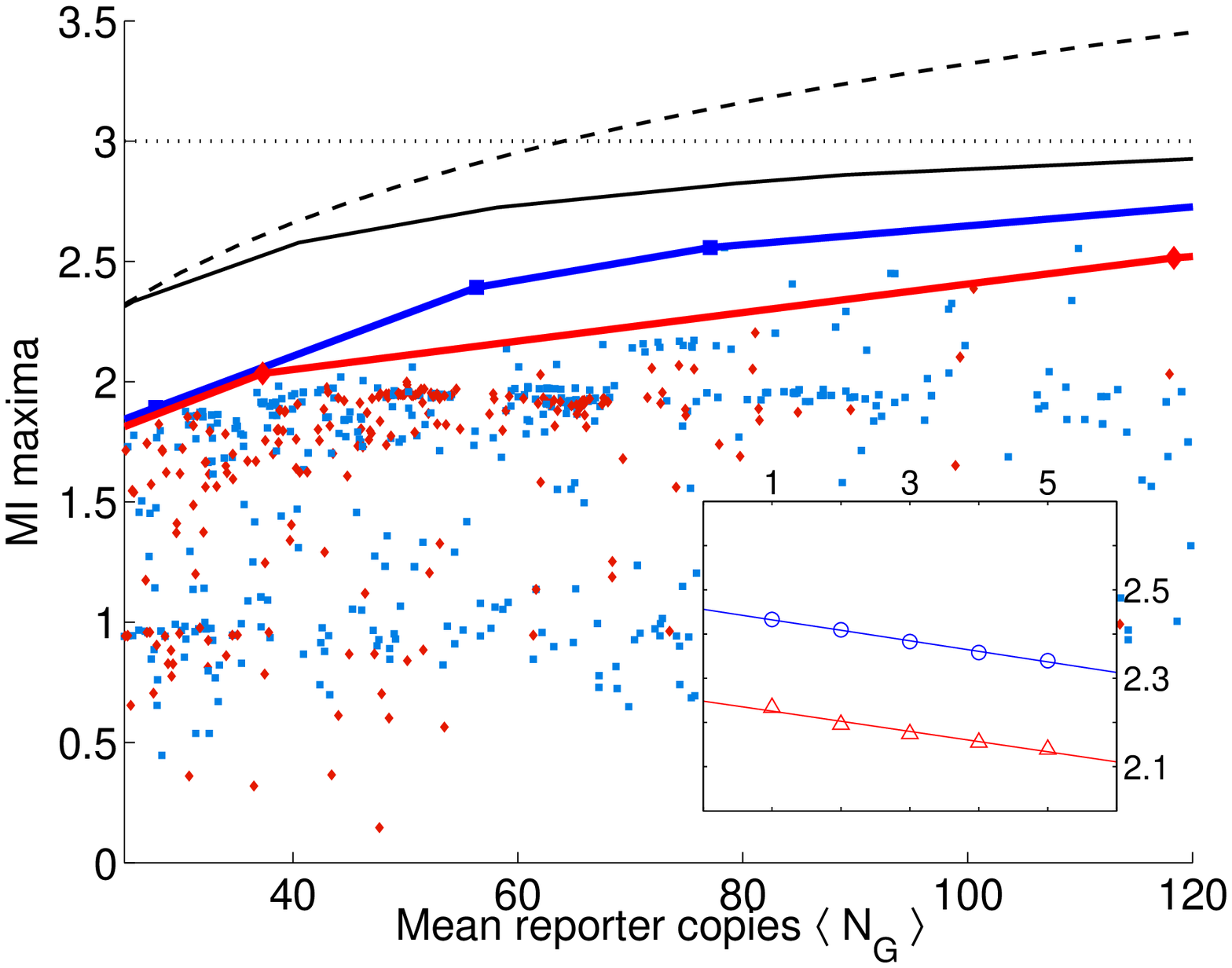} } & {\includegraphics[width=3in] {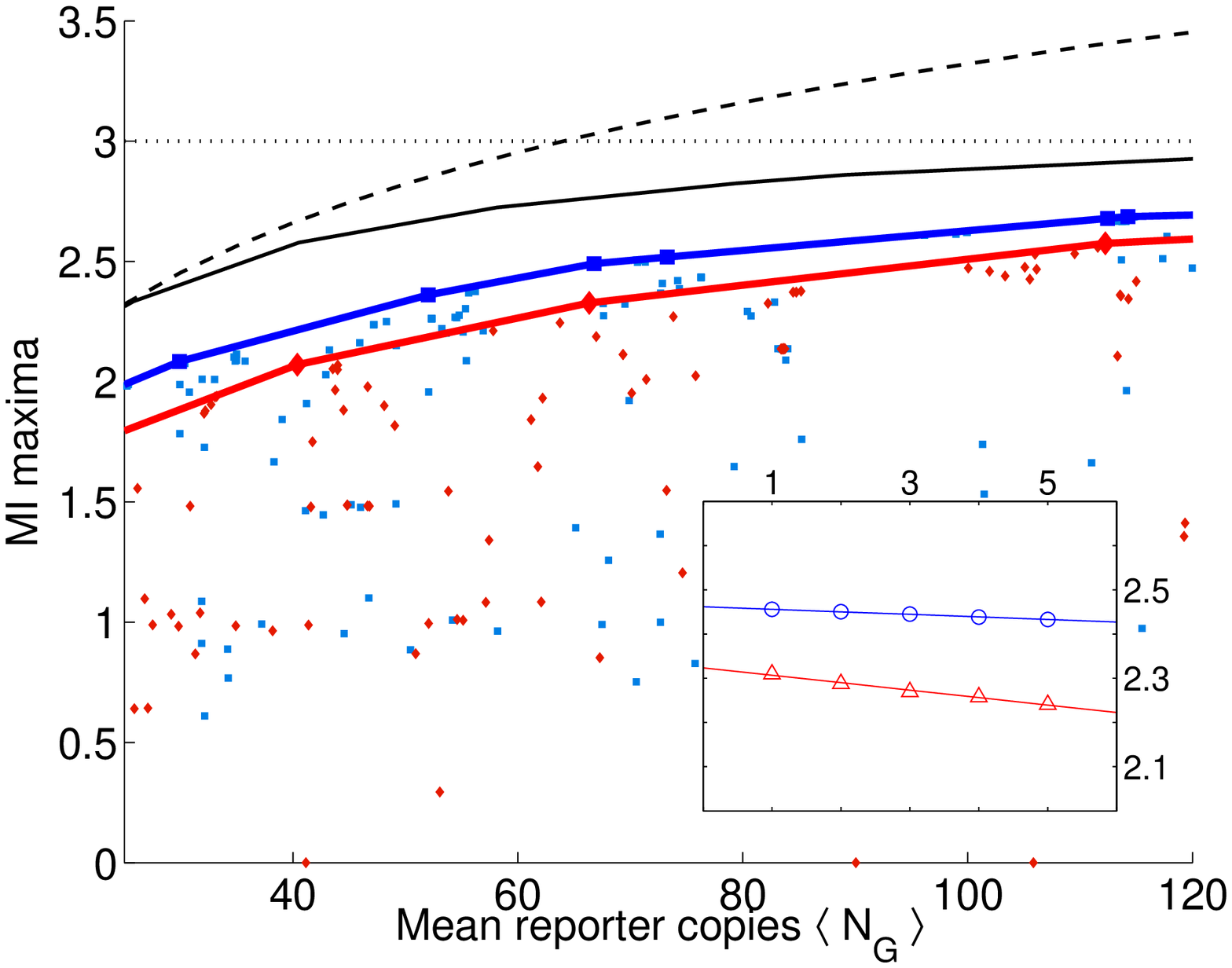} }\\
	& \\

 Circuit 6 & Circuit 7\\
{\includegraphics[width=1.1in] {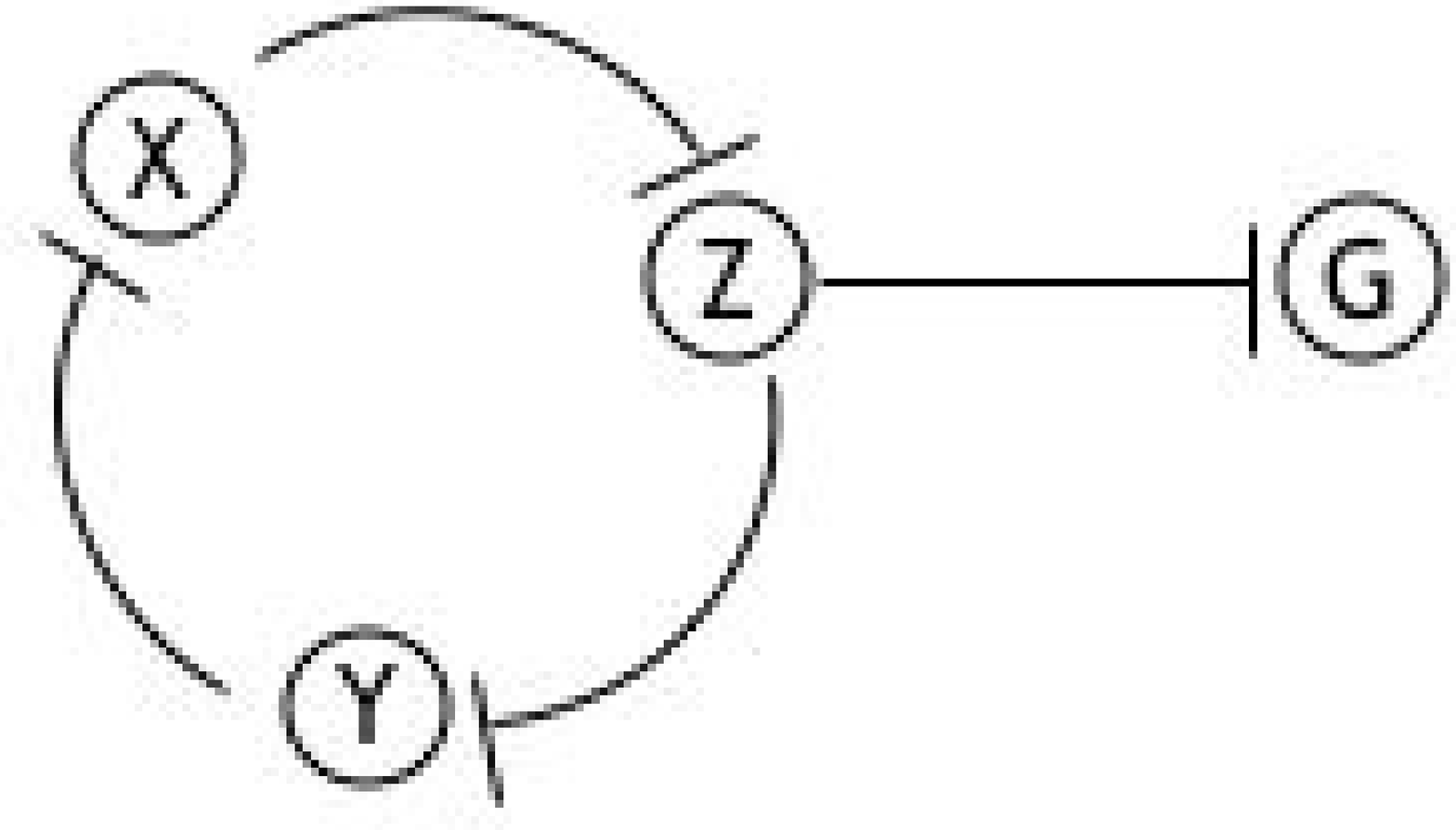} } & {\includegraphics[width=1.4in] {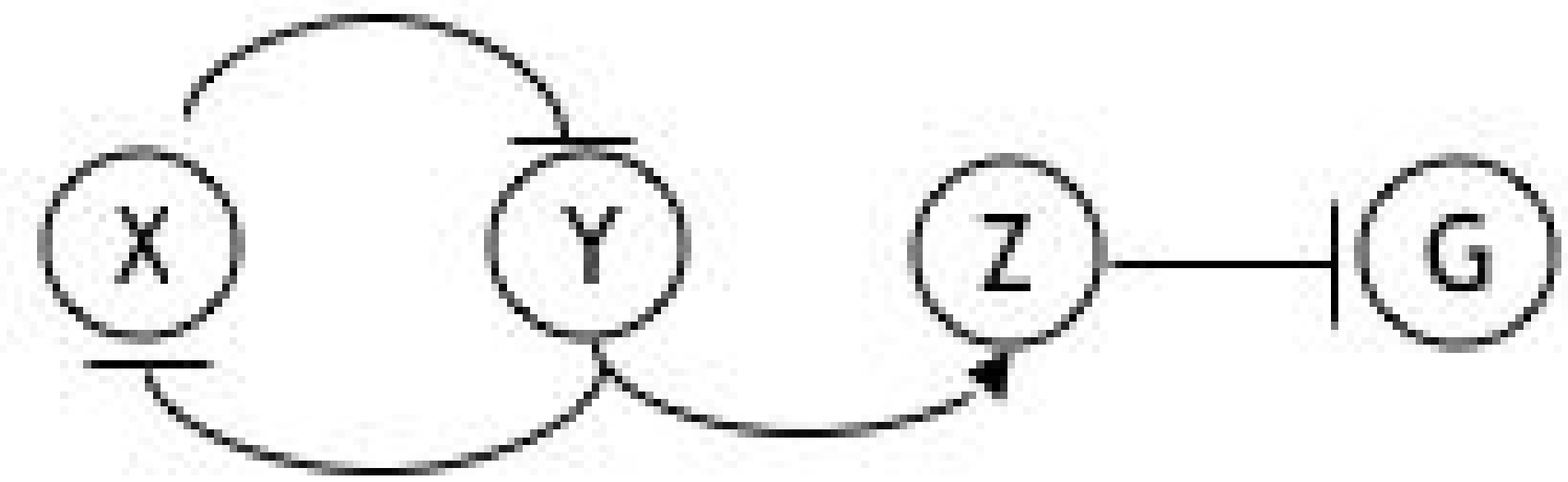} }\\
 & \\  
  {\includegraphics[width=3in] {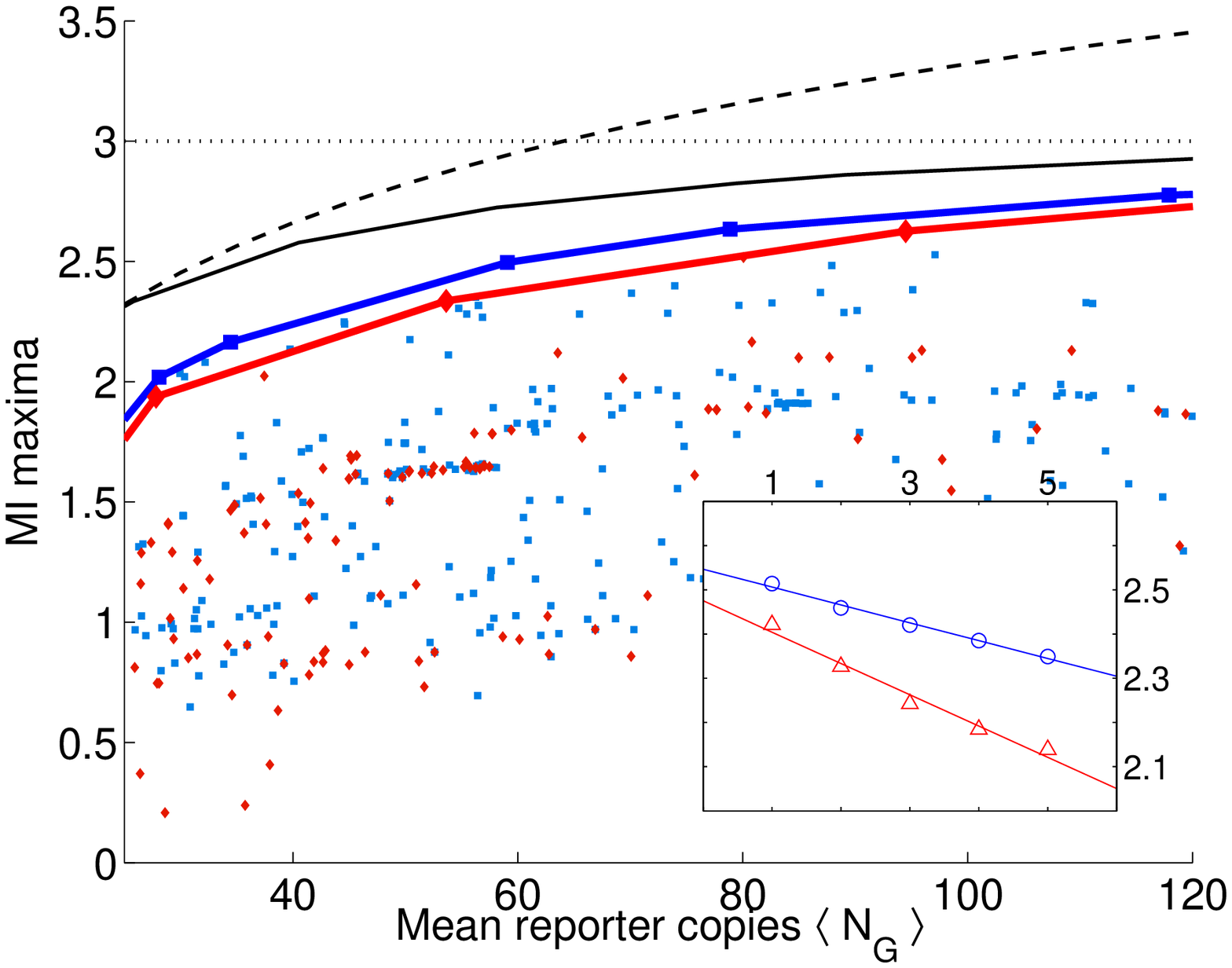} } & {\includegraphics[width=3in] {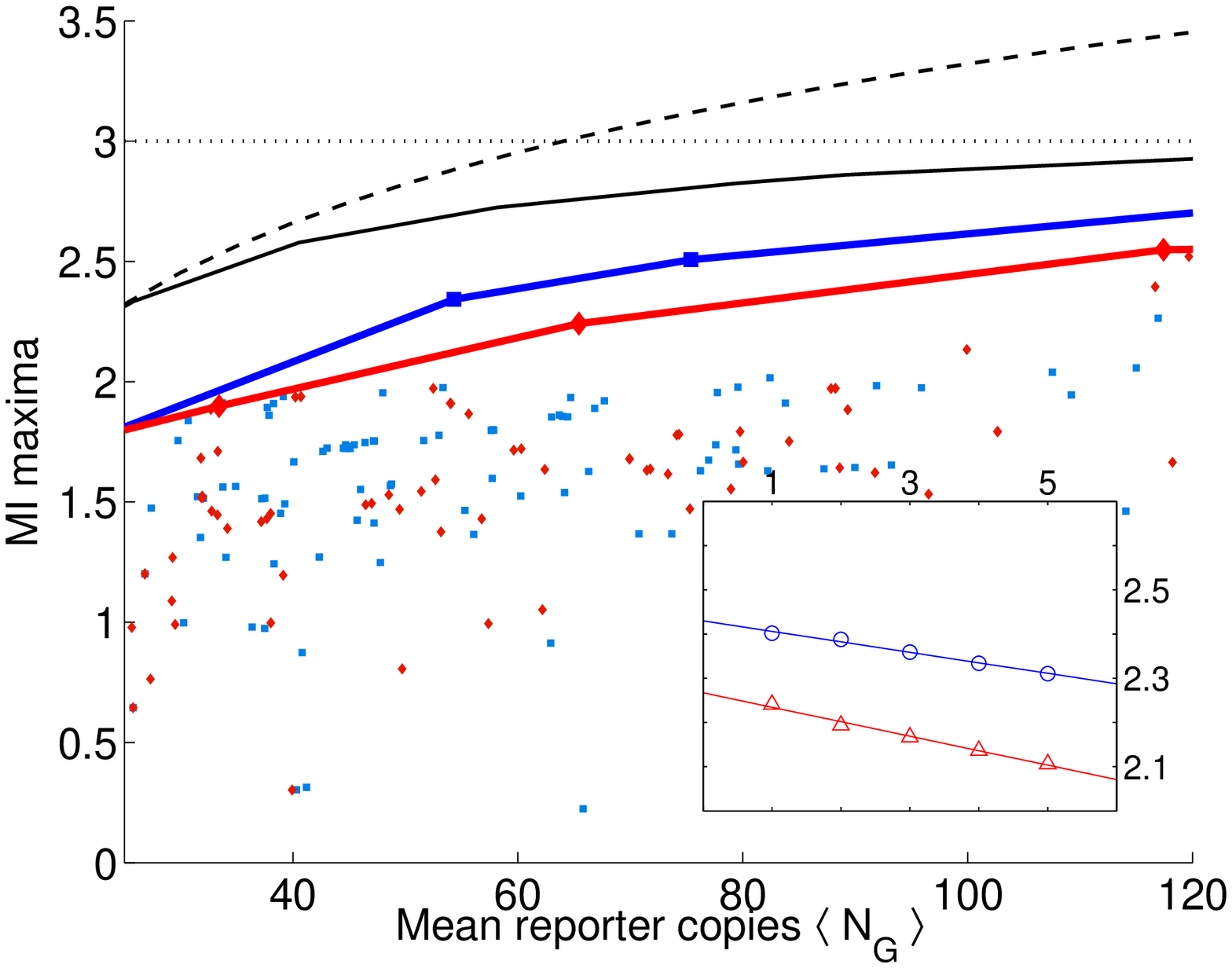} }\\
  	& \\

   Circuit 8 & Circuit 9\\
{\includegraphics[width=1.1in] {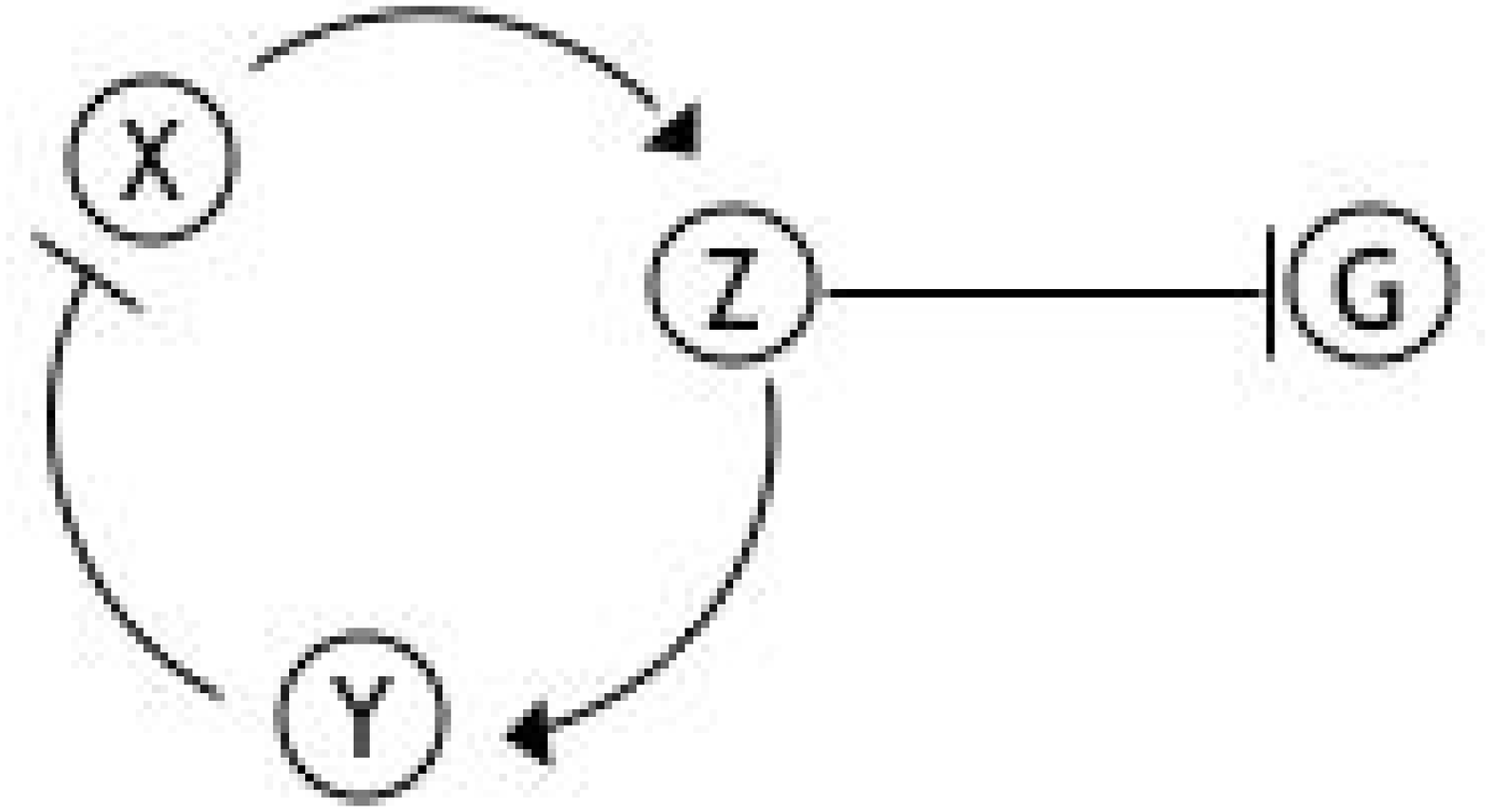} } & {\includegraphics[width=1.4in] {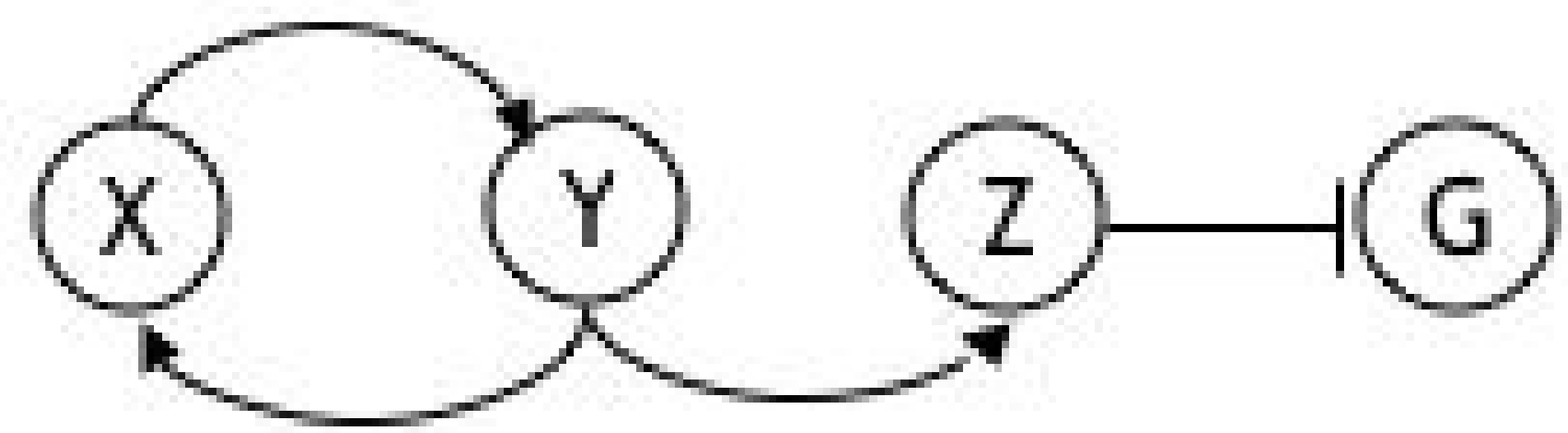} }\\
 & \\      
     
 {\includegraphics[width=3in] {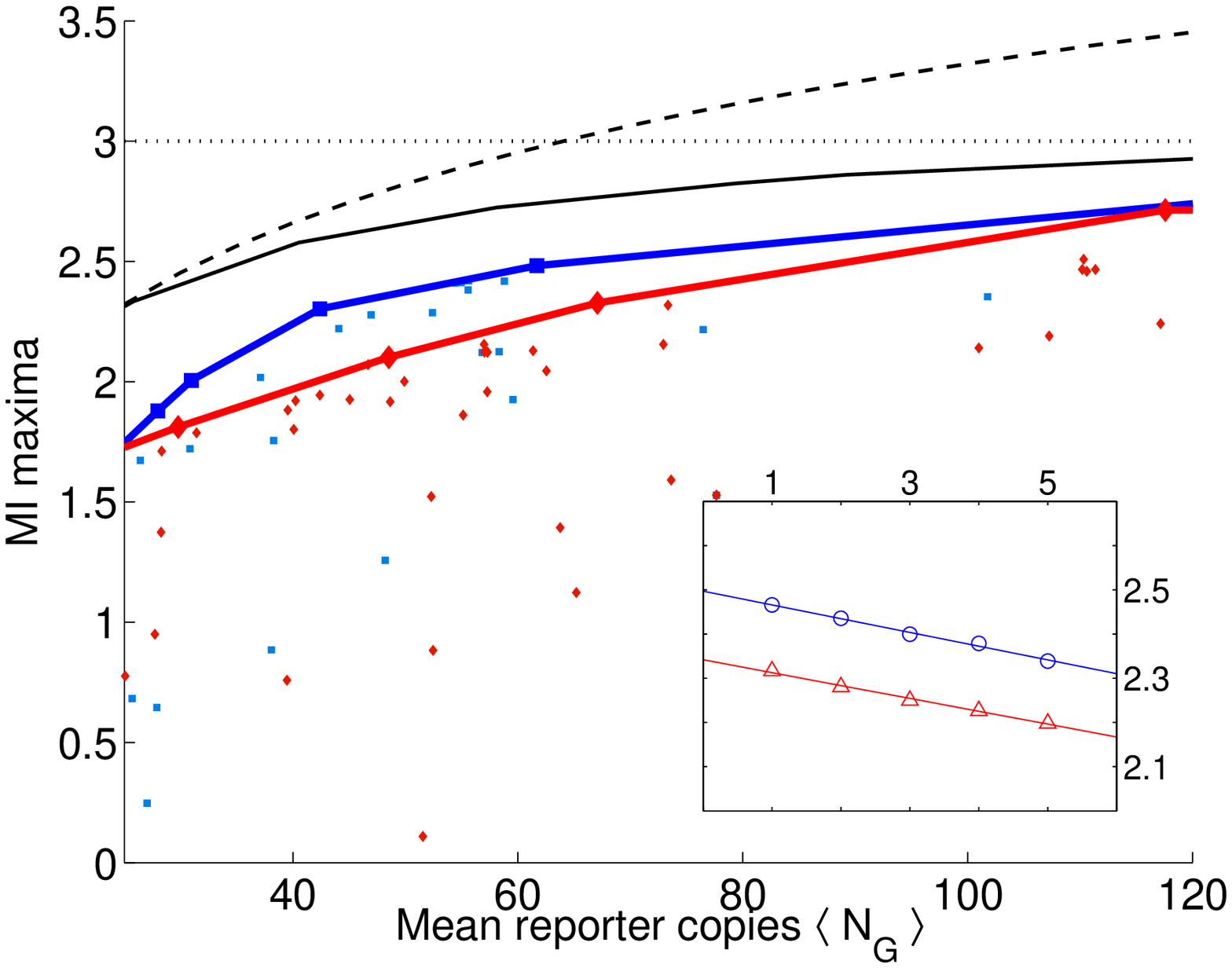} } & {\includegraphics[width=3in] {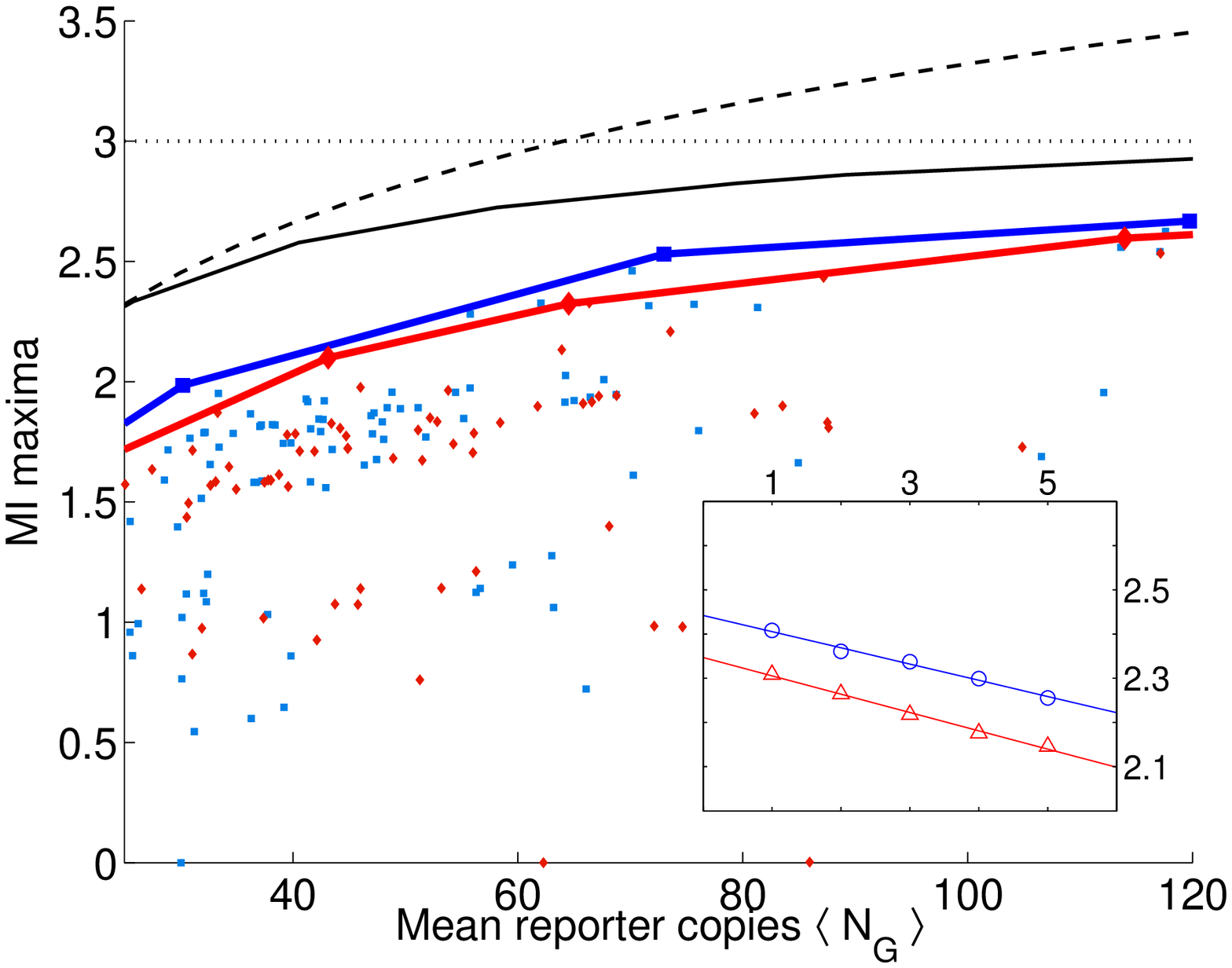} }\\ 

\newpage
 & \\
 Circuit 10 & Circuit 12 \\
 & \\ 
{\includegraphics[width=1.1in] {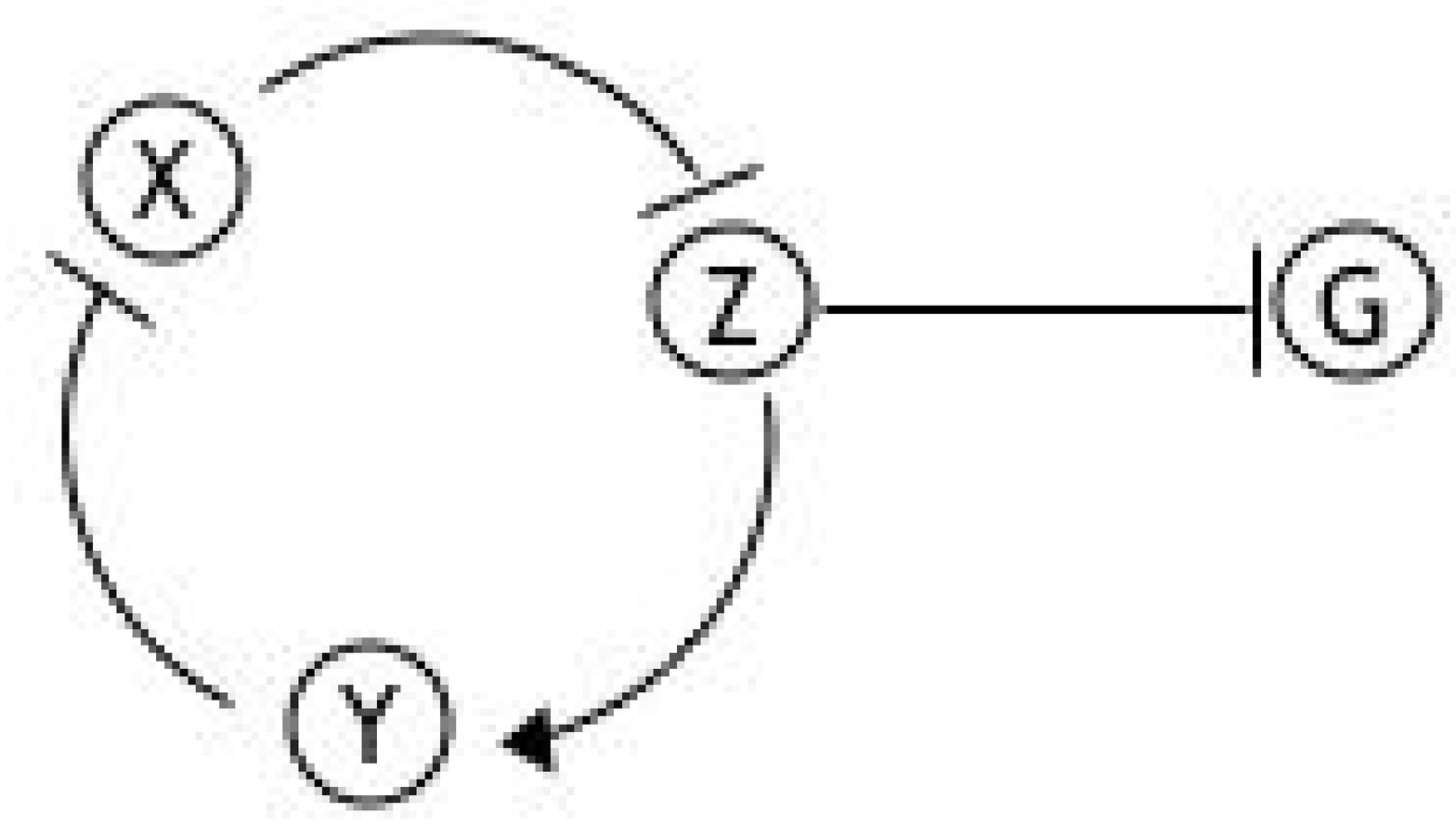} } & {\includegraphics[width=1.4in] {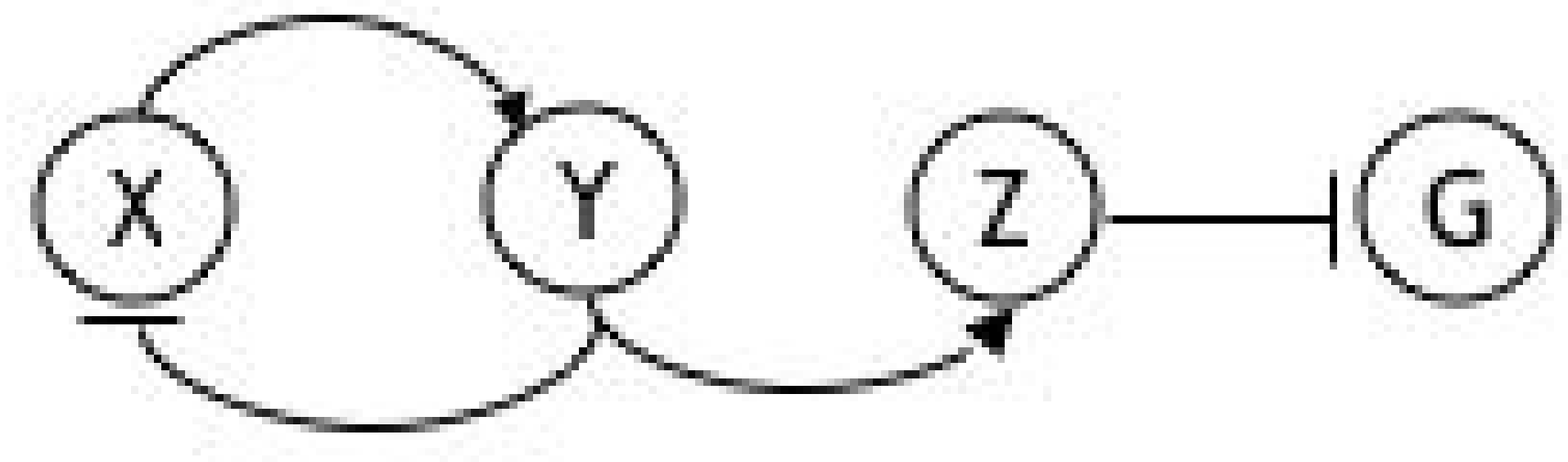} }\\
 & \\ 
{\includegraphics[width=3in] {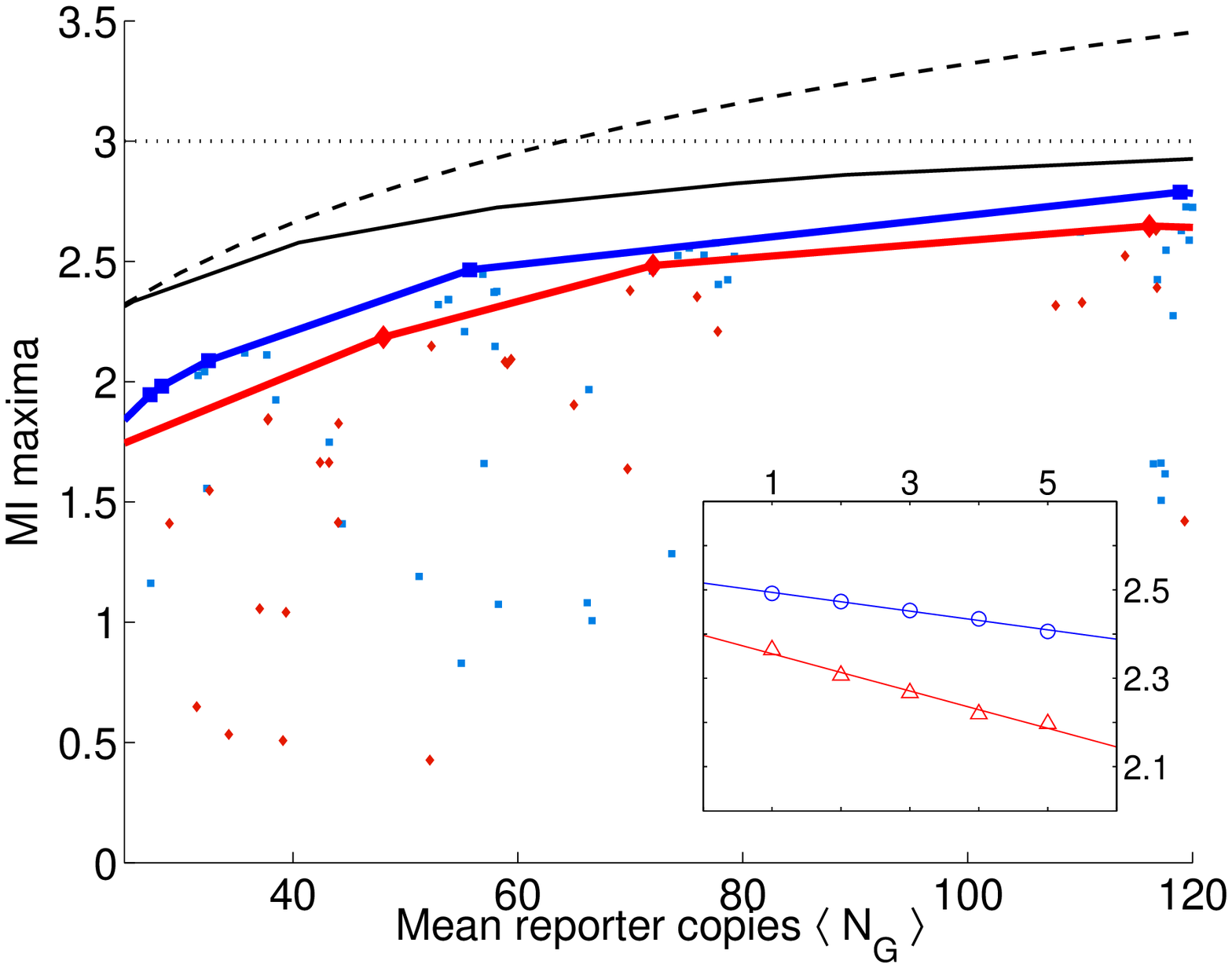} } & {\includegraphics[width=3in] {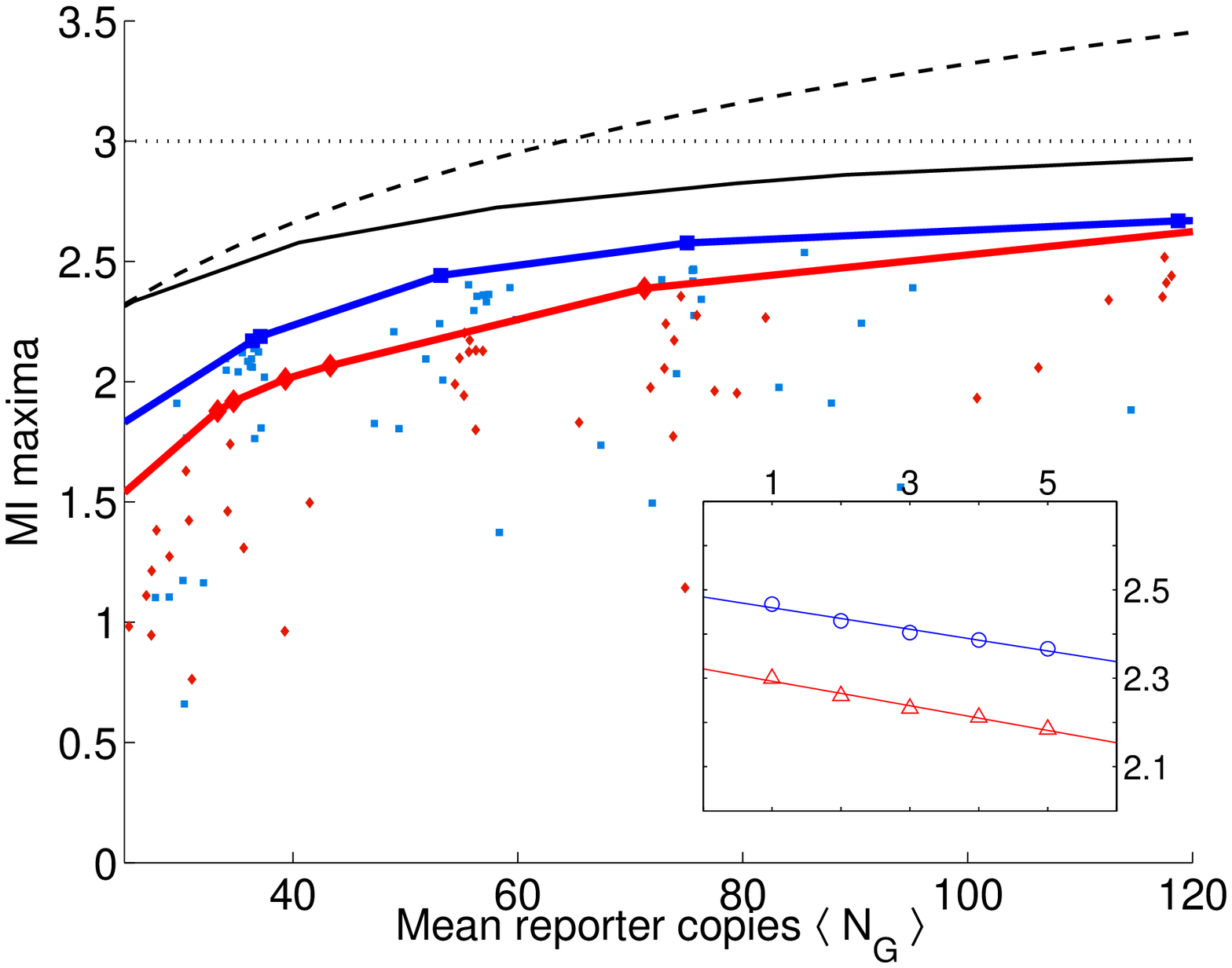} }\\

 & \\
 Circuit 14 & Circuit 15 \\
 & \\
 {\includegraphics[width=1.8in] {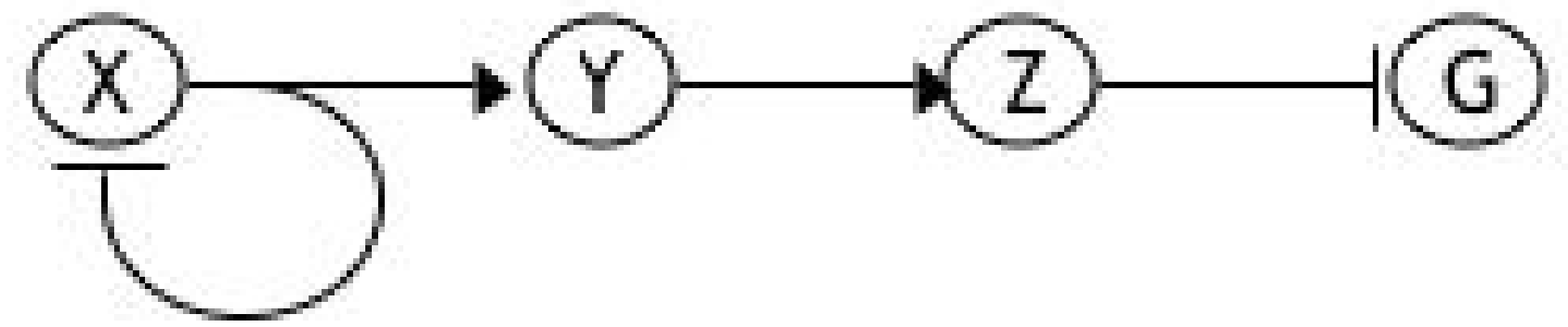} } & {\includegraphics[width=1.8in] {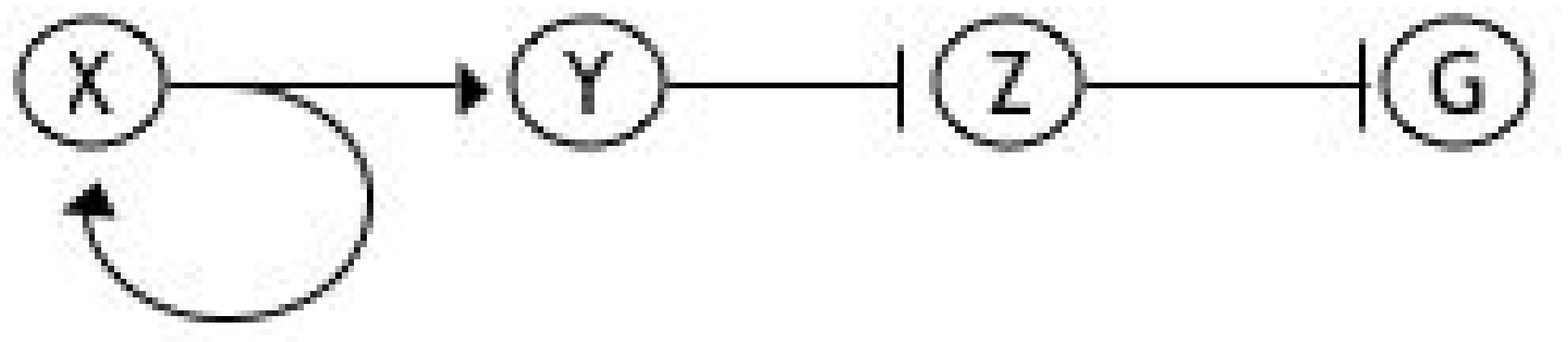} }\\
 & \\ 
 {\includegraphics[width=3in] {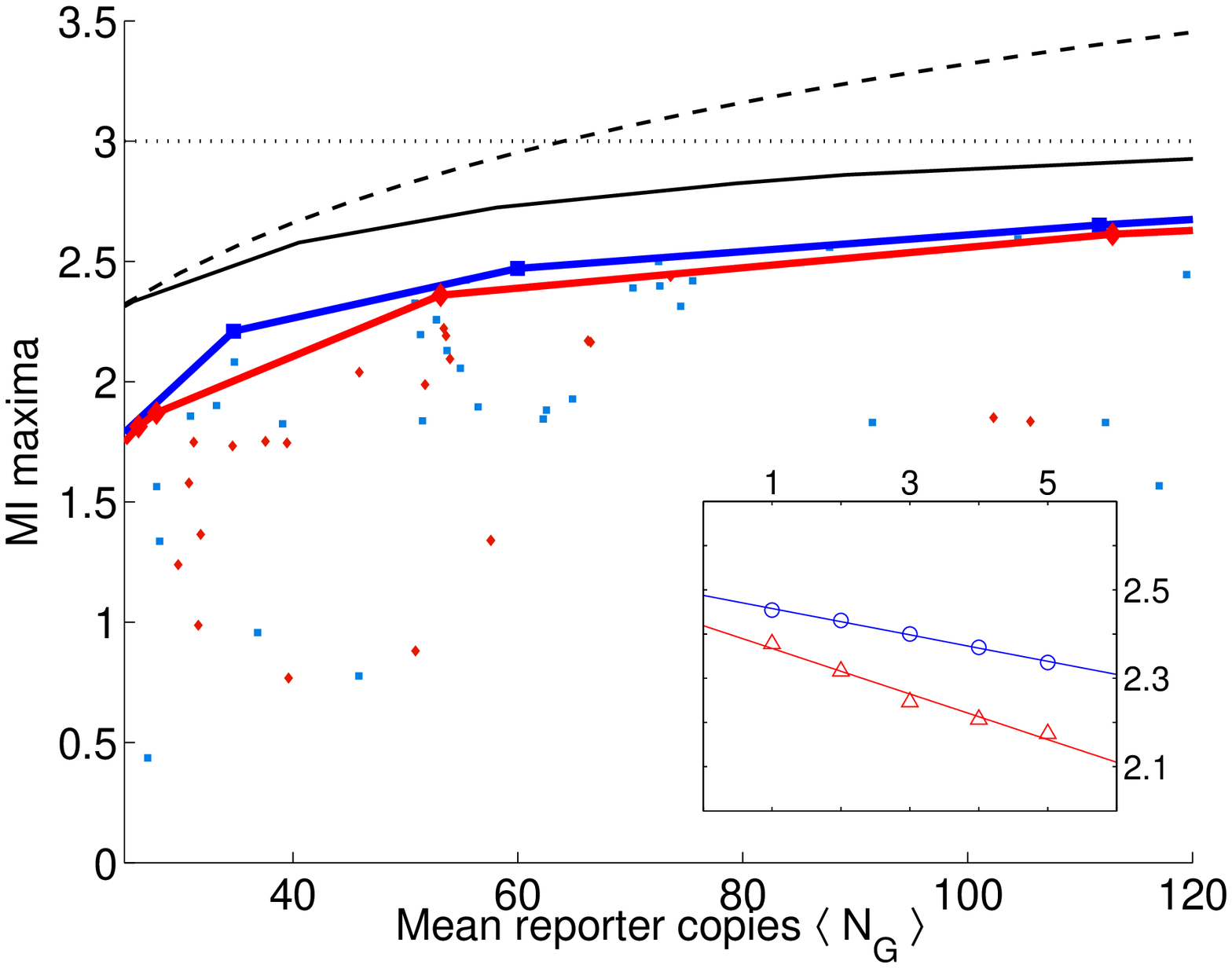} } & {\includegraphics[width=3in] {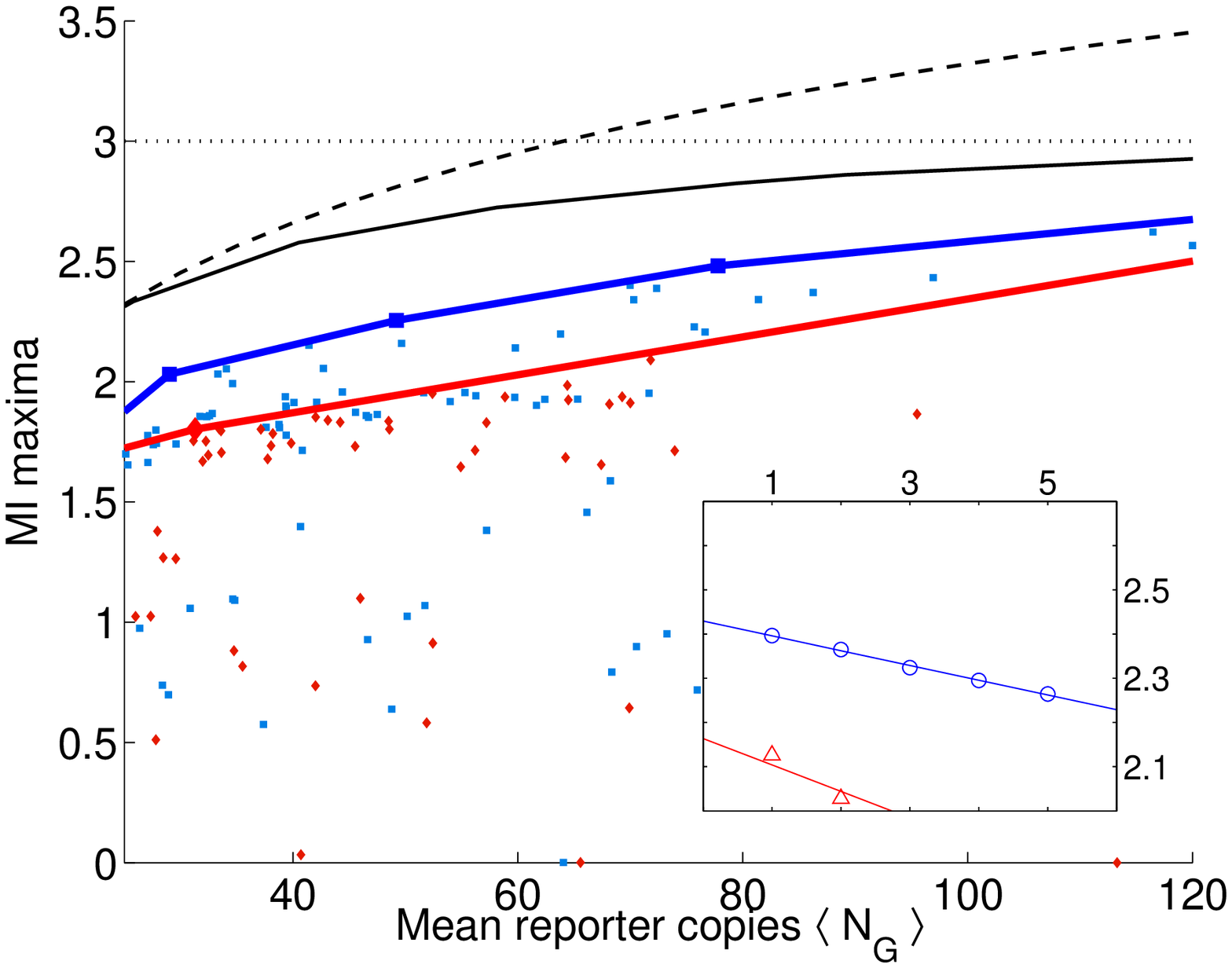} }\\
 
  \newpage
  & \\
 Circuit 16 & Circuit 18 \\
  & \\
  {\includegraphics[width=1.8in] {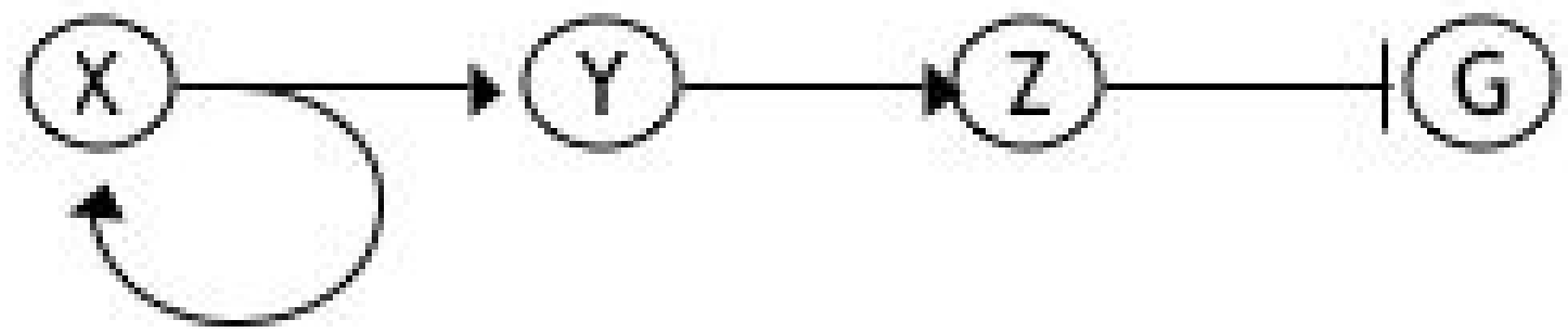} } & {\includegraphics[width=1.4in] {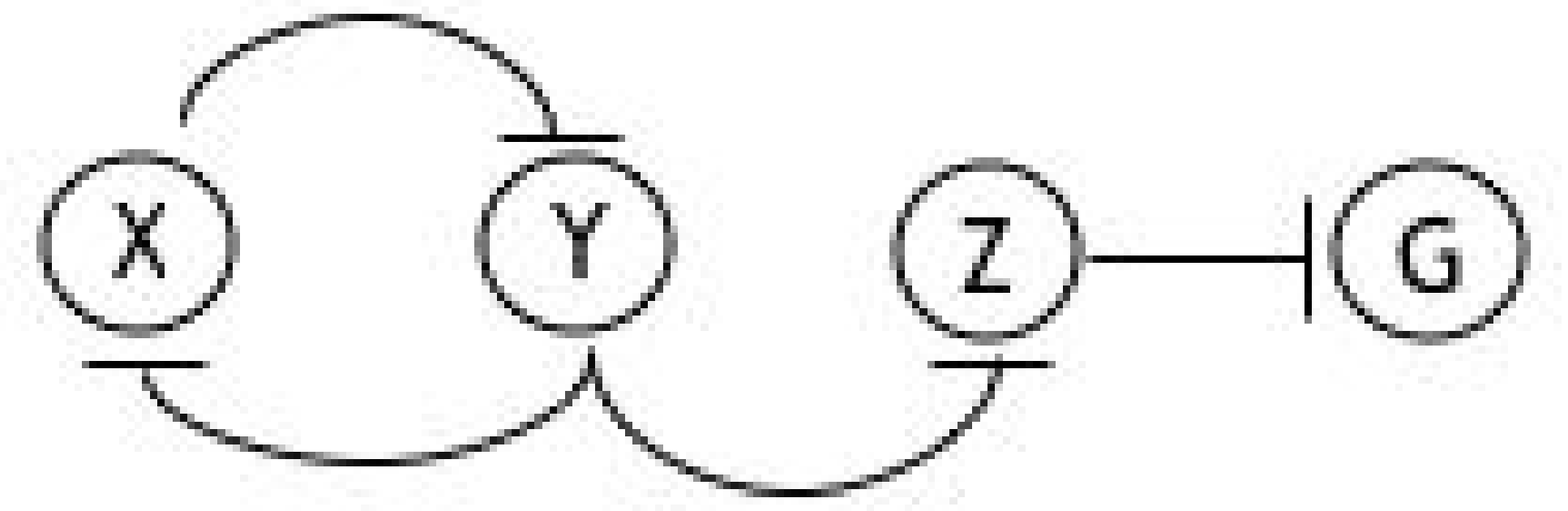} }\\
 & \\ 
  {\includegraphics[width=3in] {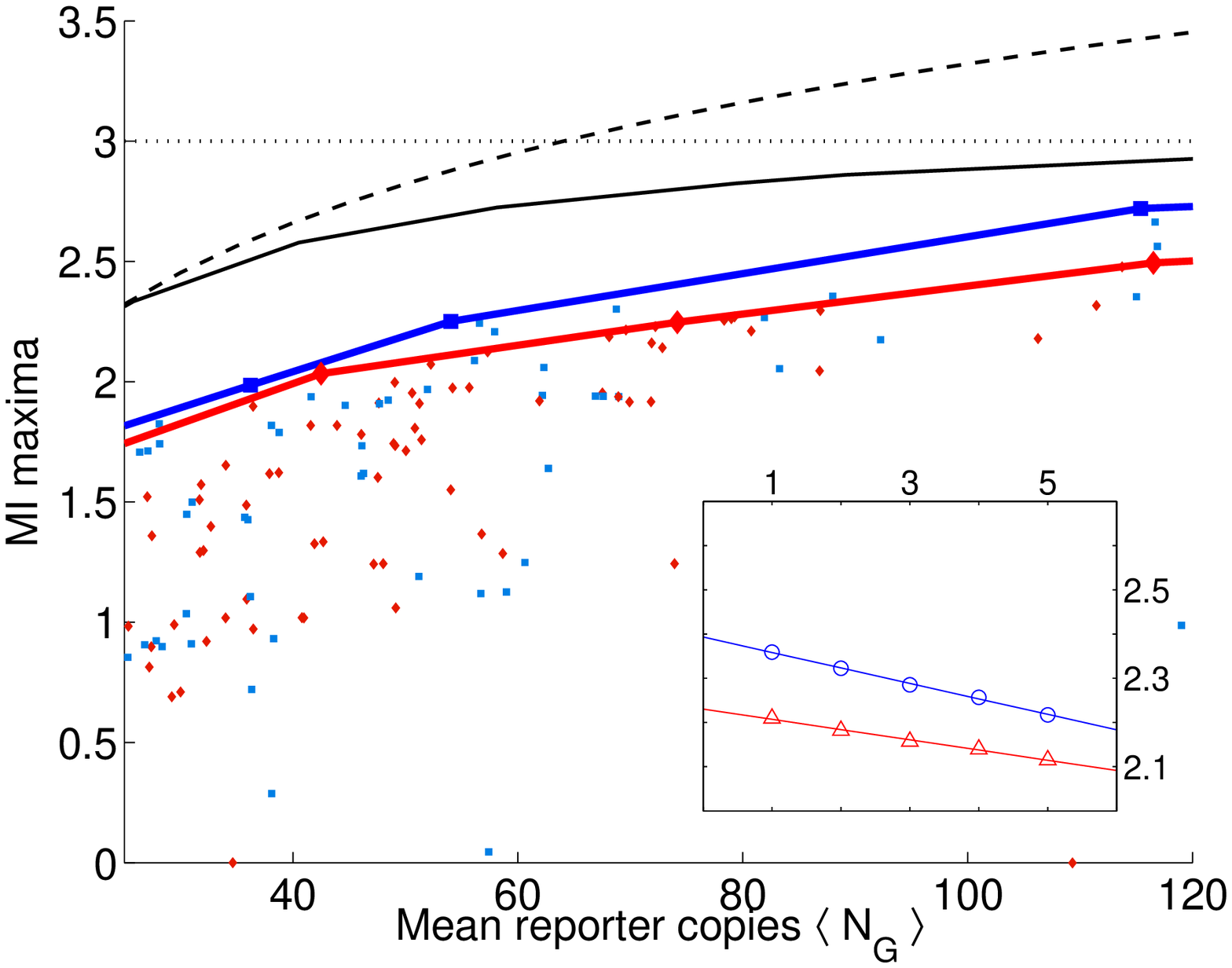} } & {\includegraphics[width=3in] {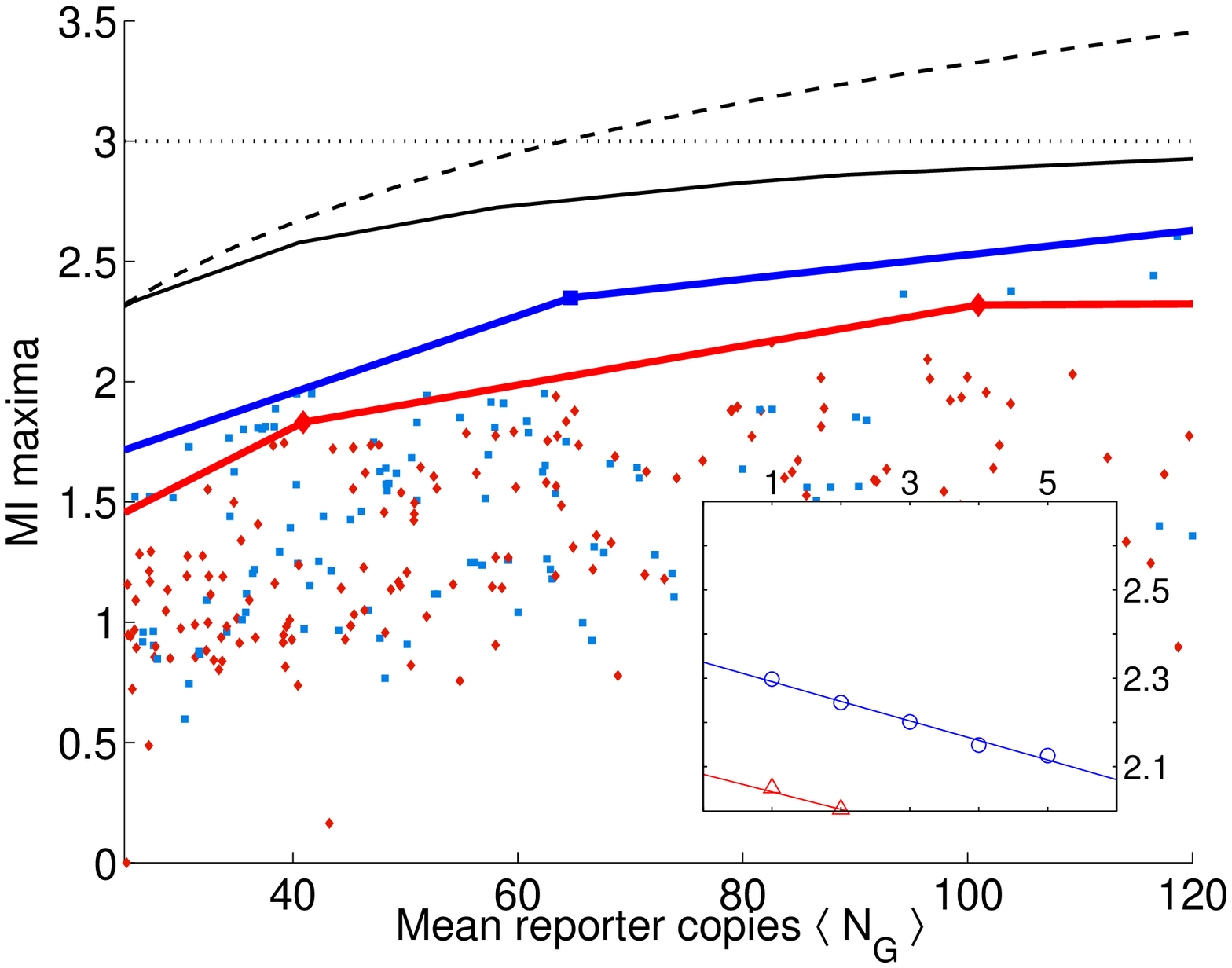} }\\
  
   & \\
 Circuit 20 & Circuit 21 \\
 & \\
 {\includegraphics[width=1.4 in] {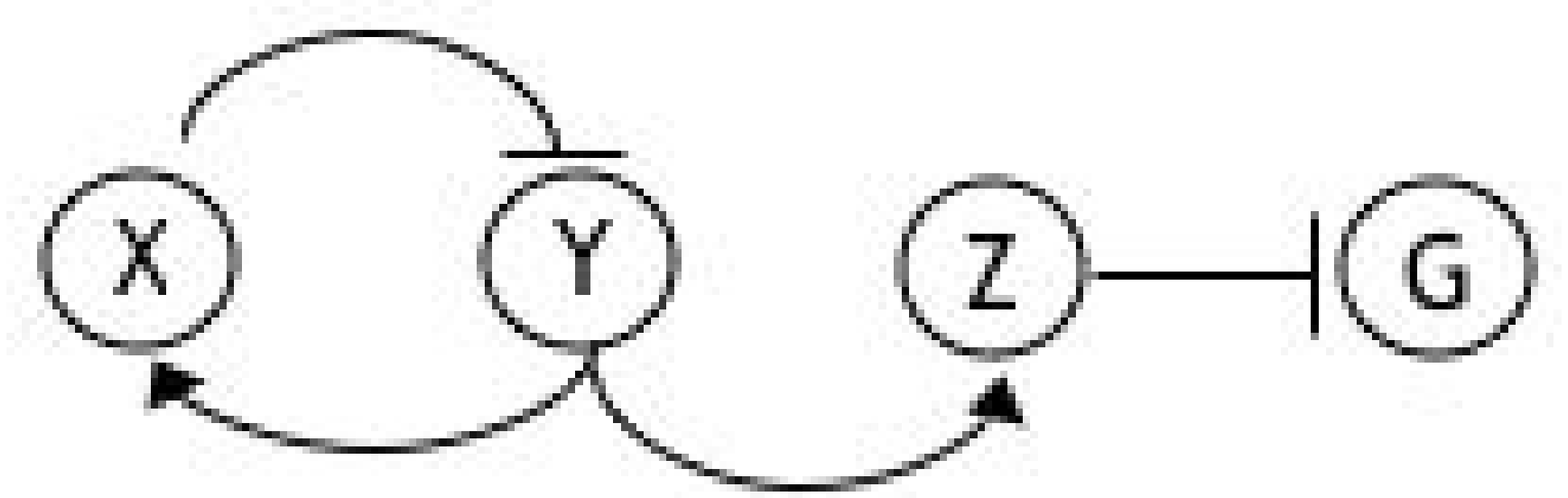} } & {\includegraphics[width=1.8in] {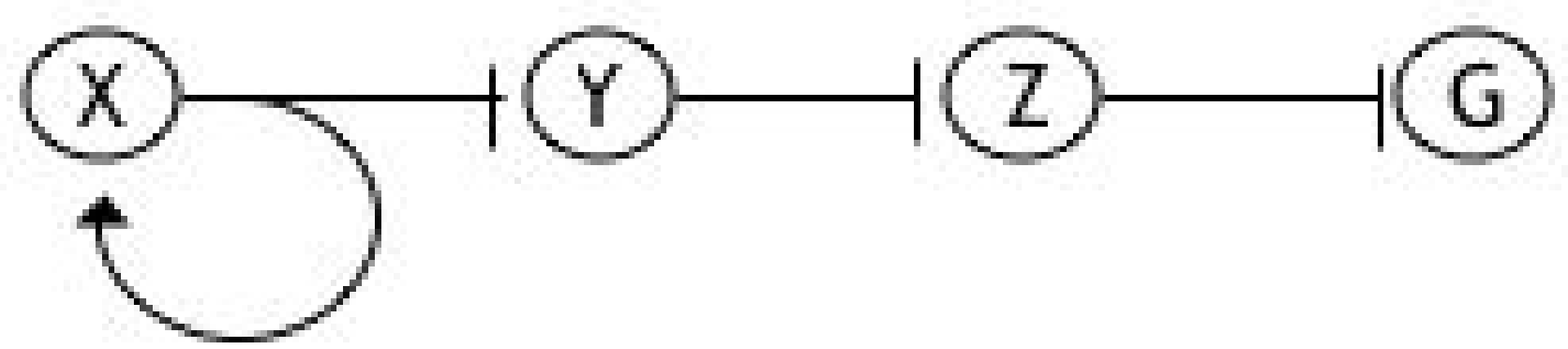} }\\
 & \\ 
  {\includegraphics[width=3in] {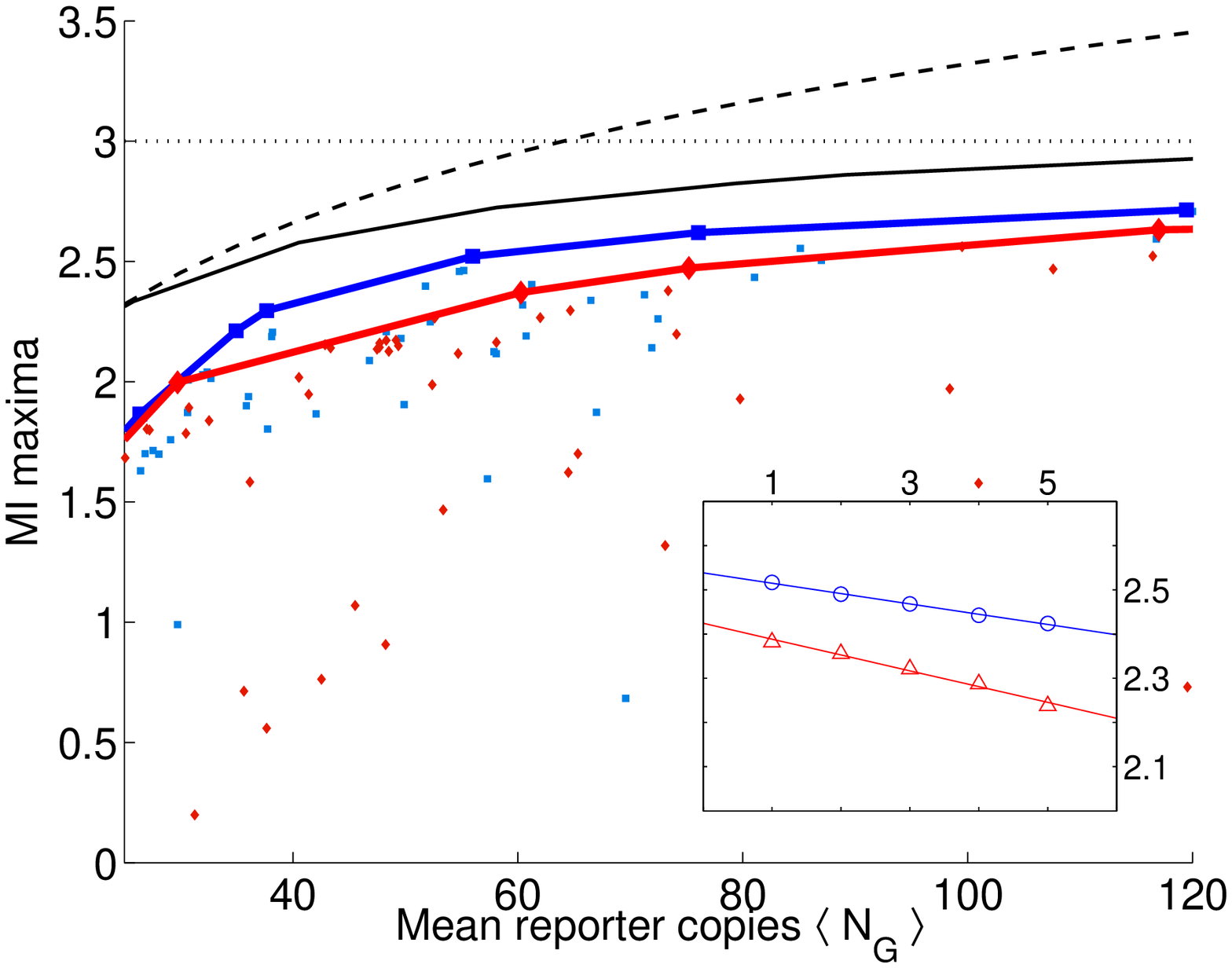} } & {\includegraphics[width=3in] {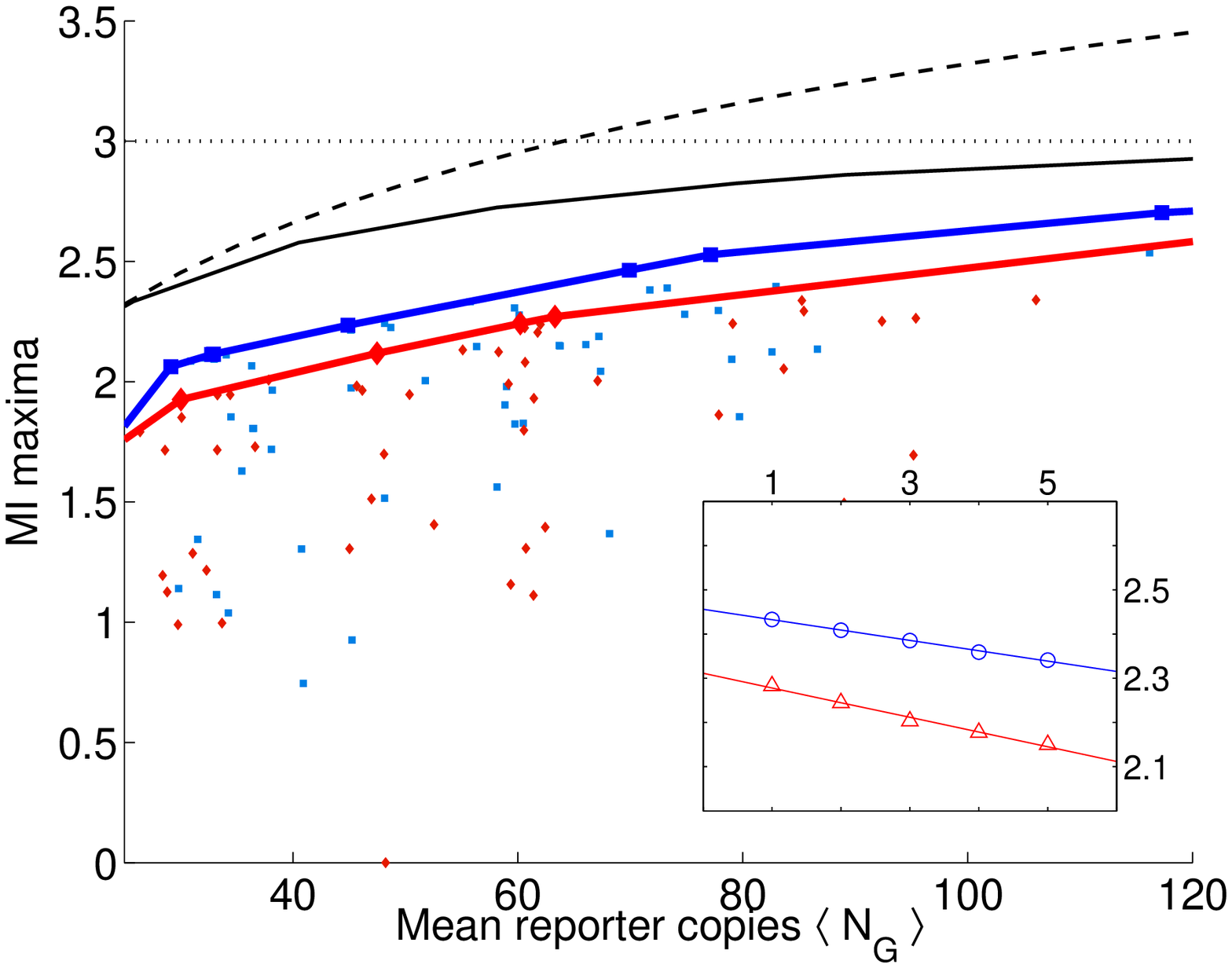} }\\

  \newpage
   & \\
 Circuit 22 & Circuit 24 \\
 & \\
 {\includegraphics[width=1.1in] {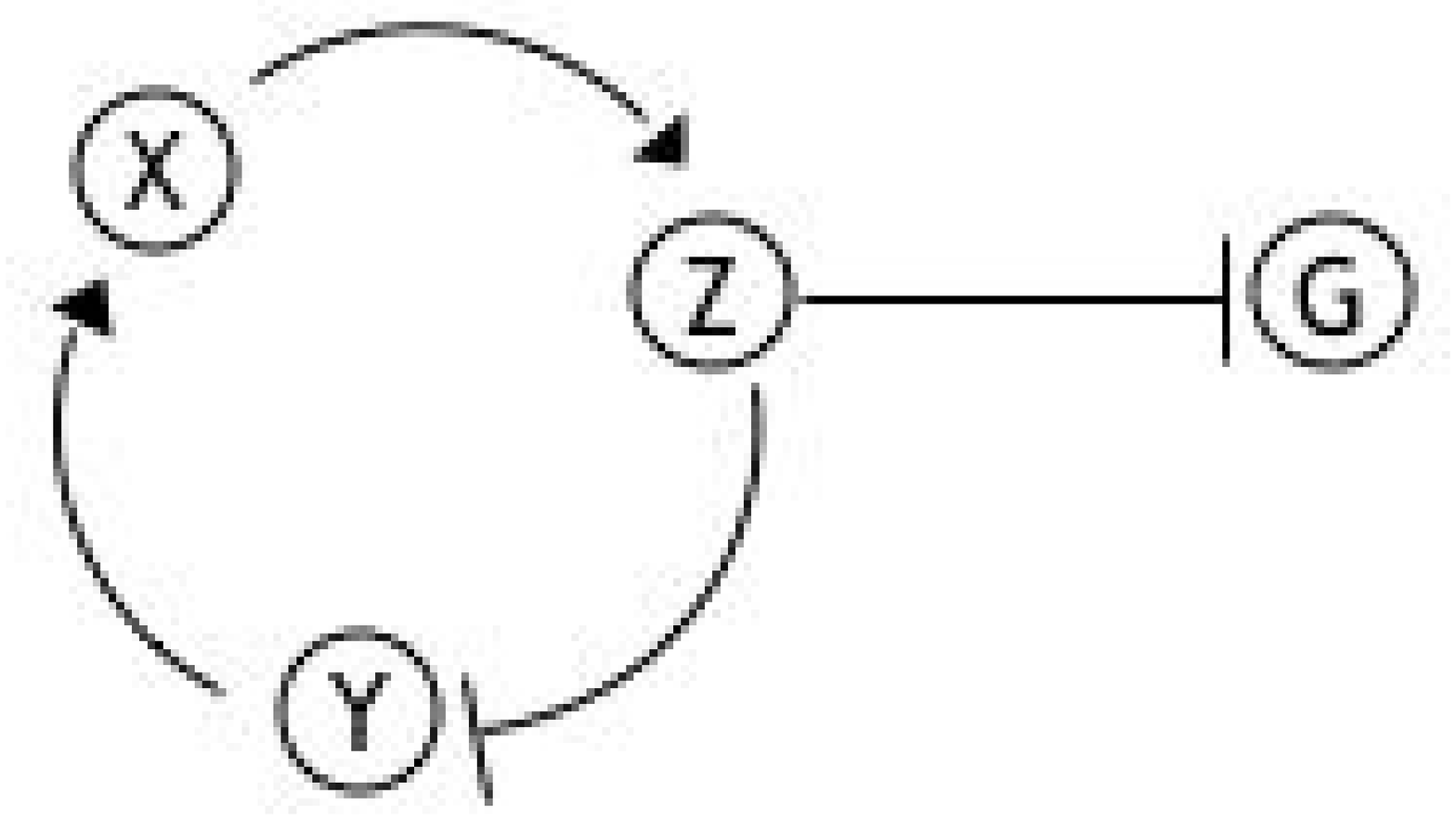} } & {\includegraphics[width=1.1in] {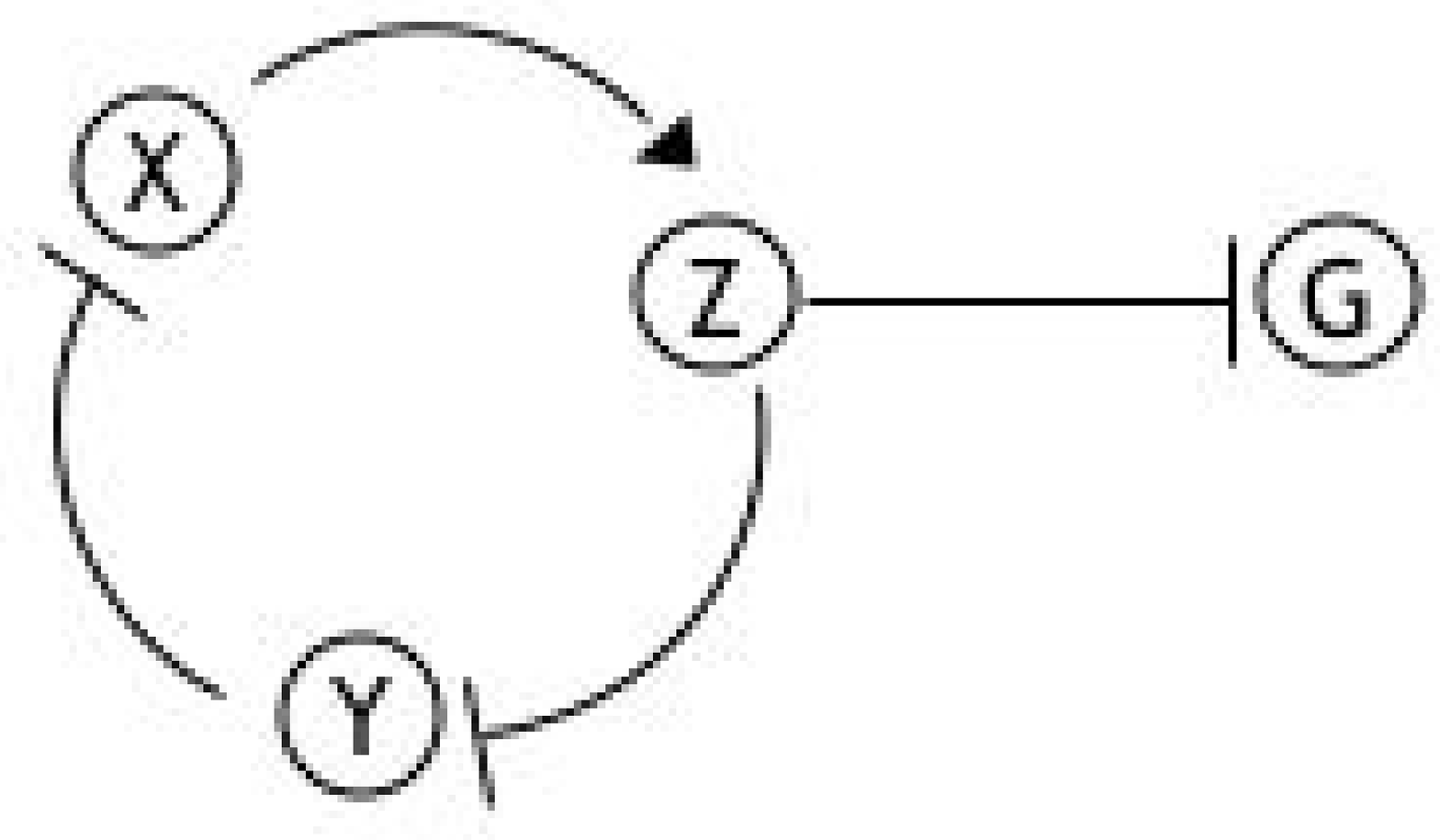} }\\
 & \\ 
{\includegraphics[width=3in] {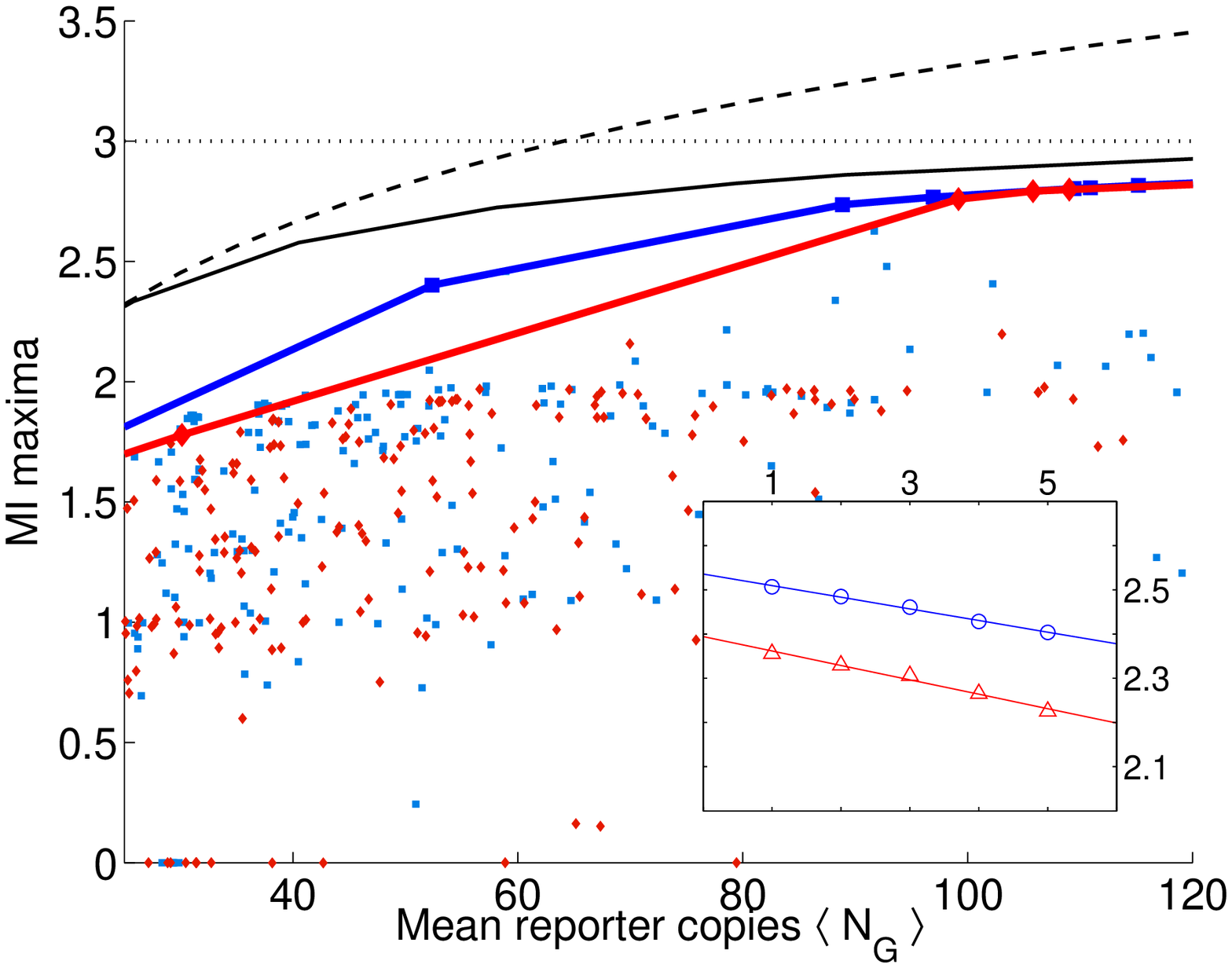} } & {\includegraphics[width=3in] {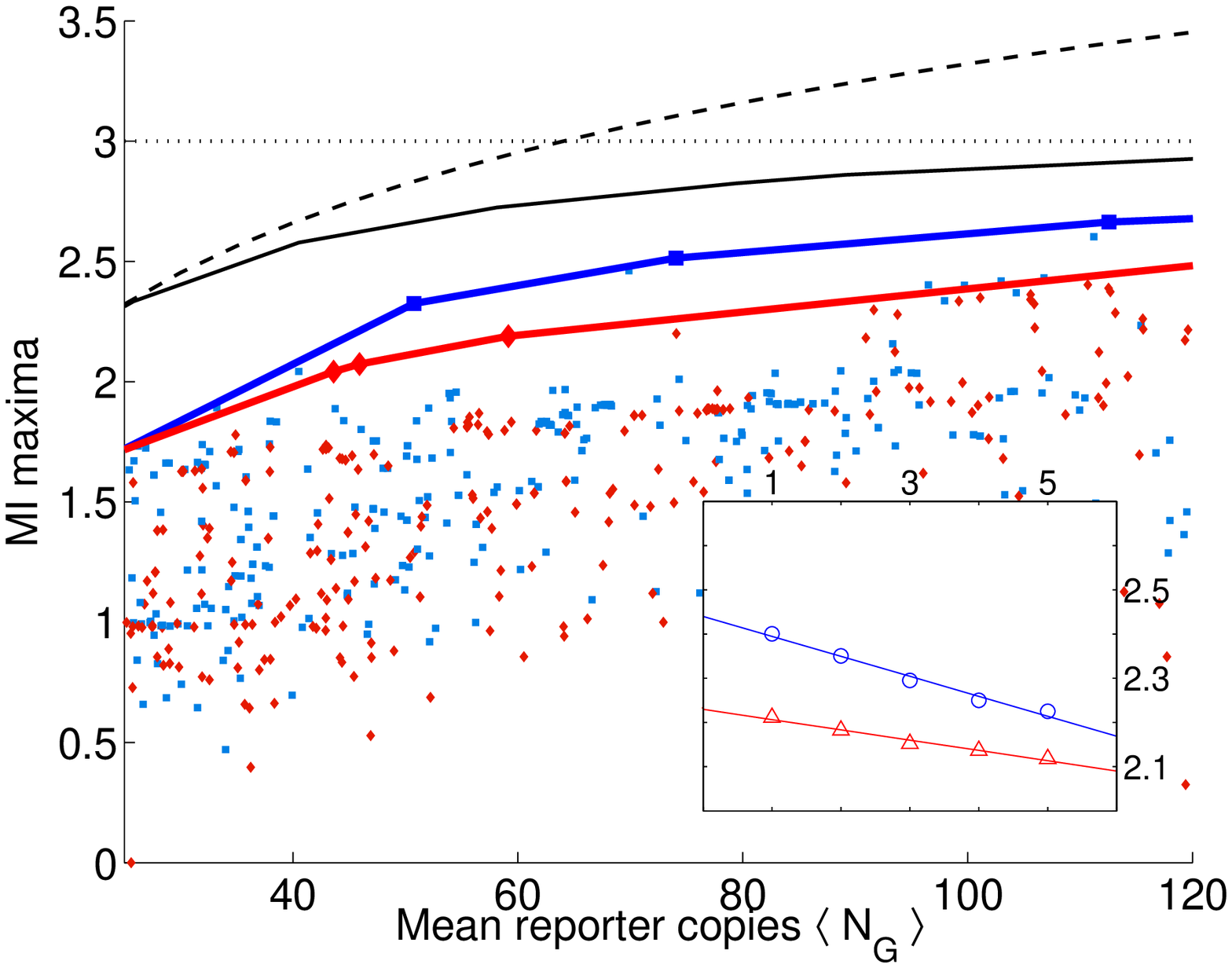} }\\

\label{ivn_all}
\end{longtable}
\end{center}

 \begin{table}[h]
\begin{center}
\small{
\begin{tabular}{|c|c|c|c|}
\hline
\bf{Number}&\bf{Topology}&\bf{$\gamma=0.001$}&\bf{$\gamma=0.01$}\\ \hline\hline
1&
\includegraphics[width=1.1 in] {top1.eps}
\T\B
&2.5570&2.3638\\
\hline
20&
\includegraphics[width=1 in] {top20.eps}
\T\B
&2.5524&2.3970\\
\hline
6&
\T \includegraphics[width=.8in] {top6.eps}
\B&2.5451&2.4818\\
\hline
2&
\includegraphics[width=1 in] {top2.eps}
\T\B
&2.5357&2.3549\\
\hline
22&
\includegraphics[width=.8 in] {top22.eps}
\T\B
&2.5354&2.3909\\
\hline
19&
\includegraphics[width=1.1 in] {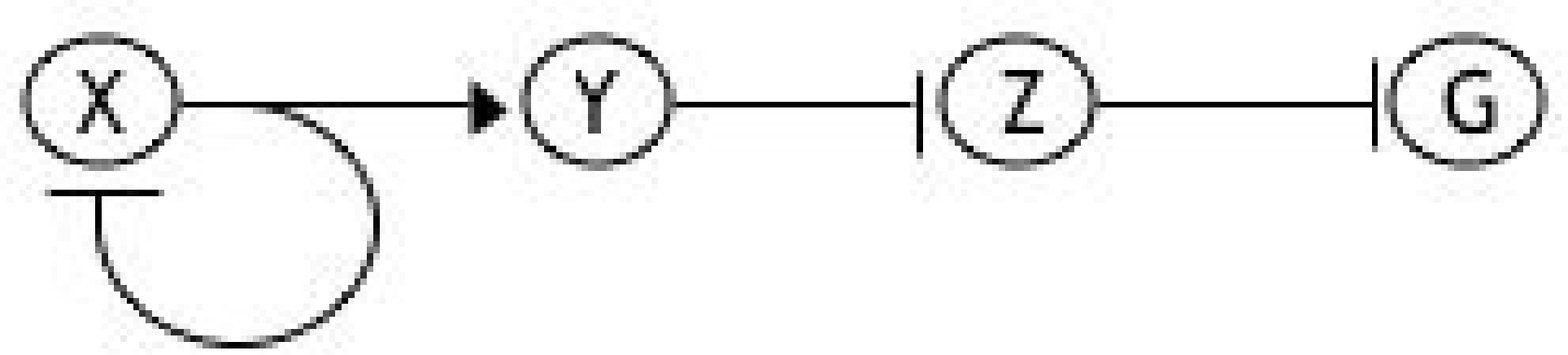}
\T\B
&2.5218&2.3718\\
\hline
10&
\includegraphics[width=.8 in] {top10.eps}
\T\B
&2.5172&2.3925\\
\hline
13&
\includegraphics[width=.8 in] {top13.eps}
\T\B
&2.5055&2.4058\\
\hline
8&
\includegraphics[width=.8 in] {top8.eps}
\T\B
&2.5002&2.3463\\
 \hline
23&
\includegraphics[width=1 in] {top23.eps}
\T\B
&2.4976&2.3831\\
\hline
14&
\includegraphics[width=1.1 in] {top14.eps}
\T\B
&2.4874&2.4251\\
\hline
12&
\includegraphics[width=1 in] {top12.eps}
 \T\B
 &2.4809&2.3219\\
\hline\hline
\end{tabular}
}
\end{center}
\caption{Table of Circuits (top $12$). Extrapolated average mutual information over range of $25$ to $120$ molecules at $\gamma=0.001$ and $\gamma=0.01$. }
\label{table2}
\end{table}
\newpage
\begin{table}[h]

\begin{center}
\small{
\begin{tabular}{| c | c | c |c |}
\hline
\bf{Number}&\bf{Circuit}&\bf{$\gamma=0.001$}&\bf{$\gamma=0.01$}\\ \hline\hline

17&
\includegraphics[width=.8 in] {top17.eps}
\T\B
&2.4695&2.2876\\
\hline

5&
\T \includegraphics[width=1 in] {top5.eps} 
\B &2.4659&2.2806\\

\hline

4&
\T \includegraphics[width=1.1 in] {top4.eps}
 \B &2.4624&2.2930\\

\hline

21&
\T\includegraphics[width=1.1 in] {top21.eps}
\B &2.4605&2.23121\\

\hline

9&
\T \includegraphics[width=1 in] {top9.eps}
\B&2.4497&2.3491\\

\hline

7&
 \T \includegraphics[width=1 in] {top7.eps}
\B&2.4420&2.2773\\

\hline

15&
\T \includegraphics[width=1.1 in] {top15.eps}
\B&2.4244&2.1587\\

\hline

24&
\T \includegraphics[width=.8 in] {top24.eps}
 \B&2.4234&2.2123\\

\hline

11&
\T \includegraphics[width=1.1 in] {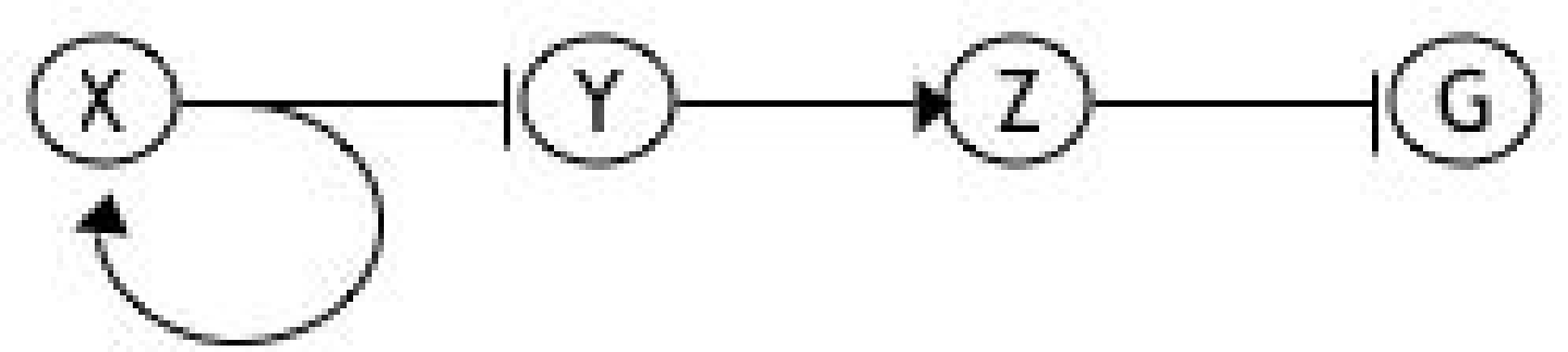}
\B&2.3958&2.2143\\

\hline

16&
\T \includegraphics[width=1.1 in] {top16.eps}
\B&2.3943&2.2281\\

\hline

18&
\T \includegraphics[width=1 in] {top18.eps}
\B&2.3603&2.0751\\

\hline

3&
\T \includegraphics[width=.8 in] {top3.eps}
\B&2.3099&2.2471\\

\hline\hline
\end{tabular}
}
\end{center}
\caption{Table of Circuits (bottom $12$). Extrapolated average mutual information over range of $25$ to $120$ molecules at $\gamma=0.001$ and $\gamma=0.01$. }
\label{table1}
\end{table}

\bibliographystyle{plain}
\bibliography{bsp_arxiv}

\end{document}